# *Constructive Galois Connections*

DAVID DARAIS
University of Vermont, USA
(*e-mail:* `David.Darais@uvm.edu`)

DAVID VAN HORN
University of Maryland, College Park, USA
(*e-mail:* `dvanhorn@cs.umd.edu`)

**Abstract**

Galois connections are a foundational tool for structuring abstraction in semantics and their use lies at the heart of the theory of abstract interpretation. Yet, mechanization of Galois connections using proof assistants remains limited to restricted modes of use, preventing their general application in mechanized metatheory and certified programming.

This paper presents *constructive Galois connections*, a variant of Galois connections that is effective both on paper and in proof assistants; is complete with respect to a large subset of classical Galois connections; and enables more general reasoning principles, including the "calculational" style advocated by Cousot.

To design constructive Galois connections we identify a restricted mode of use of classical ones which is both general and amenable to mechanization in dependently-typed functional programming languages. Crucial to our metatheory is the addition of monadic structure to Galois connections to control a "specification effect." Effectful calculations may reason classically, while pure calculations have extractable computational content. Explicitly moving between the worlds of specification and implementation is enabled by our metatheory.

To validate our approach, we provide two case studies in mechanizing existing proofs from the literature: the first uses calculational abstract interpretation to design a static analyzer; the second forms a semantic basis for gradual typing. Both mechanized proofs closely follow their original paper-and-pencil counterparts, employ reasoning principles not captured by previous mechanization approaches, support the extraction of verified algorithms, and are novel.

## 1 Introduction

Abstract interpretation is a general theory of sound approximation widely applied in programming language semantics, formal verification, and static analysis (Cousot & Cousot, 1976, 1977, 1979, 1992, 2014). In abstract interpretation, properties of programs are related between a pair of partially ordered sets: a concrete domain, $\langle \mathcal{C}, \sqsubseteq \rangle$, and an abstract domain, $\langle \mathcal{A}, \preceq \rangle$. When concrete properties have a $\preceq$-most precise abstraction, the correspondence is a *Galois connection*, formed by a pair of mappings between the domains known as *abstraction* $\alpha \in \mathcal{C} \mapsto \mathcal{A}$ and *concretization* $\gamma \in \mathcal{A} \mapsto \mathcal{C}$ such that $c \sqsubseteq \gamma(a) \iff \alpha(c) \preceq a$. Since its introduction by Cousot and Cousot in the late 1970s, this theory has formed the basis of static analyzers, type systems, model-checkers, obfuscators, program transformations, and many more applications (Cousot, 2008).

Given the remarkable set of tools contributed by this theory, an obvious desire is to incorporate its use in proof assistants to mechanically verify proofs by abstract interpretation. When embedded in a proof assistant, verified algorithms such as static analyzers can then be extracted from these proofs.

Monniaux first achieved the goal of mechanization for the theory of abstract interpretation with Galois connections in Coq (1998). However, he notes that the abstraction side ($\alpha$) of Galois connections is problematic since it requires the admission of non-constructive axioms. Use of these axioms prevents the extraction of certified programs. So while Monniaux was able to mechanically verify proofs by abstract interpretation in its full generality, certified artifacts could not be extracted in general.

Pichardie subsequently tackled the extraction problem by using a restricted formulation of abstract interpretation that only relies on the concretization ($\gamma$) side of Galois connections (2005). Doing so avoids the use of axioms and enables extraction of certified artifacts. This technique is effective and has been used to construct certified static analyzers (Pichardie, 2005; Cachera & Pichardie, 2010; Blazy *et al.*, 2013; Barthe *et al.*, 2007), most notably the Verasco static analyzer, part of the CompCert C compiler (Jourdan *et al.*, 2015; Leroy, 2009). Unfortunately, this approach sacrifices the full generality of the theory. While in principle the technique could achieve mechanization of existing soundness *theorems*, it cannot do so faithful to existing *proofs*. In particular, Pichardie writes (2005, p. 55):[1]

> The framework we have retained nevertheless loses an important property of the standard framework: being able to derive a correct approximation $f^\sharp$ from the specification $\alpha \circ f \circ \gamma$. Several examples of such derivations are given by Cousot (1999). It seems interesting to find a framework for this kind of symbolic manipulation, while remaining easily formalizable in Coq.

This important property is the so-called "calculational" style, whereby an abstract interpreter ($f^\sharp$) is derived in a correct-by-construction manner from a concrete interpreter ($f$) composed with abstraction and concretization ($\alpha \circ f \circ \gamma$). This calculational method detailed in Cousot's monograph (1999), which concludes:

> The emphasis in these notes has been on the correctness of the design by calculus. The mechanized verification of this formal development using a proof assistant can be foreseen with automatic extraction of a correct program from its correctness proof.

In the subsequent 17 years, this vision has remained unrealized, and clearly the paramount technical challenge in achieving it is obtaining both *generality* and *constructivity* in a single framework.

[1] Translated from French by the present authors.

This paper contributes *constructive Galois connections*, a framework for mechanized abstract interpretation with Galois connections that achieves both generality and constructivity, thereby enabling calculational style proofs which make use of both abstraction ($\alpha$) and concretization ($\gamma$), while also maintaining the ability to extract certified static analyzers.

We develop constructive Galois connections from the insight that many classical Galois connections used in practice are of a particular restricted form, which is reminiscent of a direct-style verification. Constructive Galois connections are the general abstraction theory for this restricted setting and can be mechanized effectively.

Our constructive Galois connections consist of analogs to abstraction and concretization, which we call *extraction* and *interpretation* and notate $\eta$ and $\mu$. Whereas classical Galois connections map between posets $\alpha \in \mathcal{C} \mapsto \mathcal{A}$ and $\gamma \in \mathcal{A} \mapsto \mathcal{C}$, constructive Galois connections differ only in that they carry a powerset on the codomain of interpretation, so $\eta \in \mathcal{C} \mapsto \mathcal{A}$ and $\mu \in \mathcal{A} \mapsto \wp(\mathcal{C})$. This simple change supports all of the benefits of abstract interpretation with classical Galois connections, while also supporting mechanized verification of executable algorithms.

We observe that constructive Galois connections contain monadic structure which isolates classical specifications from constructive algorithms. Within the effectful fragment, all of classical Galois connection reasoning can be employed, while within the pure fragment, functions must carry computational content. Remarkably, calculations can move between these modalities and verified programs may be extracted from the end result of calculation.

To support the utility of our theory we build a library for constructive Galois connections in Agda (Norell, 2007) and mechanize two existing abstract interpretation proofs from the literature. The first is drawn from Cousot's monograph (1999), which derives a correct-by-construction analyzer from a specification induced by a concrete interpreter and Galois connection. The second is drawn from Garcia, Clark and Tanter's "Abstracting Gradual Typing" (2016), which uses abstract interpretation to derive static and dynamic semantics for gradually typed languages from traditional static types. Both proofs use the "important property of the standard framework" identified by Pichardie, which is not handled by prior mechanization approaches. The mechanized proofs closely follow the original pencil-and-paper proofs, which use both abstraction and concretization, while still enabling the extraction of certified algorithms. Neither of these papers have been previously mechanized. Moreover, we know of no existing mechanized proof involving calculational abstract interpretation.

Next, we make precise the relationship between constructive Galois connections and classical Galois connections, and prove them sound and complete. These metatheory results are also mechanized; claims are marked with "AGDA✓" whenever they are proved in Agda. (All claims are marked.)

Finally, we explore the relationship between classical and constructive Galois connections in much more detail. We do this through defining constructive analogs to classical Galois connection primitives and connectives, and through two examples drawn from our first case study worked out in full detail. In this part of the paper, we compare and contrast the differences between abstraction-directed and concretization-directed calculations, and between sound and complete calculations, for both classical and constructive styles. The outcome of this study is a better understanding of how constructive calculations interact with classical Galois connections, how the mechanics of optimality change between each framework, and how to calculate multivalued algorithms in the constructive setting.

**Contributions** This paper contributes the following:

- A foundational theory of constructive Galois connections which is both general and amenable to mechanization using a dependently typed functional programming language;
- A proof library and two case studies from the literature for mechanized abstract interpretation; and
- The first mechanization of calculational abstract interpretation; and
- A detailed discussion of the relationship between constructive and classical Galois connections, and their interaction.

Relative to Darais & Van Horn (2016), we have expanded the description of constructive Galois connections (Section 3) and the second case study (Section 5), created a new section which provides details about the mechanization (Section 6), created four new sections which discuss the relationship between constructive and classical Galois connections (Sections 8, 9, 10 and 11), and created a new short section which discusses perspectives on foundations and connections to category theory (Section 13).

The remainder of the paper is organized as follows. First we give a tutorial on verifying a simple analyzer from two different perspectives: direct verification (§ 2.1) and abstract interpretation with Galois connections (§ 2.2), highlighting mechanization issues along the way. We then present constructive Galois connections as a marriage of the two approaches (§ 3). We provide two case studies: the mechanization of an abstract interpreter from Cousot's calculational monograph (§ 4), and the mechanization of Garcia, Clark and Tanter's work on gradual typing *via* abstract interpretation (§ 5). Next we discuss Agda-specific details of our mechanization framework (§ 6). Next, we formalize the metatheory of constructive Galois connections (§ 7), define constructive analogs of common classical Galois connection primitives and connectives (§ 8), and work through two extended examples in detail: the first to compare and contrast calculation styles (§ 9) and discuss deriving optimal interpreters (§ 10), and the second to explore multivalued constructive calculations (§ 11). Finally, we relate our work to the literature (§ 12), share perspectives on foundations (§ 13), and conclude (§ 14).

## 2 Verifying a Simple Static Analyzer

In this section we contrast two perspectives on verifying a static analyzer: using a direct approach, and using the theory of abstract interpretation with Galois connections. The direct approach is simple but lacks the benefits of a general abstraction framework. Abstract interpretation provides these benefits, but at the cost of added complexity and resistance to mechanized verification. In Section 3 we present an alternative perspective: abstract interpretation with *constructive* Galois connections—the topic of this paper. Constructive Galois connections marry the worlds presented in this section, providing the simplicity of direct verification, the benefits of a general abstraction framework, and support for mechanized verification.

To demonstrate both verification perspectives we design a parity analyzer in each style. For example, a parity analysis discovers that 2 has parity EVEN, $succ(1)$ has parity EVEN, and $n+n$ has parity EVEN if $n$ has parity ODD. Rather than sketch the high-level details of a complete static analyzer, we instead zoom into the low-level details of a tiny fragment: analyzing the successor arithmetic operation $succ(n)$. At this level of detail the differences, advantages and disadvantages of each approach become apparent.

### *2.1 The Direct Approach*

Using the direct approach to verification, one first designs the analyzer, then defines what it means for the analyzer to be sound, and finally completes a proof of soundness. Each step is done from scratch, and in the simplest way possible.

This approach should be familiar to most readers, and exemplifies how most researchers approach formalizing soundness for static analyzers: first posit the analyzer and soundness framework, then attempt the proof of soundness. One limitation of this approach is that the setup—which gives lots of room for error—isn't known to be correct until after completing the final proof. However, benefits of this approach are that it is simple and can easily be mechanized.

**Analyzing Successor** A parity analysis answers questions like: "what is the parity of $succ(n)$, given that $n$ is even?" To answer these questions, imagine replacing $n$ with the symbol EVEN, a stand-in for an arbitrary even number. This hypothetical expression $succ(\text{EVEN})$ is interpreted by defining a successor function over parities, rather than numbers, which we call $succ^\sharp$. This successor operation on parities is designed such that if $p$ is the parity for $n$, $succ^\sharp(p)$ will be the parity of $succ(n)$:

$$\mathbb{P} := \{\text{EVEN}, \text{ODD}\} \qquad succ^\sharp(\text{EVEN}) := \text{ODD}$$
$$succ^\sharp : \mathbb{P} \to \mathbb{P} \qquad succ^\sharp(\text{ODD}) := \text{EVEN}$$

**Soundness** The soundness of $succ^\sharp$ is defined using an interpretation for parities, which we notate $[\![p]\!]$:

$$[\![_]\!] : \mathbb{P} \to \wp(\mathbb{N}) \qquad \begin{aligned} [\![\text{EVEN}]\!] &:= \{n \mid even(n)\} \\ [\![\text{ODD}]\!] &:= \{n \mid odd(n)\} \end{aligned}$$

Given this interpretation, a parity $p$ is a valid analysis result for a number $n$ if the interpretation for $p$ contains $n$, that is $n \in [\![p]\!]$. The analyzer $succ^\sharp(p)$ is then sound if, when $p$ is a valid analysis result for some number $n$, $succ^\sharp(p)$ is a valid analysis

result for $succ(n)$:

$$n \in [\![p]\!] \implies succ(n) \in [\![succ^\sharp(p)]\!] \qquad \text{(DA-Snd)}$$

The proof is by case analysis on $p$; we show the case $p = \text{EVEN}$:

$$\begin{aligned}
& n \in [\![\text{EVEN}]\!] \\
\Leftrightarrow\ & even(n) && \wr \text{ defn. of } [\![_]\!] \ \int \\
\Leftrightarrow\ & odd(succ(n)) && \wr \text{ defn. of } even/odd \ \int \\
\Leftrightarrow\ & succ(n) \in [\![\text{ODD}]\!] && \wr \text{ defn. of } [\![_]\!] \ \int \\
\Leftrightarrow\ & succ(n) \in [\![succ^\sharp(\text{EVEN})]\!] && \wr \text{ defn. of } succ^\sharp \ \int
\end{aligned}$$

**An Even Simpler Setup** There is another way to define and prove soundness: use a function which computes the parity of a number in the definition of soundness. This approach is even simpler, and will foreshadow the constructive Galois connection setup.

$$parity : \mathbb{N} \to \mathbb{P} \qquad \begin{aligned} parity(0) &:= \text{EVEN} \\ parity(succ(n)) &:= flip(parity(n)) \end{aligned}$$

where $flip(\text{EVEN}) := \text{ODD}$ and $flip(\text{ODD}) := \text{EVEN}$. This gives an alternative and equivalent way to relate a number and a parity, due to the following correspondence:

$$n \in [\![p]\!] \iff parity(n) = p \qquad \text{(DA-Corr)}$$

The soundness of the analyzer is then restated:

$$parity(n) = p \implies parity(succ(n)) = succ^\sharp(p)$$

or by substituting $parity(n) = p$:

$$parity(succ(n)) = succ^\sharp(parity(n)) \qquad \text{(DA-Snd*)}$$

Both this statement for soundness and its proof are simpler than before. The proof follows directly from the definition of *parity* and the fact that $succ^\sharp$ is identical to *flip*.

**The Main Idea** Correspondences like (DA-Corr)—between an interpretation for analysis results ($[\![p]\!]$) and a function which computes analysis results ($parity(n)$)—are central to the constructive Galois Connection framework we will describe in Section 3. Using correspondences like these, we build a general theory of abstraction that recovers this direct approach to verification, mirrors all of the benefits of abstract interpretation with classical Galois connections, supports mechanized verification, and in some cases simplifies the proof effort. We also observe that many classical Galois connections used in practice can be ported to this simpler setting.

**Mechanized Verification** This direct approach to verification is amenable to mechanization using proof assistants like Coq and Agda. These tools are founded on constructive logic in part to support verified program extraction. In constructive logic, functions $f : A \to B$ are computable and often defined by primitive recursion over inductively defined datatypes to ensure they can be extracted and executed as programs. Analogously, propositions $P : \wp(A)$ are encoded constructively as potentially undecidable predicates $P : A \to prop$ where $x \in P \Leftrightarrow P(x)$.

To mechanize the verification of $succ^\sharp$ we first translate its definition to a constructive setting unmodified. Next we translate $[\![p]\!]$ to a relation $I(p,n)$ defined inductively *via* inference rules:

$$\frac{}{I(\textsc{even},0)} \qquad \frac{I(p,n)}{I(flip(p),succ(n))}$$

The mechanized proof of (DA-Snd) using $I$ is analogous to the one we sketched, and the mechanized proof of (DA-Snd*) follows directly by computation. The proof term for (DA-Snd*) in both Coq and Agda is simply `refl`, the reflexivity judgment for syntactic equality modulo computation in constructive logic.

**Wrapping Up** Each approach to verification we will present is distinguished by which parts are postulated and which parts are derived. Using the direct approach, the analysis ($succ^\sharp$), the interpretation for parities ($[\![p]\!]$) and the definition of soundness (DA-Snd) are all postulated up-front. When the soundness setup is correct but the analyzer is wrong, the proof at the end will not go through and the analyzer must be redesigned. Even worse, when the soundness setup and the analyzer are both wrong, the proof might actually succeed, giving a false assurance in the soundness of the analyzer. However, the direct approach is attractive because it is simple and supports mechanized verification.

### *2.2 Classical Abstract Interpretation*

To verify an analyzer using abstract interpretation with Galois connections, one first designs *abstraction* and *concretization* mappings between sets $\mathbb{N}$ and $\mathbb{P}$. These mappings are used to synthesize an optimal specification for $succ^\sharp$. One then proves that a postulated $succ^\sharp$ meets this synthesized specification, or alternatively derives the definition of $succ^\sharp$ directly from the optimal specification.

In contrast to the direct approach, rather than design the definition of *soundness*, one instead designs the definition of *abstraction* within a structured framework. Soundness is then not designed, rather it is derived from the definition of abstraction. Finally, there is added boilerplate in the abstract interpretation approach, which requires lifting definitions, specifications and proofs to powersets $\wp(\mathbb{N})$ and $\wp(\mathbb{P})$.

**Abstracting Sets** Powersets are introduced in abstraction and concretization functions to support relational mappings, like mapping the symbol EVEN to the set of all even numbers. The mappings are therefore between *powersets* $\wp(\mathbb{N})$ and $\wp(\mathbb{P})$. The abstraction and concretization mappings must also satisfy correctness criteria, detailed below, at which point they are called a *Galois connection*.

The abstraction mapping from $\wp(\mathbb{N})$ to $\wp(\mathbb{P})$ is notated $\alpha$, and is defined as the pointwise lifting of $parity(n)$:

$$\alpha : \wp(\mathbb{N}) \to \wp(\mathbb{P}) \qquad \alpha(N) := \{parity(n) \mid n \in N\}$$

The concretization mapping from $\wp(\mathbb{P})$ to $\wp(\mathbb{N})$ is notated $\gamma$, and is defined as the flattened pointwise lifting of $[\![p]\!]$:

$$\gamma : \wp(\mathbb{P}) \to \wp(\mathbb{N}) \qquad \gamma(P) := \{n \mid p \in P \wedge n \in [\![p]\!]\}$$

The correctness criteria for $\alpha$ and $\gamma$ is the following correspondence:

$$N \subseteq \gamma(P) \iff \alpha(N) \subseteq P \tag{GC-Corr}$$

The correspondence means that, to relate elements of different sets—in this case $\wp(\mathbb{N})$ and $\wp(\mathbb{P})$—it is equivalent to relate them through either $\alpha$ or $\gamma$. Mappings like $\alpha$ and $\gamma$ which share this correspondence are called Galois connections.

An equivalent correspondence to (GC-Corr) is two laws relating compositions of $\alpha$ and $\gamma$, called *expansive* and *reductive*:

$$N \subseteq \gamma(\alpha(N)) \tag{GC-Exp}$$

$$\alpha(\gamma(P)) \subseteq P \tag{GC-Red}$$

Property (GC-Red) ensures $\alpha$ is the best abstraction possible w.r.t. $\gamma$. For example, a hypothetical definition $\alpha(N) := \{\text{EVEN}, \text{ODD}\}$ is expansive but not reductive with respect to $\gamma$ as defined above because $\alpha(\gamma(\{\text{EVEN}\})) \not\subseteq \{\text{EVEN}\}$.

In general, Galois connections are defined for arbitrary posets $\langle C, \sqsubseteq^C\rangle$ and $\langle A, \sqsubseteq^A\rangle$. The correspondence (GC-Corr) and its expansive/reductive variants are generalized in this setting to use partial orders $\sqsubseteq^C$ and $\sqsubseteq^A$ instead of subset ordering. We are omitting monotonicity requirements for $\alpha$ and $\gamma$ at this point in our presentation, although these requirements are essential in the complete approach. Our example instantiates this general framework with powersets $\wp(\mathbb{N})$ and $\wp(\mathbb{P})$ in place of $C$ and $A$, and the subset operation $\subseteq$ in place of $\sqsubseteq^C$ and $\sqsubseteq^A$. Although Galois connections are often instantiated with powersets (typically for the concrete domain, and sometimes also for the abstract domain, as in our example), this need not always be the case.

**Powerset Lifting** The original functions *succ* and $succ^\sharp$ cannot be related through $\alpha$ and $\gamma$ because they are not functions between powersets. To remedy this they are lifted pointwise:

$$\uparrow succ : \wp(\mathbb{N}) \to \wp(\mathbb{N}) \qquad \uparrow succ(N) := \{succ(n) \mid n \in N\}$$
$$\uparrow succ^\sharp : \wp(\mathbb{P}) \to \wp(\mathbb{P}) \qquad \uparrow succ^\sharp(P) := \{succ^\sharp(p) \mid p \in P\}$$

These lifted operations are called the *concrete interpreter* and *abstract interpreter*, because the former operates over the *concrete domain* $\wp(\mathbb{Z})$ and the latter over the *abstract domain* $\wp(\mathbb{P})$. In the framework of abstract interpretation, static analyzers are just abstract interpreters. Lifting *succ* and $succ^\sharp$ to powersets is necessary to use the abstract interpretation framework because the abstraction and concretization functions ($\alpha$ and $\gamma$) are defined as mappings between powersets. This has the negative effect of adding boilerplate to definitions and proofs of soundness.

**Soundness** The definition of soundness for $succ^\sharp$ is synthesized by relating $\uparrow succ^\sharp$ to $\uparrow succ$ composed with $\alpha$ and $\gamma$:

$$\alpha(\uparrow succ(\gamma(P))) \subseteq \uparrow succ^\sharp(P) \qquad \text{(GC-Snd)}$$

The left-hand side of the ordering is an optimal specification for any abstraction of $\uparrow succ$ (a consequence of (GC-Corr)), and the subset ordering says $\uparrow succ^\sharp$ is an over-approximation of this optimal specification. The reason to over-approximate is because the specification is a mathematical description, and the abstract interpreter is usually expected to be an algorithm, and there may not always exist an algorithm which can match the specification precisely. The proof of (GC-Snd) is by case analysis on $P$. We do not show the proof, rather we demonstrate a proof later in this section which also synthesizes the definition of $succ^\sharp$.

One advantage of the abstract interpretation framework is that it provides a choice between four soundness properties, all of which are equivalent and generated by $\alpha$ and $\gamma$:

$$\alpha(\uparrow succ(\gamma(P))) \subseteq \uparrow succ^\sharp(P) \qquad \text{(GC-Snd/}\alpha\gamma\text{)}$$
$$\uparrow succ(\gamma(P)) \subseteq \gamma(\uparrow succ^\sharp(P)) \qquad \text{(GC-Snd/}\gamma\gamma\text{)}$$
$$\alpha(\uparrow succ(N)) \subseteq \uparrow succ^\sharp(\alpha(N)) \qquad \text{(GC-Snd/}\alpha\alpha\text{)}$$
$$\uparrow succ(N) \subseteq \gamma(\uparrow succ^\sharp(\alpha(N))) \qquad \text{(GC-Snd/}\gamma\alpha\text{)}$$

Because each soundness property is equivalent (a consequence of GC-Corr), one can choose whichever variant is easiest to prove. The soundness setup (GC-Snd) is the $\alpha\gamma$ rule, however any of the other rules can also be used. For example, one could choose $\alpha\alpha$ or $\gamma\alpha$; in these cases the proof considers four disjoint cases for $N$: $N$ is empty, $N$ contains only even numbers, $N$ contains only odd numbers, and $N$ contains both even and odd numbers.

**Completeness** The mappings $\alpha$ and $\gamma$ also synthesize an *optimality* statement for $\uparrow succ^\sharp$, by stating that it *under*-approximates the optimal specification:

$$\alpha(\uparrow succ(\gamma(P))) \supseteq \uparrow succ^\sharp(P)$$

Typically we are only interested in sound abstract interpreters, which are those that over-approximate the optimal specification. A sound and optimal interpreter is then one that both over-approximates ($\subseteq$) *and* under-approximates ($\supseteq$) the optimal specification, which is equivalent to being equal to it. For this reason, we re-state optimality as an *equality* between the abstract interpreter and the optimal specification:

$$\alpha(\uparrow succ(\gamma(P))) = \uparrow succ^\sharp(P) \qquad \text{(GC-Opt)}$$

Not all analyzers are optimal, however optimality helps identify those which approximate too much. Consider the analyzer $\uparrow succ^{\sharp\prime}$:

$$\uparrow succ^{\sharp\prime} \;:\; \wp(\mathbb{P}) \to \wp(\mathbb{P}) \qquad\qquad \uparrow succ^{\sharp\prime}(P) \;\coloneqq\; \{\textsc{even}, \textsc{odd}\}$$

This analyzer reports that $succ(n)$ could have any parity regardless of the parity for $n$; it's the analyzer that always says "I don't know." This analyzer is perfectly sound but non-optimal because $\uparrow succ^{\sharp\prime}(\{\textsc{even}\}) = \{\textsc{even}, \textsc{odd}\} \neq \alpha(\uparrow succ(\gamma(\{\textsc{even}\})))$.

Just like soundness, four completeness statements are generated by $\alpha$ and $\gamma$, however the following statements are *not* all equivalent:

$$\text{[optimal]} \qquad \alpha(\uparrow succ(\gamma(P))) = \uparrow succ^{\sharp}(P) \qquad \text{(GC-Cmp}/\alpha\gamma)$$

$$\uparrow succ(\gamma(P)) = \gamma(\uparrow succ^{\sharp}(P)) \qquad \text{(GC-Cmp}/\gamma\gamma)$$

$$\alpha(\uparrow succ(N)) = \uparrow succ^{\sharp}(\alpha(N)) \qquad \text{(GC-Cmp}/\alpha\alpha)$$

$$\text{[precise]} \qquad \uparrow succ(N) = \gamma(\uparrow succ^{\sharp}(\alpha(N))) \qquad \text{(GC-Cmp}/\gamma\alpha)$$

Abstract interpreters which satisfy the $\alpha\gamma$ variant are called *optimal* because they lose no more information than necessary, and those which satisfy the $\gamma\alpha$ variant are called *precise* because they lose no information *at all*. The abstract interpreter $succ^{\sharp}$ is optimal, but not precise because $\gamma(\uparrow succ^{\sharp}(\alpha(\{1\}))) \neq \uparrow succ(\{1\})$

To overcome mechanization issues with Galois connections, the state-of-the-art is restricted to use $\gamma\gamma$ rules only for soundness (GC-Snd/$\gamma\gamma$) and completeness (GC-Cmp/$\gamma\gamma$). This is unfortunate for completeness properties because unlike soundness, each completeness variant is not equivalent.

**Calculational Derivation of Abstract Interpreters** Rather than posit $\uparrow succ^{\sharp}$ and prove it correct directly, one can instead derive its definition through a calculational process. The process begins with the optimal specification on the left-hand-side of (GC-Opt), and reasons equationally towards the definition of an algorithm. In this way, $\uparrow succ^{\sharp}$ is not postulated, rather it is derived by calculation, and the result is both sound and optimal by construction.

The derivation is by case analysis on $P$ which has four cases: $\{\}$, $\{\textsc{even}\}$, $\{\textsc{odd}\}$ and $\{\textsc{even}, \textsc{odd}\}$; we show $P = \{\textsc{even}\}$:

$$\begin{aligned}
&\alpha(\uparrow succ(\gamma(\{\textsc{even}\}))) \\
&= \alpha(\uparrow succ(\{n \mid even(n)\})) && \wr\ \text{defn. of } \gamma\ \int \\
&= \alpha(\{succ(n) \mid even(n)\}) && \wr\ \text{defn. of } \uparrow succ\ \int \\
&= \alpha(\{n \mid odd(n)\}) && \wr\ \text{defn. of } even/odd\ \int \\
&= \{\textsc{odd}\} && \wr\ \text{defn. of } \alpha\ \int \\
&\triangleq \uparrow succ^{\sharp}(\{\textsc{even}\}) && \wr\ \text{defining } \uparrow succ^{\sharp}\ \int
\end{aligned}$$

The derivations for the other cases are analogous, and together they define the implementation of $\uparrow succ^{\sharp}$.

Deriving analyzers by calculus is attractive because it is systematic, and because it prevents the issue where an analyzer is postulated and discovered to be unsound only after failing to complete its soundness proof. However, this calculational style of abstract interpretation is not amenable to mechanized verification with program extraction because $\alpha$ is often non-constructive, an issue we describe later in this section.

**Added Complexity** The abstract interpretation approach requires a Galois connection up-front which necessitates the introduction of powersets $\wp(\mathbb{N})$ and $\wp(\mathbb{P})$. This results in powerset-lifted definitions and adds boilerplate set-theoretic reasoning to the proofs.

This is in contrast to the direct approach which never mentions powersets of parities. Not using powersets results in more understandable soundness criteria, requires no boilerplate set-theoretic reasoning, and results in fewer cases for the proof of soundness. This boilerplate becomes magnified in a mechanized setting where all details must be spelled out to a proof assistant. Furthermore, the simpler proof of (DA-Snd*)—which was immediate from the definition of *parity*—cannot be recovered within the general abstract interpretation framework, rather it must be formulated as a special case. Therefore, in the current state of affairs, one is required to abandon potentially simpler proof techniques in exchange for the benefits of the abstract interpretation framework.

**Resistance to Mechanized Verification** Despite the beauty and utility of abstract interpretation with Galois connections, advocates of the approach have yet to reconcile their use with advances in mechanized reasoning: *every mechanized verification of an executable abstract interpreter to-date has resisted the use of Galois connections, even when initially designed to take advantage of the framework.*

The issue in mechanizing Galois connections amounts to a conflict between supporting both classical set-theoretic reasoning and executable static analyzers. Supporting executable analyzers calls for constructive mathematics, which is a problem for $\alpha$ functions because they are often non-constructive, an observation first made by Monniaux (1998). To work around this limitation, Pichardie (2005) advocates for designing abstract interpreters which are merely inspired by Galois connections, but ultimately avoid their use in verification, which he terms the "$\gamma$-only" approach. Successful verification projects such as Verasco adopt this "$\gamma$-only" technique (Jourdan *et al.*, 2015; Leroy, 2009), despite the use of Galois connections in the design of Astrée (Blanchet *et al.*, 2003), the analyzer upon which Verasco is based.

While it is possible to verify abstract interpreters using Galois connections within tools based on classical mathematics (*e.g.*, Isabelle/HOL, or Coq extended with classical axioms), this approach requires a strict separation between the logical and algorithmic fragments of the system. If extraction of certified algorithms is not desired, this poses no issue at all, and if extraction is desired, then the use of Galois connections must be separated completely from the defined program analyzer. This prohibits use of the calculational method, where the specification induced by Galois connections is transformed *into* an algorithm, thereby crossing the barrier between logical and algorithmic fragments of the system. Furthermore, it is common for calculationally derived analyzers to mention abstraction ($\alpha$) functions directly, which again poses an issue if algorithms and definitions which rely on classical mathematics (like $\alpha$) must be kept separate for the purposes of program extraction. Overcoming this limitation—the inability to intermix classical Galois connections and algorithmic definitions—is the primary motivation for our development of constructive Galois connections.

To better understand the foundational issues with Galois connections and $\alpha$ functions, consider verifying the abstract interpretation approach to soundness for our parity analyzer using a proof assistant built on constructive logic. In this setting, the encoding of the Galois connection must support elements of infinite powersets—like the set of all even numbers—as well as executable abstract interpreters which manipulate elements of finite powersets—like $\{\text{EVEN}, \text{ODD}\}$. To support representing infinite sets, the powerset $\wp(\mathbb{N})$ is modeled constructively as a predicate $\mathbb{N} \rightarrow \mathit{prop}$. To support defining executable analyzers that manipulate sets of parities, the powerset $\wp(\mathbb{P})$ is modeled as an enumeration of its inhabitants, which we call $\mathbb{P}^c$:

$$\mathbb{P}^c := \{\text{EVEN}, \text{ODD}, \bot, \top\}$$

where $\bot$ and $\top$ represent $\{\}$ and $\{\text{EVEN}, \text{ODD}\}$. This enables a definition for $\uparrow\!\mathit{succ}^\sharp : \mathbb{P}^c \rightarrow \mathbb{P}^c$ which can be extracted and executed. The consequence of this design is a Galois connection between $\mathbb{N} \rightarrow \mathit{prop}$ and $\mathbb{P}^c$; the issue is now $\alpha$:

$$\alpha : (\mathbb{N} \rightarrow \mathit{prop}) \rightarrow \mathbb{P}^c$$

This version of $\alpha$ cannot be defined constructively, as doing so requires deciding predicates over $\phi : \mathbb{N} \rightarrow \mathit{prop}$. To define $\alpha$ one must perform case analysis on predicates like $\exists n, \phi(n) \wedge \mathit{even}(n)$ to *compute* an element of $\mathbb{P}^c$, which is not possible for arbitrary $\phi$. (The exercise also fails if powersets are modeled with decidable predicates $\phi : \mathbb{N} \rightarrow \mathbb{B}$.) $\alpha$ functions are often used directly in the definition of calculated abstract interpreters (as is the case in Cousot's monograph (Cousot, 1999)), and a non-algorithmic $\alpha$ function will prevent extraction for these interpreters.

However, $\gamma$ *can* be defined constructively as a relation (2-arity proposition):

$$\gamma : \mathbb{P}^c \rightarrow (\mathbb{N} \rightarrow \mathit{prop})$$

In general, any *theorem* of soundness using Galois connections can be rewritten to use only $\gamma$, making use of (GC-Corr); this is the essence of the "$\gamma$-only" approach, embodied by the soundness variant (GC-Snd/$\gamma\gamma$). However, this principle does not apply to all *proofs* of soundness using Galois connections, many of which mention $\alpha$ in practice. For example, the $\gamma$-only setup does not support calculation in the style advocated by Cousot (1999). Furthermore, not all *completeness* theorems can be translated to $\gamma$-only style, such as (GC-Cmp/$\gamma\alpha$) [precise] which is used to show an abstract interpreter is fully precise.

**Wrapping Up** Abstract interpretation differs from the direct approach in which parts of the design are postulated and which parts are derived. The direct approach requires postulating the analyzer and definition of soundness. Using abstract interpretation, a Galois connection between sets is postulated instead, and definitions for soundness and completeness are synthesized from the Galois connection. Because soundness and completeness are synthesized rather than designed directly, it is more likely that they will be correct. This high-assurance for the specification of correctness helps prevent situations where a proof is completed successfully against a buggy specification, resulting in a buggy analyzer with false assurance. Finally, abstract interpretation supports deriving the definition of a static analyzer directly from its proof of correctness. The derivation process will reject buggy implementation fragments early, because every step of the derivation is checked for correctness.

The downside of abstract interpretation is that it requires lifting *succ* and $succ^\sharp$ into powersets, which results in boilerplate set-theoretic reasoning in the proof of soundness. Finally, due to foundational issues, the abstract interpretation framework is not amenable to mechanized verification while also supporting program extraction using constructive logic.

## 3 Constructive Galois Connections

In this section we describe abstract interpretation with constructive Galois connections. Constructive Galois connections are a parallel universe of Galois connections analogous to classical ones. The framework enjoys all the benefits of abstract interpretation, but like the direct approach avoids the pitfalls of added complexity and resistance to mechanized verification.

We will describe the framework of constructive Galois connections between sets $C$ and $A$. When instantiated to $\mathbb{N}$ and $\mathbb{P}$, the framework recovers exactly the direct approach from Section 2.1. We will initially describe constructive Galois connections in the absence of partial orders, or more precisely, we will assume the discrete partial order: $x \sqsubseteq y \Leftrightarrow x = y$. (Partial orders didn't appear in our demonstration of classical abstract interpretation, but they are essential to the general theory.) At the end of this section we describe generalizing from sets to posets, generalizing from abstract and concrete functions to relations, and how to recover classical soundness results from constructive ones. The fully general theory of constructive Galois connections is described in Section 7 where it is again compared side-by-side to classical Galois connections.

**Abstracting Sets** A constructive Galois connection between sets $C$ and $A$ contains two mappings: the first is called *extraction*, notated $\eta$, and the second is called *interpretation*, notated $\mu$:

$$\eta \ : \ C \to A \qquad\qquad \mu \ : \ A \to \wp(C)$$

$\eta$ and $\mu$ are analogous to classical Galois connection mappings $\alpha$ and $\gamma$. In the parity analysis described in Section 2.1, the extraction function was *parity* and the interpretation function was $[\![_]\!]$.

Constructive Galois connection mappings $\eta$ and $\mu$ must form a correspondence similar to (GC-Corr):

$$x \in \mu(y) \iff \eta(x) = y \qquad \text{(CGC-Corr)}$$

The intuition behind the correspondence is the same as before: to compare an element $x$ in $C$ to an element $y$ in $A$, it is equivalent to compare them through either $\eta$ or $\mu$.

Like classical Galois connections, the correspondence between $\eta$ and $\mu$ is stated equivalently through two composition laws. Extraction functions $\eta$ which form a constructive Galois connection are also a "best abstraction", analogously to $\alpha$ in the classical setup:

$$x \in \mu(\eta(x)) \qquad \text{(CGC-Exp)}$$

$$x \in \mu(y) \implies \eta(x) = y \qquad \text{(CGC-Red)}$$

In general, it is possible to induce $\mu$ as the inverse-image of any function $\eta$ (just like in the classical framework where any $\alpha$ can induce a corresponding $\gamma$):

$$\mu(y) \coloneqq \{x \mid \eta(x) = y\}$$

This induced $\mu$ is guaranteed to satisfy (CGC-Corr). However, this inverse-image definition can be cumbersome to work with, and there are practical benefits to defining $\mu$ directly for the purposes of proofs and calculations.

*Aside* We use the term *extraction function* and symbol $\eta$ from Nielson *et al* (1999) where $\eta$ is used to simplify the definition of an abstraction function $\alpha$. We recover $\alpha$ functions from $\eta$ in a similar way. However, their treatment of $\eta$ is a side-note to simplifying the definition of $\alpha$ and nothing more. We take this simple idea much further to realize an entire theory of abstraction around $\eta/\mu$ functions and their correspondences. In this "lowered" theory of $\eta/\mu$ we describe soundness/optimality criteria and calculational derivations analogous to that of $\alpha/\gamma$ while also supporting mechanized verification, none of which is true of Nielson *et al*'s use of $\eta$.

**Induced Specifications** Four equivalent soundness criteria are generated by $\eta$ and $\mu$ just like in the classical framework. Each soundness statement uses $\eta$ and $\mu$ in a different but equivalent way (assuming CGC-Corr). For a concrete $f : C \to C$ and abstract $f^\sharp : A \to A$, $f^\sharp$ is sound *iff* any of the following properties hold:

$$x \in \mu(y) \implies \eta(f(x)) = f^\sharp(y) \qquad \text{(CGC-Snd/}\eta\mu\text{)}$$

$$x \in \mu(y) \implies f(x) \in \mu(f^\sharp(y)) \qquad \text{(CGC-Snd/}\mu\mu\text{)}$$

$$\eta(f(x)) = f^\sharp(\eta(x)) \qquad \text{(CGC-Snd/}\eta\eta\text{)}$$

$$f(x) \in \mu(f^\sharp(\eta(x))) \qquad \text{(CGC-Snd/}\mu\eta\text{)}$$

In the direct approach to verifying an example parity analysis described in Section 2.1, the first soundness property (DA-Snd) is generated by the $\mu\mu$ variant, and the second soundness property (DA-Snd*) which enjoyed a simpler proof is generated by the $\eta\eta$ variant. We discuss completeness equations in Section 3.1.

**Calculational Derivation of Abstract Interpreters** The constructive Galois connection framework also supports deriving abstract interpreters through calculation, analogously to the calculation we demonstrated in Section 2.2. To support calculational reasoning, the four logical soundness criteria are rewritten into statements about subsumption between powerset elements:

$$\{\eta(f(x)) \mid x \in \mu(y)\} \subseteq \{f^\sharp(y)\} \qquad \text{(CGC-Snd}/\eta\mu^*)$$

$$\{f(x) \mid x \in \mu(y)\} \subseteq \mu(f^\sharp(y)) \qquad \text{(CGC-Snd}/\mu\mu^*)$$

$$\{\eta(f(x))\} \subseteq \{f^\sharp(\eta(x))\} \qquad \text{(CGC-Snd}/\eta\eta^*)$$

$$\{f(x)\} \subseteq \mu(f^\sharp(\eta(x))) \qquad \text{(CGC-Snd}/\mu\eta^*)$$

Using the $\eta\mu^*$ soundness rule, one calculates towards a definition for $f^\sharp$ starting from the left-hand-side, which is the optimal specification for abstract interpreters of $f$.

To demonstrate calculation using constructive Galois connections, we show the derivation of $succ^\sharp$ from its induced specification, the result of which is sound by construction; we show $p = \text{EVEN}$:

$$\begin{aligned}
&\{parity(succ(n)) \mid n \in [\![\text{EVEN}]\!]\} \\
&= \{flip(parity(n)) \mid n \in [\![\text{EVEN}]\!]\} && \wr \text{ defn. of } parity \ \int \\
&= \{flip(\text{EVEN})\} && \wr \text{ Eq. (DA-Corr) } \int \\
&= \{\text{ODD}\} && \wr \text{ defn. of } flip \ \int \\
&\triangleq \{succ^\sharp(\text{EVEN})\} && \wr \text{ defining } succ^\sharp \ \int
\end{aligned}$$

Technically the result of the derivation is a singleton set lifting of $succ^\sharp(\text{EVEN})$, and the abstraction for *succ* must be "unlifted" from this singleton set. We will show another perspective on this calculation later in this section, where the derivation of $succ^\sharp$ is not only sound by construction, but computable by construction as well.

**Mechanized Verification** In addition to the benefits of a general abstraction framework, constructive Galois connections are amenable to mechanization in a way that classical Galois connections are not. In our Agda library and case studies we mechanize constructive Galois connections in full generality, as well as proofs that use both mapping functions, such as calculational derivations.

As we discussed in Sections 2.1 and 2.2, the constructive encoding for infinite powersets $\wp(A)$ is $A \to prop$. This results in the following types for $\eta$ and $\mu$ when encoded constructively:

$$\eta : \mathbb{N} \to \mathbb{P} \qquad\qquad \mu : \mathbb{P} \to \mathbb{N} \to prop$$

In constructive logic, the arrow type $\mathbb{N} \to \mathbb{P}$ classifies computable functions, and the arrow type $\mathbb{P} \to \mathbb{N} \to prop$ classifies potentially undecidable relations. (CGC-Corr) is then mechanized without issue:

$$\mu(p,n) \iff \eta(n) = p$$

See the mechanization details in Section 2.1 for how $\eta$ and $\mu$ are defined constructively for the example parity analysis.

**Wrapping Up** Constructive Galois connections are a general abstraction framework similar to classical Galois connections. At the heart of the constructive Galois connection framework is a correspondence (CGC-Corr) analogous to its classical counterpart. From this correspondence, soundness and completeness criteria are synthesized for abstract interpreters. Constructive Galois connections also support calculational derivations of abstract interpreters which are sound and optimal by construction. In addition to these benefits of a general abstraction framework, constructive Galois connections are amenable to mechanized verification. Both extraction ($\eta$) and interpretation ($\mu$) can be mechanized effectively, as well as proofs of soundness, completeness, and calculational derivations.

### *3.1 Partial Orders, Monotonicity, and Relations*

The full theory of constructive Galois connections generalizes to posets $\langle C, \sqsubseteq^C \rangle$ and $\langle A, \sqsubseteq^A \rangle$ by making the following changes:

- Powersets must be downward-closed, that is for $X : \wp(C)$:

  $$x \in X \wedge x' \sqsubseteq x \implies x' \in X \qquad \text{(PowerMon)}$$

  Singleton sets $\{x\}$ are reinterpreted to mean $\{x' \mid x' \sqsubseteq x\}$. For mechanization, this means $\wp(C)$ is encoded as an *antitonic* function, notated with a down-right arrow $C \searrow prop$, where the partial ordering on $prop$ is by implication.
- Functions must be monotonic, that is for $f : C \to C$:

  $$x \sqsubseteq x' \implies f(x) \sqsubseteq f(x') \qquad \text{(FunMon)}$$

  We notate monotonic functions $f : C \nearrow C$. Monotonicity is required for mappings $\eta$ and $\mu$, and concrete and abstract interpreters $f$ and $f^\sharp$.
- The constructive Galois connection correspondence is generalized to partial orders in place of equality, that is for $\eta$ and $\mu$:

  $$x \in \mu(y) \iff \eta(x) \sqsubseteq y \qquad \text{(CGP-Corr)}$$

  or alternatively, by generalizing the reductive property:

  $$x \in \mu(y) \implies \eta(x) \sqsubseteq y \qquad \text{(CGP-Red)}$$
- Soundness criteria are also generalized to partial orders:

  $$x \in \mu(y) \implies \eta(f(x)) \sqsubseteq f^\sharp(y) \qquad \text{(CGP-Snd/}\eta\mu\text{)}$$
  $$x \in \mu(y) \implies f(x) \in \mu(f^\sharp(y)) \qquad \text{(CGP-Snd/}\mu\mu\text{)}$$
  $$\eta(f(x)) \sqsubseteq f^\sharp(\eta(x)) \qquad \text{(CGP-Snd/}\eta\eta\text{)}$$
  $$f(x) \in \mu(f^\sharp(\eta(x))) \qquad \text{(CGP-Snd/}\mu\eta\text{)}$$

and completeness criteria are as follows:

$$\text{[optimal]} \qquad x \in \mu(y) \wedge y' \sqsubseteq \eta(f(x)) \iff y' \sqsubseteq f^\sharp(y) \qquad \text{(CGP-Cmp/}\eta\mu\text{)}$$

$$x \in \mu(y) \wedge x' \sqsubseteq f(x) \iff x' \in \mu(f^\sharp(y)) \qquad \text{(CGP-Cmp/}\mu\mu\text{)}$$

$$y \sqsubseteq \eta(f(x)) \iff y \sqsubseteq f^\sharp(\eta(x)) \qquad \text{(CGP-Cmp/}\eta\eta\text{)}$$

$$\text{[precise]} \qquad x' \sqsubseteq f(x) \implies x' \in \mu(f^\sharp(\eta(x))) \qquad \text{(CGP-Cmp/}\mu\eta\text{)}$$

The $x$ on the left-hand-side of the first completeness rule is implicitly existentially quantified, *i.e.*, with explicit quantifiers the rule is:

$$\forall y \ \ y'. \ (\exists x. \ x \in \mu(y) \wedge y' \sqsubseteq \eta(f(x))) \iff y' \sqsubseteq f^\sharp(y)$$

Soundness criteria are merely simplifications of the left-to-right implication direction of the completeness criteria. Each of the completeness criteria are not equivalent, as was also the case for classical Galois connections. Following the terminology of classical Galois connections, we call abstract interpreters $f^\sharp$ which satisfy the $\eta\mu$ variant *optimal* and those which satisfy the $\mu\eta$ variant *precise*.

To demonstrate when partial orders and monotonicity are necessary, consider designing a parity analyzer for the *max* operator:

$$max^\sharp : \mathbb{P} \times \mathbb{P} \to \mathbb{P} \qquad \begin{aligned} max^\sharp(\textsc{even},\textsc{even}) &:= \textsc{even} \qquad & max^\sharp(\textsc{even},\textsc{odd}) &:= ? \\ max^\sharp(\textsc{odd},\textsc{odd}) &:= \textsc{odd} \qquad & max^\sharp(\textsc{odd},\textsc{even}) &:= ? \end{aligned}$$

The last two cases for $max^\sharp$ cannot be defined because the maximum of an even and odd number could be either even or odd, and there is no representative for "any number" in $\mathbb{P}$. To remedy this, we add $\textsc{any}$ to the set of parities: $\mathbb{P}^+ := \mathbb{P} \cup \{\textsc{any}\}$; the new element $\textsc{any}$ is interpreted: $[\![\textsc{any}]\!] := \{n \mid n \in \mathbb{N}\}$; the partial order on $\mathbb{P}^+$ becomes: $\textsc{even},\textsc{odd} \sqsubseteq \textsc{any}$; and the correspondence continues to hold using this partial order: $n \in [\![p^+]\!] \iff parity(n) \sqsubseteq p^+$. $max^\sharp$ is then defined using the abstraction $\mathbb{P}^+$ and proven sound and optimal following the abstract interpretation paradigm.

**Generalizing to Relations** The above soundness rules are stated for concrete *functions* $f : C \nearrow C$. However, they generalize easily to *relations* $R : \wp(C \times C)$ and *predicate transformers* $F : \wp(C) \nearrow \wp(C)$ (*i.e.*, collecting semantics). In both cases, we consider $f : C \nearrow \wp(C)$ defined by:

$$f(x) := \{y \mid R(x,y)\}$$

in the case of relations, and

$$f(x) := F(\{x\})$$

in the case of predicate transformers. Given a candidate abstraction $f^\sharp : A \nearrow \wp(A)$, the four (equivalent) soundness criteria are as follows, (which we write as set-subsumptions rather than implications due to the number of existentially quantified

variables involved):

$$\{\eta(x') \mid x \in \mu(y), x' \in f(x)\} \subseteq f^\sharp(y) \quad \text{(CGP-Snd-R/}\eta\mu\text{)}$$
$$\{x' \mid x \in \mu(y), x' \in f(x)\} \subseteq \{x \mid y' \in f^\sharp(y), x \in \mu(y')\} \quad \text{(CGP-Snd-R/}\mu\mu\text{)}$$
$$\{\eta(x') \mid x' \in f(x)\} \subseteq f^\sharp(\eta(x)) \quad \text{(CGP-Snd-R/}\eta\eta\text{)}$$
$$f(x) \subseteq \{x' \mid y \in f^\sharp(\eta(x)), x' \in \mu(y)\} \quad \text{(CGP-Snd-R/}\mu\eta\text{)}$$

The completeness criteria are analogous, but with set equality ($=$) in place of subsumption ($\subseteq$). These equations come from the adjunction framework, which we describe in more detail in Section 7. In particular, the shape of the set comprehensions and existentially quantified variables arise from monadic composition in the powerset monad.

### 3.2 Relationship to Classical Galois Connections

We clarify the relationship between constructive and classical Galois connections in three ways:

- Any constructive Galois connection can be lifted to obtain an equivalent classical Galois connection, and likewise for soundness and completeness proofs.
- Any classical Galois connection which can be recovered by a constructive one contains no additional expressive power, rendering it an equivalent theory with added boilerplate reasoning.
- Not all classical Galois connections can be recovered by constructive ones.

From these relationships we conclude that one benefits from using constructive Galois connections whenever possible, classical Galois connections when no constructive one exists, and both theories together as needed. We make these claims precise in Section 7, and explore the subtleties of their relationship and interaction in detail in Sections 9, 10 and 11. We point out connections to more general categorical settings in Section 13.

Aside We call the standard Galois connection framework “classical” because it is not amenable to mechanization, and our proposed framework “constructive” because it is amenable to mechanized verification. This is not to be confused with the classical or constructive nature of the mathematics used to interpret either framework. It is possible to use both frameworks side-by-side, each interpreted either using classical or constructive mathematics. However, classical Galois connections are less *useful* when interpreted constructively, and likewise for constructive Galois connections interpreted classically.

A classical Galois connection is recovered from a constructive one through the following lifting:

$$\alpha : \wp(C) \to \wp(A) \qquad \alpha(X) := \{\eta(x) \mid x \in X\}$$
$$\gamma : \wp(A) \to \wp(C) \qquad \gamma(Y) := \{x \mid y \in Y \wedge x \in \mu(y)\}$$

When a classical Galois connection can be written in this form for some $\eta$ and $\mu$, then one can use the simpler setting of abstract interpretation with constructive

Galois connections without any loss of generality. We also observe that many classical Galois connections in practice can be written in this form, and therefore can be mechanized effectively using constructive Galois connections. The case studies in presented in Sections 4 and 5 are two such cases, although the original authors of those works did not initially write their classical Galois connections in this explicitly lifted form.

An example of a classical Galois connection which is not recovered by lifting a constructive Galois is the Independent Attributes (IA) abstraction, which abstracts relations $R : \wp(A \times B)$ with their component-wise splitting $\langle R_l, R_r \rangle : \wp(A) \times \wp(B)$:

$$\begin{aligned} \alpha &: \wp(A \times B) \to \wp(A) \times \wp(B) & \alpha(R) &:= \langle \{x \mid \exists y. \langle x, y \rangle \in R\}, \{y \mid \exists x. \langle x, y \rangle \in R\} \rangle \\ \gamma &: \wp(A) \times \wp(B) \to \wp(A \times B) & \gamma(R_l, R_r) &:= \{\langle x, y \rangle \mid x \in R_l, y \in R_r\} \end{aligned}$$

This Galois connection *is* amenable to mechanized verification. In a constructive setting, $\alpha$ and $\gamma$ are maps between $A \times B \to prop$ and $(A \to prop) \times (B \to prop)$, and can be defined directly using logical connectives $\exists$ and $\wedge$:

$$\begin{aligned} \alpha(R) &:= \langle \lambda x. \exists y. R(x, y), \lambda y. \exists x. R(x, y) \rangle \\ \gamma(R_l, R_r) &:= \lambda \langle x, y \rangle. R_l(x) \wedge R_r(y) \end{aligned}$$

IA can be mechanized effectively because the Galois connection consists of mappings between specifications, and the foundational issue of constructing values from specifications does not appear. IA is not a constructive Galois connection because there is no pure function $\eta$ underlying the abstraction function $\alpha$.

Because constructive Galois connections can be lifted to classical ones, a constructive Galois connection can interact directly with IA through its lifting, even in a mechanized setting. However, once a constructive Galois connection is lifted it loses its computational properties and cannot be extracted and executed. In practice, IA is used to weaken ($\sqsubseteq$) an induced optimal specification after which the calculated interpreter is shown to be optimal ($=$) up-to-IA. IA never appears in the final calculated interpreter, so not having a constructive Galois connection formulation poses no issue. We explore how a constructive Galois connection derivation interacts with IA in detail in Sections 9 and 10.

### *3.3 The "Specification Effect"*

The machinery of constructive Galois connections follow a *monadic effect* discipline, where the effect type is the classical powerset $\wp(_)$; we call this a *specification effect*. First we will describe the monadic structure of powersets $\wp(_)$ and what we mean by "specification effect." Then we will recast the theory of constructive Galois connections in this monadic style, giving insights into why the theory supports mechanized verification, and foreshadowing key fragments of the metatheory we develop in Section 7.

The monadic structure of classical powersets is standard, and is analogous to the nondeterminism monad familiar to Haskell programmers. However, the model $\wp(A) := A \to prop$ is the uncomputable nondeterminism monad and mirrors the use of set-comprehensions on paper to describe uncomputable sets (specifications), rather than the use of monad comprehensions in Haskell to describe computable sets (constructed values).

We generalize $\wp(_)$ to a *monotonic* monad, similarly to how we generalized powersets to posets in Section 3.1. This results in monotonic versions of monad operators *ret* and *bind*:

$$\begin{aligned} \mathit{ret} &: A \nearrow \wp(A) & \mathit{bind} &: \wp(A) \times (A \nearrow \wp(B)) \nearrow \wp(B) \\ \mathit{ret}(x) &:= \{x' \mid x' \sqsubseteq x\} & \mathit{bind}(X,f) &:= \{y \mid x \in X \wedge y \in f(x)\} \end{aligned}$$

We adopt Moggi's notation (Moggi, 1989) for monadic extension where $\mathit{bind}(X,f)$ is written $f^*(X)$, or just $f^*$ for $\lambda X.f^*(X)$. The monad and functor laws hold for downward-closed powersets (despite the contravariant occurrence of $A$ in the definition of $\wp(A)$), and we mechanize these proofs in our Agda development.

We call the powerset type $\wp(A)$ a specification effect because it has monadic structure, supports encoding arbitrary properties over values in $A$, and cannot be "escaped from" in constructive logic, similar to the ***IO*** monad in Haskell. In classical mathematics, there is an isomorphism between singleton powersets $\wp^1(A)$ and the set $A$. However, no such constructive mapping exists for $\wp^1(A) \rightarrow A$. Such a function would decide arbitrary predicates in $A \rightarrow \mathit{prop}$ to *compute* the $A$ inside the singleton set. This observation, that you can program inside $\wp(_)$ monadically in constructive logic, but you can't escape the monad, is why we call it a specification effect.

Given the monadic structure for powersets, and the intuition that they encode a specification effect in constructive logic, we can recast the theory of constructive Galois connections using monadic operators. To do this we define a helper operator which injects "pure" functions into the "effectful" function space:

$$\mathit{pure} : (A \nearrow B) \nearrow (A \nearrow \wp(B)) \qquad \mathit{pure}(f)(x) := \mathit{ret}(f(x))$$

We then rewrite (CGC-Corr) using *ret* and *pure*:

$$\mathit{ret}(x) \subseteq \mu(y) \iff \mathit{pure}(\eta)(x) \subseteq \mathit{ret}(y) \qquad \text{(CGM-Corr)}$$

and we rewrite the expansive and reductive variant of the correspondence using *ret*, *bind* (notated $_^*$) and *pure*:

$$\mathit{ret}(x) \subseteq \mu^*(\mathit{pure}(\eta)(x)) \qquad \text{(CGM-Exp)}$$

$$\mathit{pure}(\eta)^*(\mu(y)) \subseteq \mathit{ret}(y) \qquad \text{(CGM-Red)}$$

The four soundness and completeness conditions can also be written in monadic style; we show the $\eta\mu$ soundness property here:

$$\mathit{pure}(\eta)^*(\mathit{pure}(f)^*(\mu(y))) \subseteq \mathit{pure}(f^\sharp)(y) \qquad \text{(CGM-Snd/}\eta\mu\text{)}$$

The left-hand-side of the ordering is the optimal specification for $f^\sharp$, just like (CGP-Snd/$\eta\mu$) but using monadic operators. The right-hand-side of the ordering is $f^\sharp$ lifted to the monadic function space. The constructive calculation of $\mathit{succ}^\sharp$ we showed earlier in this section is a calculation of this form.

Both sides of the ordering (CGM-Snd/$\eta\mu$) have the monadic type $\wp(\mathbb{P})$, however they differ in whether or not they contain specification effects. The specification on the left *has effects*—because it makes use of the interpretation function $\mu$—meaning it uses classical reasoning and can't be executed. (The monadic bind operation isn't contributing any effects; it merely propagates them.) The abstract interpreter on the right *has no effects*—because it is simply the injection of a "pure" function into the monadic function space—meaning it *can* be extracted and executed. The calculated abstract interpreter is thus not only sound and optimal by construction, *it is computable by construction.*

Constructive Galois connections are empowering because they treat specification like an effect, which optimal specifications *ought to have*, and which algorithmic abstract interpreters *ought not to have.* Using a monadic effect discipline we support calculations which start with a specification effect, and where the "effect" is eliminated through the process of calculation. The monad laws are crucial in canceling uses of *ret* with *bind* to arrive at a final pure computation. For example, the first step in a derivation for (CGM-Snd/$\eta\mu$) can immediately simplify using monad laws from:

$$pure(\eta)^*(pure(f)^*(\mu(y))) \subseteq pure(f^\sharp)(y)$$

to:

$$pure(\eta \circ f)^*(\mu(y)) \subseteq pure(f^\sharp)(y)$$

## 4 Case Study 1: Calculational AI

In this section we apply constructive Galois connections to the *Calculational Design of a Generic Abstract Interpreter* from Cousot's monograph (1999). To our knowledge, *we achieve the first mechanically verified abstract interpreter derived by calculus.*

The key challenge in mechanizing the interpreter is supporting both abstraction ($\alpha$) and concretization ($\gamma$) mappings, which are required by the calculational approach. Classical Galois connections do not support mechanization of $\alpha$ without the use of axioms, and these required axioms block computation, preventing the extraction of verified algorithms. In particular, the analysis algorithm that Cousot derives *via* calculation mentions $\alpha$ directly in its definition, making it even more critical to move to a constructive framework if extraction of an executable algorithm is desired.

To verify Cousot's generic abstract interpreter we use constructive Galois connections, which we described in Section 3 and formalize in Section 7. Using constructive Galois connections we encode extraction ($\eta$) and interpretation ($\mu$) mappings as constructive analogs to $\alpha$ and $\gamma$, calculate an abstract interpreter for an imperative programming language which is sound and computable by construction, and recover the original classical Galois connection results through a systematic lifting.

First we describe the setup for the analyzer: the abstract syntax, the concrete semantics, and the constructive Galois connections involved. Following the abstract interpretation paradigm with constructive Galois connections we design abstract interpreters for denotation functions and semantics relations. We show a fragment

$$
\begin{array}{rrcll}
i \in & \mathbb{Z} & \coloneqq & \{\ldots,-1,0,1,\ldots\} & \text{integers} \\
b \in & \mathbb{B} & \coloneqq & \{true, false\} & \text{booleans} \\
x \in & \texttt{var} & ::= & \ldots & \text{variables} \\
\oplus \in & \texttt{aop} & ::= & + \mid - \mid \times \mid / & \text{arithmetic op.} \\
\otimes \in & \texttt{cmp} & ::= & < \mid = & \text{comparison op.} \\
\oslash \in & \texttt{bop} & ::= & \vee \mid \wedge & \text{boolean op.} \\
ae \in & \texttt{aexp} & ::= & i \mid x \mid \texttt{rand} \mid ae \oplus ae & \text{arithmetic exp.} \\
be \in & \texttt{bexp} & ::= & b \mid ae \otimes ae \mid be \oslash be & \text{boolean exp.} \\
ce \in & \texttt{cexp} & ::= & \texttt{skip} \mid ce \,;\, ce & \text{skip \& sequence exp.} \\
 & & \mid & x \coloneqq ae & \text{assignment exp.} \\
 & & \mid & \texttt{if}\ be\ \texttt{then}\ ce\ \texttt{else}\ ce & \text{conditional exp.} \\
 & & \mid & \texttt{while}\ be\ \texttt{do}\ ce & \text{while loop exp.}
\end{array}
$$

Fig. 1. Case Study 1: WHILE Abstract Syntax

of our Agda mechanization which closely mirrors the pencil-and-paper proof, as well as Cousot's original derivation. See Section 6 for a more in-depth tutorial on our mechanization approach, *e.g.*, our encodings for posets, monotonic functions, and proof combinators in Agda.

### *4.1 Concrete Semantics*

The WHILE language is an imperative programming language with arithmetic expressions, variable assignment and while-loops. We show the syntax for this language in Figure 1. WHILE syntactically distinguished arithmetic, boolean and command expressions. rand is an arithmetic expression which can evaluate to any integer. Syntactic categories $\oplus$, $\otimes$ and $\oslash$ range over arithmetic, comparison and boolean operators, and are introduced to simplify the presentation. The WHILE language is taken from Cousot's monograph (1999).

The concrete semantics of WHILE is sketched without full definition in Figure 2. Denotation functions $[\![_]\!]^a$, $[\![_]\!]^c$ and $[\![_]\!]^b$ give semantics to arithmetic, comparison and boolean operators. The semantics of compound syntactic expressions are given operationally with relations $\Downarrow^a$, $\Downarrow^b$ and $\mapsto^c$. Relational semantics are given for arithmetic and boolean expressions due to the nondeterminism of rand, and for command expressions due to the nontermination of while. (Although other techniques for handling termination would also suffice, *e.g.*, Domains *à la* Scott (Scott, 1975).) These semantics serve as the starting point for designing an abstract interpreter.

### *4.2 Abstract Semantics with Constructive GCs*

Using abstract interpretation with constructive Galois connections, we design an abstract semantics for WHILE in the following steps:

$$
\begin{array}{rcl}
\rho \in \mathtt{env} & \coloneqq & \mathtt{var} \rightharpoonup \mathbb{Z} \\
[\![_]\!]^a & : & aop \to \mathbb{Z} \times \mathbb{Z} \rightharpoonup \mathbb{Z} \\
[\![_]\!]^c & : & cmp \to \mathbb{Z} \times \mathbb{Z} \to \mathbb{B} \\
[\![_]\!]^b & : & bop \to \mathbb{B} \times \mathbb{B} \to \mathbb{B}
\end{array}
\qquad
\begin{array}{rcl}
\varsigma \in \Sigma & \coloneqq & \mathtt{env} \times \mathtt{cexp} \\
_ \vdash _ \Downarrow^a _ & : & \wp(\mathtt{env} \times \mathtt{aexp} \times \mathbb{Z}) \\
_ \vdash _ \Downarrow^b _ & : & \wp(\mathtt{env} \times \mathtt{bexp} \times \mathbb{B}) \\
_ \mapsto^c _ & : & \wp(\Sigma \times \Sigma)
\end{array}
$$

$$
\frac{}{\rho \vdash \mathtt{rand} \Downarrow^a i}\textsc{ARand}
\qquad
\frac{\rho \vdash ae_1 \Downarrow^a i_1 \qquad \rho \vdash ae_2 \Downarrow^a i_2}{\rho \vdash ae_1 \oplus ae_2 \Downarrow^a [\![\oplus]\!]^a(i_1, i_2)}\textsc{AOp}
$$

$$
\frac{\rho \vdash ae \Downarrow^a i}{\langle \rho, x \coloneqq ae \rangle \mapsto^c \langle \rho[x \leftarrow i], \mathtt{skip} \rangle}\textsc{CAssign}
$$

$$
\frac{\rho \vdash be \Downarrow^b true}{\langle \rho, \mathtt{if}\ be\ \mathtt{then}\ ce_1\ \mathtt{else}\ ce_2 \rangle \mapsto^c \langle \rho, ce_1 \rangle}\textsc{CIf-T}
$$

$$
\frac{\rho \vdash be \Downarrow^b false}{\langle \rho, \mathtt{if}\ be\ \mathtt{then}\ ce_1\ \mathtt{else}\ ce_2 \rangle \mapsto^c \langle \rho, ce_2 \rangle}\textsc{CIf-F}
$$

$$
\frac{\rho \vdash be \Downarrow^b true}{\langle \rho, \mathtt{while}\ be\ \mathtt{do}\ ce \rangle \mapsto^c \langle \rho, ce\ ;\ \mathtt{while}\ be\ \mathtt{do}\ ce \rangle}\textsc{CWhile-T}
$$

$$
\frac{\rho \vdash be \Downarrow^b false}{\langle \rho, \mathtt{while}\ be\ \mathtt{do}\ ce \rangle \mapsto^c \langle \rho, \mathtt{skip} \rangle}\textsc{CWhile-F}
$$

Fig. 2. Case Study 1: `WHILE` Concrete Semantics

1. An abstraction for each set $\mathbb{Z}$, $\mathbb{B}$ and `env`.
2. An abstraction for each denotation function $[\![_]\!]^a$, $[\![_]\!]^c$ and $[\![_]\!]^b$.
3. An abstraction for each semantics relation $\Downarrow^a$, $\Downarrow^b$ and $\mapsto^c$.

Each abstract set forms a constructive Galois connection with its concrete counterpart. Soundness criteria is synthesized for abstract functions and relations using constructive Galois connection mappings. Finally, we verify and calculate abstract interpreters from these specifications which are sound and computable by construction. We describe the details of this process only for integers and environments (the sets $\mathbb{Z}$ and `env`), arithmetic operators (the denotation function $[\![_]\!]^a$), and arithmetic expressions (the semantics relation $\Downarrow^a$). See the Agda development accompanying this paper for the full mechanization of `WHILE`, and Sections 9, 10, and 11 for a detailed account of binary arithmetic operators and conditional command expressions.

**Abstracting Integers** We design a simple sign abstraction for integers, although more powerful abstractions are certainly possible (Cousot, 1999; Miné, 2006). The final abstract interpreter for `WHILE` is parameterized by any abstraction for integers, meaning another abstraction can be plugged in without added proof effort.

The sign abstraction begins with three representative elements: neg, zer and pos, representing negative integers, the integer 0, and positive integers. To support representing integers which could be negative or 0, negative or positive, or 0 or positive, etc., we design a set which is complete w.r.t. these logical disjunctions:

$$i^\sharp \in \mathbb{Z}^\sharp \;:=\; \{\mathtt{none}, \mathtt{neg}, \mathtt{zer}, \mathtt{pos}, \mathtt{negz}, \mathtt{nzer}, \mathtt{posz}, \mathtt{any}\}$$

$\mathbb{Z}^\sharp$ is given meaning through an interpretation function $\mu^z$, the analog of a $\gamma$ from the classical Galois connection framework:

$$\mu^z \;:\; \mathbb{Z}^\sharp \nearrow \wp(\mathbb{Z}) \qquad \begin{aligned} \mu^z(\mathtt{none}) &:= \{\} & \mu^z(\mathtt{negz}) &:= \{i \mid i \leq 0\} \\ \mu^z(\mathtt{neg}) &:= \{i \mid i < 0\} & \mu^z(\mathtt{nzer}) &:= \{i \mid i \neq 0\} \\ \mu^z(\mathtt{zer}) &:= \{0\} & \mu^z(\mathtt{posz}) &:= \{i \mid i \geq 0\} \\ \mu^z(\mathtt{pos}) &:= \{i \mid i > 0\} & \mu^z(\mathtt{any}) &:= \{i \mid i \in \mathbb{Z}\} \end{aligned}$$

The partial ordering on abstract integers coincides with subset ordering under $\mu^z$, that is, $i_1^\sharp \sqsubseteq^z i_2^\sharp \iff \mu^z(i_1^\sharp) \subseteq \mu^z(i_2^\sharp)$:

$$\mathtt{none} \sqsubseteq^z i^\sharp \sqsubseteq^z \mathtt{any} \qquad \begin{aligned} \mathtt{neg} &\sqsubseteq^z \mathtt{negz}, \mathtt{nzer} \\ \mathtt{zer} &\sqsubseteq^z \mathtt{negz}, \mathtt{posz} \\ \mathtt{pos} &\sqsubseteq^z \mathtt{nzer}, \mathtt{posz} \end{aligned}$$

and we write $i_1^\sharp \sqcup i_2^\sharp$ as the least-upper bound (join) of $i_1^\sharp$ and $i_2^\sharp$, *e.g.*, $\mathtt{neg} \sqcup \mathtt{zero} = \mathtt{negz}$. To be a constructive Galois connection, $\mu^z$ forms a correspondence with a best abstraction function $\eta^z$:

$$\eta^z \;:\; \mathbb{Z} \to \mathbb{Z}^\sharp \qquad \eta^z(n) \;:=\; \begin{cases} \mathtt{neg} & \mathit{if} \quad i < 0 \\ \mathtt{zer} & \mathit{if} \quad i = 0 \\ \mathtt{pos} & \mathit{if} \quad i > 0 \end{cases}$$

and the constructive Galois connection correspondence (CGC-Corr) easily follows:

$$i \in \mu^z(i^\sharp) \iff \eta^z(i) \sqsubseteq^z i^\sharp$$

*The Classical Design* The concretization function $\gamma$ in the classical design is identical to the interpretation function using constructive Galois connections:

$$\gamma^z \;:\; \mathbb{Z}^\sharp \nearrow \wp(\mathbb{Z}) \qquad \gamma^z(i^\sharp) \;:=\; \mu^z(i^\sharp)$$

The abstraction function is the key difference using classical Galois connections, which is recovered through a lifting of our $\eta^z$:

$$\alpha^z \;:\; \wp(\mathbb{Z}) \nearrow \mathbb{Z}^\sharp \qquad \alpha^z(I) \;:=\; \bigsqcup_{i \in I} \eta^z(i)$$

Abstraction functions of this form—$\wp(C) \nearrow A$, for some concrete set $C$ and abstract set $A$—are representative of most Galois connections used in the literature for static analyzers. However, these abstraction functions are precisely the part of classical Galois connections which inhibit mechanized verification. The extraction function $\eta^z$ does not manipulate powersets, does not inhibit mechanized verification, and recovers the original non-constructive $\alpha^z$ through this standard lifting.

**Abstracting Environments** An abstract environment maps variables to abstract integers rather than concrete integers.

$$\rho^\sharp \in \mathsf{env}^\sharp \;\coloneqq\; \mathsf{var} \to \mathbb{Z}^\sharp$$

$\mathsf{env}^\sharp$ is given meaning through an interpretation function $\mu^r$:

$$\mu^r \;:\; \mathsf{env}^\sharp \nearrow \wp(\mathsf{env}) \qquad\qquad \mu^r(\rho^\sharp) \;\coloneqq\; \{\rho \mid \forall x.\rho(x) \in \mu^z(\rho^\sharp(x))\}$$

An abstract environment represents concrete environments that agree pointwise with some represented integer in the codomain.

The order on abstract environments is the standard pointwise ordering and coincides with subset ordering under $\mu^r$, that is, $\rho_1^\sharp \sqsubseteq^r \rho_2^\sharp \iff \mu^r(\rho_1^\sharp) \subseteq \mu^r(\rho_2^\sharp)$:

$$\rho_1^\sharp \sqsubseteq^r \rho_2 \;\coloneqq\; \forall x.\rho_1^\sharp(x) \sqsubseteq^z \rho_2^\sharp(x)$$

To form a constructive Galois connection, $\mu^r$ forms a correspondence with a best abstraction function $\eta^r$:

$$\eta^r \;:\; \mathsf{env} \to \mathsf{env}^\sharp \qquad\qquad \eta^r(\rho) \;\coloneqq\; \lambda x.\eta^z(\rho(x))$$

and the constructive Galois connection correspondence (CGC-Corr) easily follows:

$$\rho \in \mu^r(\rho^\sharp) \iff \eta^r(\rho) \sqsubseteq^r \rho^\sharp$$

*The Classical Design* To contrast with Cousot's original design using classical abstract interpretation, the key difference is again the abstraction function. The abstraction function using classical Galois connections is:

$$\alpha^r \;:\; \wp(\mathsf{env}) \nearrow \mathsf{env}^\sharp \qquad\qquad \alpha^r(R) \;\coloneqq\; \lambda x.\alpha^z(\{\rho(x) \mid \rho \in R\})$$

which is also not amenable to mechanized verification.

**Abstracting Functions** After designing constructive Galois connections for $\mathbb{Z}$ and $\mathsf{env}$ we define what it means for $[\![_]\!]^{a\sharp}$, some abstract denotation for arithmetic operators, to be a sound abstraction of $[\![_]\!]^{a}$, its concrete counterpart. This is done through a specification induced by mappings $\eta$ and $\mu$, analogously to how specifications are induced using classical Galois connections.

The specification which encodes soundness and optimality for $[\![_]\!]^{a\sharp}$ is generated using the constructive Galois connection for $\mathbb{Z}$:

$$\langle i_1, i_2\rangle \in \mu^{z\times z}(i_1^\sharp, i_2^\sharp) \wedge \langle i_1^{\sharp\prime}, i_2^{\sharp\prime}\rangle \sqsubseteq \eta^z([\![ae]\!]^a(i_1,i_2)) \Leftrightarrow \langle i_1^{\sharp\prime}, i_2^{\sharp\prime}\rangle \sqsubseteq [\![ae]\!]^{a\sharp}(i_1^\sharp, i_2^\sharp)$$

(See CGP-Cmp/$\eta\mu$ [optimal] in Section 3 for the origin of this equation.) For $[\![_]\!]^{a\sharp}$, we postulate its definition and verify its correctness post-facto using the above property, although we omit the proof details here. The definition of $[\![_]\!]^{a\sharp}$ is standard, and returns $\mathsf{none}$ in the case of division by zero. We show only the definition of $+$

here:

$$[\![_]\!]^{a\sharp} : \texttt{aexp} \rightarrow \mathbb{Z}^\sharp \times \mathbb{Z}^\sharp \nearrow \mathbb{Z}^\sharp \qquad [\![+]\!]^{a\sharp}(i_1^\sharp, i_2^\sharp) := \bigsqcup \begin{cases} \texttt{pos} & \textit{if} \;\; \texttt{pos} \sqsubseteq^z i_1^\sharp \vee \texttt{pos} \sqsubseteq^z i_2^\sharp \\ \texttt{neg} & \textit{if} \;\; \texttt{neg} \sqsubseteq^z i_1^\sharp \vee \texttt{neg} \sqsubseteq^z i_2^\sharp \\ \texttt{zer} & \textit{if} \;\; \texttt{zer} \sqsubseteq^z i_1^\sharp \wedge \texttt{zer} \sqsubseteq^z i_2^\sharp \\ \texttt{zer} & \textit{if} \;\; \texttt{pos} \sqsubseteq^z i_1^\sharp \wedge \texttt{neg} \sqsubseteq^z i_2^\sharp \\ \texttt{zer} & \textit{if} \;\; \texttt{neg} \sqsubseteq^z i_1^\sharp \wedge \texttt{pos} \sqsubseteq^z i_2^\sharp \end{cases}$$

The definition follows the intuition of considering all cases of polarities for $i_1^\sharp$ and $i_2^\sharp$. The first case can be read "if either argument could be positive, then the result could be positive", and the third case "if both arguments could be zero, then the result could be zero". The join outside the cases gives the "smallest" results which is consistent with each case. *E.g.*,

$$[\![+]\!]^{a\sharp}(\texttt{posz}, \texttt{zero}) = \bigsqcup\{\texttt{pos}, \texttt{none}, \texttt{zer}, \texttt{none}, \texttt{none}\} = \texttt{posz}$$

*The Classical Design* To contrast with Cousot's original design using classical abstract interpretation, the key difference is that we avoid powerset liftings alltogether. Using classical Galois connections, the concrete denotation function must be lifted to powersets:

$$[\![_]\!]^a_\wp : \texttt{aexp} \rightarrow \wp(\mathbb{Z} \times \mathbb{Z}) \rightarrow \wp(\mathbb{Z}) \qquad [\![ae]\!]^a_\wp(II) := \{[\![ae]\!]^a(i_1, i_2) \mid \langle i_1, i_2 \rangle \in II\}$$

and then $[\![_]\!]^{a\sharp}$ is proven correct w.r.t. this lifting using $\alpha^z$ and $\gamma^z$:

$$\alpha^z([\![ae]\!]^a_\wp(\gamma^z(i_1^\sharp, i_2^\sharp))) = [\![ae]\!]^{a\sharp}(i_1^\sharp, i_2^\sharp)$$

This property cannot be mechanized without axioms because $\alpha^z$ is non-constructive. Furthermore, the proof involves additional powerset boilerplate reasoning, which is not present in our mechanization of correctness for $[\![_]\!]^{a\sharp}$ using constructive Galois connections. The state-of-the art approach of "$\gamma$-only" verification would instead mechanize the $\gamma\gamma$ variant of correctness:

$$[\![ae]\!]^a_\wp(\gamma^z(i_1^\sharp, i_2^\sharp)) = \gamma^z([\![ae]\!]^{a\sharp}(i_1^\sharp, i_2^\sharp))$$

which is similar to our $\mu\mu$ rule:

$$\langle i_1, i_2 \rangle \in \mu^{z \times z}(i_1^\sharp, i_2^\sharp) \wedge \langle i_1', i_2' \rangle = [\![ae]\!]^a(i_1, i_2) \Leftrightarrow \langle i_1', i_2' \rangle \in \mu^z([\![ae]\!]^{a\sharp}(i_1^\sharp, i_2^\sharp))$$

The benefit of our approach is that soundness and completeness properties which also mention extraction ($\eta$) can also be mechanized, like calculating abstract interpreters from their specification.

**Abstracting Relations** The verification of an abstract interpreter for relations is similar to the design for functions: induce a specification using the constructive Galois connection, and prove correctness w.r.t. the induced spec. The relations we abstract are $\Downarrow^a$, $\Downarrow^b$ and $\mapsto^c$, and we call their abstract interpreters $\mathcal{A}^\sharp$, $\mathcal{B}^\sharp$ and $\mathcal{C}^\sharp$. Rather than postulate the definitions of the abstract interpreters, we calculate them from their specifications, the results of which are sound and computable by construction. The arithmetic and boolean abstract interpreters are functions from abstract environments to abstract integers, and the abstract interpreter for commands computes the next abstract transition states of execution. (We only present select calculations for $\mathcal{A}^\sharp$; see our accompanying Agda development for each calculation in mechanized form, and Sections 9, 10 and 11 for detailed calculations of binary arithmetic operators and conditional command expressions.) $\mathcal{A}^\sharp$ has type:

$$\mathcal{A}^\sharp[_] \;:\; \mathtt{aexp} \to \mathtt{env}^\sharp \nearrow \mathbb{Z}^\sharp$$

To induce a spec for $\mathcal{A}^\sharp$, we first revisit the concrete semantics relation as a powerset-valued function, which we call $\mathcal{A}$:

$$\mathcal{A}[_] \;:\; \mathtt{aexp} \to \mathtt{env} \to \wp(\mathbb{Z}) \qquad\qquad \mathcal{A}[ae](\rho) \;:=\; \{i \mid \rho \vdash ae \Downarrow^a i\}$$

The induced spec for $\mathcal{A}^\sharp$ is generated with the monadic bind operator, which we notate using Moggi's star notation $_^*$:

$$pure(\eta^z)^*(\mathcal{A}[ae]^*(\mu^r(\rho^\sharp))) \subseteq pure(\mathcal{A}^\sharp[ae])(\rho^\sharp)$$

which unfolds to:

$$\{\eta^z(i) \mid \rho \in \mu^r(\rho^\sharp) \wedge \rho \vdash ae \Downarrow^a i\} \subseteq \{\mathcal{A}^\sharp[ae](\rho^\sharp)\}$$

To calculate $\mathcal{A}^\sharp$ we reason equationally from the spec on the left towards the singleton set on the right, and declare the result the definition of $\mathcal{A}^\sharp$. We do this by case analysis on $ae$; we show the cases for $ae = \mathtt{rand}$ and $ae = x$ in Figure 3. Each calculation can also be written in monadic form, which is the style we mechanize; we repeat the variable case in monadic form in the figure.

**Mechanized Calculation** Our Agda calculation of $\mathcal{A}^\sharp$ strongly resembles the on-paper monadic one. We show the Agda proof code for abstract variable references in Figure 4. The first line is the top level definition site for the derivation of $\mathcal{A}^\sharp$ for the `Var` case. The `proof-mode` term is part of our "proof-mode" library which gives support for calculational reasoning in the form of Agda proof combinators with mixfix syntax. Statements surrounded by double square brackets $[[e]]$ restate the current proof state, which Agda will check is correct. Reasoning steps are employed through ≀ e ∫ terms, which transform the proof state from the previous form to the next. Equality steps (which do not lose precision) are notated ≀ e ∫[≈], whereas ordered steps (which may lose precision) are notated ≀ e ∫[⊑]. The term [`focus-right` [·] *of e*] focuses the goal to the right of the outermost application, scoped between `begin` and `end`.

The mechanized proof proceeds by focusing to $\mathcal{A}[x]^*(\mu^r(\rho^\sharp))$ (Line 03). The proof state is rewritten *via* an equality based on the definition of $\mathcal{A}[x]$ (Line 05), which corresponds to the first step of the on-paper derivation. (The Agda expression `pure lookup[` $x$ `]` is identical to $pure(\lambda\rho.\rho(x))$. We don't write the literal lambda in the Agda because each lambda used in the calculation must come with a proof of monotonicity, which we instead provide at the definition site of the helper operation `lookup[_]`.) The next step of the Agda calculation (Line 07) replaces `pure · lookup[` $x$ `]` $\odot$ $\mu^r$ with an over-approximation $\mu^z$ $\odot$ `pure · lookup♯[` $x$ `]` (where $\odot$ is the monadic composition operator) justified by a separate small proof

**Case** $ae = \mathsf{rand}$**:**

$$
\begin{aligned}
&\{\eta^z(i) \mid \rho \in \mu^r(\rho^\sharp) \wedge \rho \vdash \mathsf{rand} \Downarrow^a i\} \\
&= \{\eta^z(i) \mid \rho \in \mu^r(\rho^\sharp) \wedge i \in \mathbb{Z}\} && \wr\ \text{defn. of } \rho \vdash \mathsf{rand} \Downarrow^a i\ \int \\
&\subseteq \{\eta^z(i) \mid i \in \mathbb{Z}\} && \wr\ \varnothing \text{ when } \mu^r(\rho^\sharp) = \varnothing\ \int \\
&\subseteq \{\mathsf{any}\} && \wr\ \{\mathsf{any}\} \text{ mon. w.r.t. } \sqsubseteq^z\ \int \\
&\triangleq \{\mathcal{A}^\sharp[\mathsf{rand}](\rho^\sharp)\} && \wr\ \text{defining } \mathcal{A}^\sharp[\mathsf{rand}]\ \int
\end{aligned}
$$

**Case** $ae = x$**:**

$$
\begin{aligned}
&\{\eta^z(i) \mid \rho \in \mu^r(\rho^\sharp) \wedge \rho \vdash x \Downarrow^a i\} \\
&= \{\eta^z(\rho(x)) \mid \rho \in \mu^r(\rho^\sharp)\} && \wr\ \text{defn. of } \rho \vdash x \Downarrow^a i\ \int \\
&= \{\eta^z(i) \mid i \in \mu^z(\rho^\sharp(x))\} && \wr\ \text{defn. of } \mu^r(\rho^\sharp)\ \int \\
&\subseteq \{\rho^\sharp(x)\} && \wr\ \text{Eq. (CGC-Red)}\ \int \\
&\triangleq \{\mathcal{A}^\sharp[x](\rho^\sharp)\} && \wr\ \text{defining } \mathcal{A}^\sharp[x]\ \int
\end{aligned}
$$

**Case** $ae = x$ **(Monadic):**

$$
\begin{aligned}
&pure(\eta^z)^*(\mathcal{A}[x]^*(\mu^r(\rho^\sharp))) \\
&= pure(\eta^z)^*(pure(\lambda\rho.\rho(x))^*(\mu^r(\rho^\sharp))) && \wr\ \text{defn. of } \mathcal{A}[x]\ \int \\
&\subseteq pure(\eta^z)^*(\mu^{z*}(\rho^\sharp(x))) && \wr\ \text{defn. of } \mu^r(\rho^\sharp)\ \int \\
&\subseteq ret(\rho^\sharp(x)) && \wr\ \text{Eq. (CGC-Red)}\ \int \\
&\triangleq pure(\mathcal{A}^\sharp[x])(\rho^\sharp) && \wr\ \text{defining } \mathcal{A}^\sharp[x]\ \int
\end{aligned}
$$

Fig. 3. Case Study 1: Select Constructive Galois Connection Calculations

named `lookup`/$\mu^r$/`defn`, and which corresponds to the second step of the on-paper derivation. The Agda version includes an extra step (lines 09–10) to explicitly reduce the monadic expression:

$$\mu^z \ * \ \cdot \ (\mathtt{pure} \ \cdot \ \mathtt{lookup}\sharp[\ x\ ] \ \cdot \ \rho^\sharp) \quad \equiv \quad \mu^z \ \cdot \ (\mathtt{lookup}\sharp[\ x\ ] \ \cdot \ \rho^\sharp)$$

using one of the monad laws (the right-unit law, which is named $*$/`right-unit` in Agda), whereas this step is implicit in the on-paper derivation. The last step is to apply the reductive property of the constructive Galois connection (Line 14), after which we define $\mathcal{A}\sharp[_]$ through unification in Agda with the resulting definition.

Using constructive Galois connections, our mechanized calculation closely follows Cousot's classical one, uses both $\eta$ and $\mu$ mappings, and results in a verified, executable static analyzer. Such a result is not possible using classical Galois connections, due to the appearance of $\alpha$ inside the calculated algorithm, and the inability to encode $\alpha$ functions constructively.

We complete the full calculation of Cousot's generic abstract interpreter for `WHILE` in Agda as supplemental material to this paper, where the resulting interpreter is both sound and computable by construction. We also provide our "proof-mode" library which supports general calculational reasoning with posets.

```
-- Agda Calculation of Case ae = x:
α[𝒜] (Var x) ρ♯ = [proof-mode]
  do [[ (pure · η^z) ∗ · (𝒜[ Var x ] ∗ · (μ^r · ρ♯)) ]]
   ▪ [focus-right [·] of (pure · η^z) ∗ ] begin
     do [[ 𝒜[ Var x ] ∗ · (μ^r · ρ♯) ]]
      ▪ ⟅ 𝒜[Var]/defn ⟆[≈]
      ▪ [[ (pure · lookup[ x ]) ∗ · (μ^r · ρ♯) ]]
      ▪ ⟅ lookup/μ^r/defn ⟆[⊑]
      ▪ [[ μ^z ∗ · (pure · lookup♯[ x ] · ρ♯) ]]
      ▪ [[ μ^z ∗ · (ret · (lookup♯[ x ] · ρ♯)) ]]
      ▪ ⟅ ∗/right-unit ⟆[≈]
      ▪ [[ μ^z · (lookup♯[ x ] · ρ♯) ]]
     end
   ▪ [[ (pure · η^z) ∗ · (μ^z · (lookup♯[ x ] · ρ♯)) ]]
   ▪ ⟅ reductive[ημ] ⟆[⊑]
   ▪ [[ ret · (lookup♯[ x ] · ρ♯) ]]
   ▪ [[ pure · 𝒜♯[ Var x ] · ρ♯ ]] □
```

Fig. 4. Case Study 1: Constructive Galois Connection Calculations in Agda

*The Classical Design* Classically, one first designs a powerset lifting of the concrete semantics, called a *collecting semantics*:

$$\mathcal{A}_{\wp}[_] : \mathsf{aexp} \to \wp(\mathsf{env}) \nearrow \wp(\mathbb{Z}) \qquad \mathcal{A}_{\wp}[ae](R) := \{i \mid \rho \in R \wedge \rho \vdash ae \Downarrow^{a} i\}$$

The classical soundness specification for $\mathcal{A}^\sharp[ae](\rho^\sharp)$ is then:

$$\alpha^z(\mathcal{A}_{\wp}[ae](\gamma^r(\rho^\sharp))) \sqsubseteq \mathcal{A}^\sharp[ae](\rho^\sharp)$$

However, as usual, the abstraction $\alpha^z$ cannot be mechanized effectively, preventing a mechanized derivation of $\mathcal{A}^\sharp$ by calculus.

## 5 Case Study 2: Gradual Type Systems

Recent work in metatheory for gradual type systems (Garcia *et al.*, 2016) shows how a Galois connection discipline can guide the design of gradual typing systems. Starting with a Galois connection between precise and gradual types, both the static and dynamic semantics of the gradual language are derived systematically. This technique is called Abstracting Gradual Typing (AGT).

The design presented by Garcia *et al* is to begin with a precise type system, like the simply typed lambda calculus, and add a new type (?) which functions as the top element ($\top$) in the lattice of type precision. The precise typing rules are presented with meta-operators for subtyping (<:) and for the join operator in the subtyping lattice ($\ddot{\vee}$). The gradual type system is then written using abstract variants of

$$\boxed{\Gamma \vdash e : \tau}$$

$$\frac{\Gamma \vdash e_1 : \tau_1 \quad \tau_1 <: \mathbb{B} \quad \Gamma \vdash e_2 : \tau_2 \quad \Gamma \vdash e_3 : \tau_3}{\Gamma \vdash \texttt{if}\ e_1\ \texttt{then}\ e_2\ \texttt{else}\ e_3 : \tau_2 \mathbin{\ddot{\vee}} \tau_3}\ \textsc{If} \qquad \frac{\Gamma \vdash e_1 : \tau_1 \quad \tau_1 <: \tau_{11} \to \tau_{21} \quad \Gamma \vdash e_2 : \tau_2 \quad \tau_2 <: \tau_{11}}{\Gamma \vdash e_1(e_2) : \tau_{21}}\ \textsc{App}$$

$$\frac{\Gamma \vdash e : \tau_1 \quad \tau_1 <: \tau_2}{\Gamma \vdash e :: \tau_2 : \tau_2}\ \textsc{Coe}$$

Fig. 5. Case Study 2: Syntax Directed Precise Type System

subtyping and join ($<:^\sharp$ and $\ddot{\vee}^\sharp$) which are proven correct w.r.t. specifications induced by the Galois connection.

**The Precise Type System** The AGT paper describes two designs for gradual type systems in increasing complexity. We chose to mechanize a hybrid of the two which is simple, like the first design, yet still exercises key challenges addressed by the second. We also made slight modifications to the design at parts to make mechanization easier, but without changing the nature of the system.

The precise type system we mechanized is the simply typed lambda calculus with booleans, and top and bottom elements for a subtyping lattice, which we call `any` and `none`:

$$\tau \in \texttt{type} ::= \texttt{none} \mid \mathbb{B} \mid \tau \to \tau \mid \texttt{any}$$

Terms are standard boolean terms with if/then/else conditionals, lambda expressions, and a type ascription term $e :: \tau$:

$$e \in \texttt{exp} ::= \texttt{true} \mid \texttt{false} \mid \texttt{if}\ e\ \texttt{then}\ e\ \texttt{else}\ e \mid x \mid \lambda x.e \mid e(e) \mid e :: \tau$$

The first design in the AGT paper does not involve subtyping, and their second design incorporates record types with width and depth subtyping. By just focusing on `none` and `any`, we exercise the subtyping machinery of their approach without the blowup in complexity from formalizing record types.

The typing rules in AGT are written in strictly syntax-directed form, with explicit use of subtyping in rule hypotheses. In Figure 5 we show three precise typing rules for if-statements, application and coercion. The subtyping lattice in the precise system is the "safe for substitution" lattice, and well typed programs enjoy progress and preservation.

**Gradual Types** The essence of AGT is to design a gradual type system by *abstract interpretation* of the precise type system. To do this, a new top element is added to the precise type system, although rather than representing the top of the *subtyping* lattice like any, it represents the top of the *precision* lattice, and is notated ?:

$$\tau^\sharp \in \mathtt{type}^\sharp ::= \mathtt{none} \mid \mathbb{B} \mid \tau^\sharp \to \tau^\sharp \mid \mathtt{any} \mid ?$$

The partial ordering has ? at the top ($\tau^\sharp \sqsubseteq ?$) and is otherwise discrete, and arrow types are monotonic (covariant) in both the domain and codomain:

$$(\tau^\sharp_{11} \to \tau^\sharp_{21}) \sqsubseteq (\tau^\sharp_{12} \to \tau^\sharp_{22}) \iff \tau^\sharp_{11} \sqsubseteq \tau^\sharp_{12} \ \wedge\ \tau^\sharp_{21} \sqsubseteq \tau^\sharp_{22}$$

Just as in our other designs by abstract interpretation, $\mathtt{type}^\sharp$ is given meaning by an interpretation function $\mu$, which is the constructive analog of a classical concretization function $\gamma$:

$$\mu : \mathtt{type}^\sharp \nearrow \wp(\mathtt{type}) \qquad \begin{aligned} \mu(\tau^\sharp) &:= \tau \quad \mathit{when} \quad \tau^\sharp = \tau \in \{\mathtt{none}, \mathbb{B}, \mathtt{any}\} \\ \mu(\tau^\sharp_1 \to \tau^\sharp_2) &:= \{\tau_1 \to \tau_2 \mid \tau_1 \in \mu(\tau^\sharp_1) \wedge \tau_2 \in \mu(\tau^\sharp_2)\} \\ \mu(?) &:= \{\tau \mid \tau \in \mathit{type}\} \end{aligned}$$

The extraction function $\eta$ is, remarkably, the identity function:

$$\eta : \mathtt{type} \to \mathtt{type}^\sharp \qquad\qquad \eta(\tau) = \tau$$

and that the constructive Galois correspondence (CGC-Corr) easily follows:

$$\tau \in \mu(\tau^\sharp) \iff \eta(\tau) \sqsubseteq \tau^\sharp$$

**Constructive GCs in Agda** In Agda, the interpretation function $\mu$ takes the form of an inductively defined relation:

```
data _∈μᵗ[_] : type → type♯ → Set where
  ⊤ : ∀ {τ} → τ ∈μᵗ[ ⊤ ]
  Any : Any ∈μᵗ[ Any ]
  None : None ∈μᵗ[ None ]
  ⟨𝔹⟩ : ⟨𝔹⟩ ∈μᵗ[ ⟨𝔹⟩ ]
  _⟨→⟩_ : ∀ {τ₁ τ₂ τ₁♯ τ₂♯}
    → τ₁ ∈μᵗ[ τ₁♯ ]
    → τ₂ ∈μᵗ[ τ₂♯ ]
    → (τ₁ ⟨→⟩ τ₂) ∈μᵗ[ τ₁♯ ⟨→⟩ τ₂♯ ]
```

and the extraction function $\eta$ is the identity *injection* from precise types to gradual types, because in Agda type is not a subtype of type♯, rather they are disjoint types:

```
ηᵗ : type → type♯
ηᵗ Any = Any
ηᵗ None = None
ηᵗ ⟨𝔹⟩ = ⟨𝔹⟩
ηᵗ (τ₁ ⟨→⟩ τ₂) = ηᵗ τ₁ ⟨→⟩ ηᵗ τ₂
```

**Gradual Operators** Given the constructive Galois connection between gradual and precise types, we synthesize specifications for abstract analogs of subtyping ($<:$) and the subtyping join operator ($\ddot{\vee}$), and relate them to their abstractions ($<:^\sharp$ and $\ddot{\vee}^\sharp$). In the AGT paper, the specification for abstract subtyping is generated by predicate lifting on the RHS of the following bi-implication:

$$\tau_1^\sharp <:^\sharp \tau_2^\sharp \iff \tau_1 <: \tau_2 \text{ for some } \langle \tau_1, \tau_2 \rangle \in \langle \mu(\tau_1^\sharp), \mu(\tau_2^\sharp) \rangle$$

The specification for abstract joins is generated *via* standard pre and post composition with extraction ($\eta$) and interpretation ($\mu$) functions on the RHS of the following equality:

$$\tau_1^\sharp \, \ddot{\vee}^\sharp \, \tau_2^\sharp = \bigsqcup \{ \eta(\tau_1 \, \ddot{\vee} \, \tau_2) \mid \tau_1 \in \mu(\tau_1^\sharp), \tau_2 \in \mu(\tau_2^\sharp) \}$$

In Agda we define abstract subtyping and abstract join following the AGT paper, and prove them sound w.r.t. their induced specifications. In particular, the above specifications guarantee that the gradual type ? will satisfy the standard rules for gradual subtyping and join:

$$? <:^\sharp \tau^\sharp \qquad\qquad \tau^\sharp <:^\sharp ? \qquad\qquad ? \, \ddot{\vee}^\sharp \, \tau^\sharp = \tau^\sharp \, \ddot{\vee}^\sharp \, ? = ?$$

The first two gradual subtyping rules for ? are surprising to those unfamiliar with the literature on gradual typing. In the context of AGT, they are justified by the specification induced by Galois connection.

$$\begin{array}{ll}
? <:^\sharp \tau^\sharp & \\
\Leftrightarrow \tau_1 <: \tau_2 \text{ for some } \langle \tau_1, \tau_2 \rangle \in \langle \mu(?), \mu(\tau^\sharp) \rangle & \wr \ \text{specification for } <:^\sharp \ \int \\
\Leftarrow \text{there exists } \tau \text{ s.t. } \langle \tau, \tau \rangle \in \langle \mu(?), \mu(\tau^\sharp) \rangle & \wr \ \tau <: \tau \text{ for all } \tau \ \int \\
\Leftrightarrow \text{there exists } \tau \text{ s.t. } \tau \in \mu(\tau^\sharp) & \wr \ \tau \in \mu(?) \text{ for all } \tau \ \int \\
\Leftrightarrow \texttt{true} & \wr \ \mu(\tau^\sharp) \neq \varnothing \text{ for all } \tau^\sharp \ \int
\end{array}$$

**Gradual Metatheory** Using AGT, the gradual type system is a syntactic analog to the precise one but with gradual types and operators, which we show in Figure 6. Using this system, and constructive Galois connections, we mechanize in Agda three key metatheory results from the AGT paper. We mechanize (1) equivalence for fully-annotated terms (FAT), which states that any term $e$ which is typeable at $\tau$ under the precise system is also typeable at $\tau$ under the gradual system. We mechanize (2) embedding of dynamic language terms (EDL), which states that any closed untyped term is typeable under the gradual system at type ? *via* an embedding $\lceil _ \rceil$ that annotates sub-terms with ?. Finally, we mechanize (3) the gradual guarantee (GG), which states that decreasing the precision of types (by going higher in the lattice) does not affect the typeability of any term under the gradual system:

$$\vdash e : \tau \iff \vdash^G e : \tau \tag{FAT}$$

$$\texttt{closed}(e) \implies \vdash^G \lceil e \rceil : ? \tag{EDL}$$

$$\vdash^G e_1^\sharp : \tau_1^\sharp \ \wedge \ e_1^\sharp \sqsubseteq e_2^\sharp \implies \vdash^G e_2^\sharp : \tau_2^\sharp \ \wedge \ \tau_1^\sharp \sqsubseteq \tau_2^\sharp \tag{GG}$$

$$\boxed{\Gamma^\sharp \vdash e^\sharp : \tau^\sharp}$$

$$\frac{\begin{array}{ll}\Gamma^\sharp \vdash^G e_1^\sharp : \tau_1^\sharp & \tau_1^\sharp <:^\sharp \mathbb{B} \\ \Gamma^\sharp \vdash^G e_2^\sharp : \tau_2^\sharp & \\ \Gamma^\sharp \vdash^G e_3^\sharp : \tau_3^\sharp & \end{array}}{\Gamma^\sharp \vdash^G \texttt{if}\ e_1\ \texttt{then}\ e_2\ \texttt{else}\ e_3 : \tau_2^\sharp \mathbin{\ddot{\vee}^\sharp} \tau_3^\sharp}\ \textsc{G-If}$$

$$\frac{\begin{array}{ll}\Gamma^\sharp \vdash^G e_1^\sharp : \tau_1^\sharp & \tau_1^\sharp <:^\sharp \tau_{11}^\sharp \to \tau_{21}^\sharp \\ \Gamma^\sharp \vdash^G e_2^\sharp : \tau_2^\sharp & \tau_2^\sharp <:^\sharp \tau_{11}^\sharp \end{array}}{\Gamma^\sharp \vdash^G e_1^\sharp(e_2^\sharp) : \tau_{21}^\sharp}\ \textsc{G-App} \qquad \frac{\Gamma^\sharp \vdash^G e^\sharp : \tau_1^\sharp \qquad \tau_1^\sharp <:^\sharp \tau_2^\sharp}{\Gamma^\sharp \vdash^G e^\sharp :: \tau_2^\sharp : \tau_2^\sharp}\ \textsc{G-Coe}$$

Fig. 6. Case Study 2: Systematically Constructed Gradual Type System

## 6 Mechanization in Agda

In this section we guide the reader through the details of our mechanization approach in Agda, and highlight areas of mechanization that were challenging or otherwise of interest. The mechanization can be found at `github.com/plum-umd/cgc`.

Our mechanization consists of five Agda modules organized into folders: a custom core library (folder `/Prelude`), a module for manipulating partially ordered sets and downward closed powersets (folder `/Poset`), a module for Galois connections, both classical and constructive (`/Poset/GaloisConnection`), the first case study of a calculating a generic abstract interpreter (folder `/CDGAI`), and the second case study of verifying a gradual type system *via* abstract interpretation with constructive Galois connections (folder `/ADI`).

In the rest of this section we show code snippets which are slightly simplified from the exact code in the project for the purposes of presentation, *e.g.*, we inline some definitions and omit universe annotations to Agda datatypes.

**Posets** We define partial orders in Agda first as pre orders. These pre orders induce an equivalence relation, and the partial order we work with is the one induced by the pre order w.r.t. its induced equivalence relation. We encode pre orders as an Agda dependent record which contains the carrier relation as well as proofs of pre-order laws, as shown in Figure 7.

The line `open PreOrder {{…}} public` makes this record a candidate for typeclass resolution, meaning a canonical instance of `PreOrder X` will be selected automatically when an implicit argument is needed at that type. Implicit arguments which trigger typeclass resolution are written `{{_ : PreOrder X}}`. We show the induced equivalence relation `_≍_` and the antisymmetry law for the partial order induced by `_≼_` w.r.t. `_≍_`.

We encode posets in Agda as a wrapper datatype `Poset` around a carrier set and its `PreOrder` record where the wrapping is witnessed by `⇑`, also shown in Figure 7. Elements of partially ordered sets are encoded as a wrapper around elements of the

```
record PreOrder (A : Set) : Set where
  field
    -- carrier relation
    _≼_ : A → A → Set
    -- reflexivity
    xRx[≼] : ∀ {x} → x ≼ x
    -- transitivity
    _⊚[≼]_ : ∀ {x y z} → y ≼ z → x ≼ y → x ≼ z
open PreOrder {{…}} public
-- induced equivalence relation
_≍_ : ∀ {A} {{_ : PreOrder A}} → A → A → Set
x ≍ y = (x ≼ y) ∧ (y ≼ x)
-- antisymmetry
_⋈[≼]_ : ∀ {A} {{_ : PreOrder A}} {x y} → (x ≼ y) → (y ≼ x) → x ≍ y
ε₁ ⋈[≼] ε₂ = ⟨ ε₁ , ε₂ ⟩

data Poset : Set where
  ⇧ : (A : Set) → {{_ : PreOrder A}} → Poset
dom : ∀ → Poset → Set
dom (⇧ A) = A
data ⟪_⟫ (A : Poset) : Set where
  ⟨_⟩ : dom A → ⟪ A ⟫
data _⊑_ {A : Poset} : relation ⟪ A ⟫ where
  ⟨_⟩ : {x y : dom A} → x ≼ y → ⟨ x ⟩ ⊑ ⟨ y ⟩
data _≈_ {A : Poset} : relation ⟪ A ⟫ where
  ⟨_,_⟩ : ∀ {x y} → x ⊑ y → y ⊑ x → x ≈ y
```

Fig. 7. Pre Orders, Partial Orders, and Posets in Agda

carrier set, that is `⟨ x ⟩ : ⟪ ⇧ A ⟫` when `x : A`. We do this wrapping to improve typechecking, *e.g.*, Agda sometimes has difficulty resolving implicit arguments when `x : A`, `y : A`, `PreOrder A` is in scope, and `x ≼ y` appears in the type of a term (*e.g*, like `_⋈[≼]_` shown previously). However these arguments are easily inferred by Agda when `x ⊑ y` appears in the type of a term and `x : ⟪ A ⟫`, `y : ⟪ A ⟫`.

**Monotonic Functions and Powersets** We encode monotonic functions as a native Agda function paired with an explicit proof of monotonicity, as shown in Figure 8. The ordering relation on functions is the pointwise ordering. We define notation for the lifting of monotonic functions from an Agda `Set` into an Agda `Poset`. Finally, we introduce notation for applying wrapped monotonic functions to wrapped elements of carrier sets.

We do this wrapping to control when native Agda function definitions are available for reduction during Agda's typechecking phase. During our abstract interpreter calculations, the Agda typechecker must unify native Agda functions for definitional equality, *e.g.*, as justification for a rewrite step. By wrapping functions, this reduction will not happen for functions supplied as parameters unless the function is explicitly unwrapped. *E.g.*, the expression `f · x · y` is syntactically a monotonic function applied to two arguments, and will remain neutral during

```
-- monotonicity proposition
mon : ∀ {A B} → (⟪ A ⟫ → ⟪ B ⟫) → Set
mon f = ∀ {x y} → x ⊑ y → f x ⊑ f y
-- monotonic function
data _↗_ (A : Poset) (B : Poset) : Set where
  [λ_] : ∀ (f : ⟪ A ⟫  → ⟪ B ⟫) {f-proper : mon f} → A ↗ B
-- ordering on monotonic functions
data _≼ᶠ_ {A : Poset} {B : Poset} : (A ↗ B) → (A ↗ B) → Set where
  ⟨_⟩ : ∀ {f f-proper g g-proper}
    → (∀ {x} → f x ⊑ g x)
    → [λ f ] {f-proper} ≼ᶠ [λ g ] {g-proper}
-- lifting from Set to Poset
_⇗_ : Poset → Poset → Poset
-- lifted function application
A ⇗ B = ⇑ (A ↗ B)
_·_ : ∀ {A B} → (A ⇗ B) → A → B
⟨ [λ f ] ⟩ · x = f x

-- antitonicity proposition
ant : ∀ {A} → (⟪ A ⟫ → Set) → Set
ant φ = ∀ {x y} → y ⊑ x → φ x → φ y
-- downward closed powerset
data pow (A : Poset) : Set where
  [ω_] : ∀ (φ : ⟪ A ⟫ → Set) {φ-proper : ant φ} → pow A
-- ordering on downward closed powerset
data _≼ᵖ_ {A : Poset} : pow A → pow A → Set where
  ⟨_⟩ : ∀ {φ φ-proper θ θ-proper}
    → (∀ {x} → φ x → θ x)
    → [ω φ ] {φ-proper} ≼ᵖ [ω θ ] {θ-proper}
-- lifting from Set to Poset
Pow : ∀ → Poset → Poset
Pow A = ⇑ (pow A)
-- lifted posetset containment proposition
_∈_ : ∀ {A : Poset} → ⟪ A ⟫ → ⟪ Pow A ⟫ → Set
x ∈ ⟨ [ω φ ] ⟩ = φ x
```

Fig. 8. Monotonic Functions and Downward Closed Powersets in Agda

typechecking because of the wrapping when `f` is a parameter. However, if we unwrap `f` *via* pattern matching `[λ f-native ] = f` (or did not used the wrapped encoding) this expression would reduce to `f-native x y`, which discards the fact that `f-native` is monotonic.

We encode downward closed powersets as Agda as an Agda characteristic function into `Set` paired with an explicit proof of antitonicity, also shown in Figure 8. The ordering relation on powersets is the pointwise ordering. We define notation for the lifting of downward closed powersets from an Agda `Set` into an Agda `Poset`. Finally, we introduce notation for element containment between a wrapped carrier set element and a wrapped powerset characteristic function.

```
record _⇄_ (A B : Poset) : Set where
  field
    α[_] : ⟪ A ↗ B ⟫
    γ[_] : ⟪ B ↗ A ⟫
    extensive[_] : id♮ ⊑ γ[_] ∘♮ α[_]
    reductive[_] : α[_] ∘♮ γ[_] ⊑ id♮
open _⇄_ public

record _⇄ᶜ_ (A B : Poset) : Set where
  field
    η[_] : ⟪ A ↗ B ⟫
    μ[_] : ⟪ B ↗ Pow A ⟫
    extensiveᶜ[_] : return ⊑ μ[_] ⊛ pure · η[_]
    reductiveᶜ[_] : pure · η[_] ⊛ μ[_] ⊑ return
open _⇄ᶜ_ public
```

Fig. 9. Classical and Constructive Galois Connections in Agda

Like monotonic functions, we do this wrapping to control when native Agda predicates are available for reduction during Agda's typechecking phase. *E.g.*, the expression `x ∈ φ` is syntactically a proposition that `x` is an element of the downward closed powerset `φ`, where `φ` is a parameter. However, if we unwrap `φ` *via* pattern matching `[ω φ-native ] = φ` this expression would reduce to `φ-native x`, which discards the fact that `φ` is antitonic.

**Galois Connections** Classical Galois connections are encoded as a dependent record containing both abstraction and concretization mapping, as well as expansive and reductive laws. Constructive Galois connections are encoded analogously, but for extraction and interpretation variants of abstraction and concretization. Both of these encodings are shown in Figure 9.

We define the identity function `id♮` and function composition `_∘♮_` as lifted to the monotonic function space `⟪ A ↗ B ⟫` (as opposed to native Agda functions `A → B`). `return` and `pure` are defined for the downward closed powerset monad, and `⊛` is monadic composition.

**Proof Mode Library** To facilitate calculational style proofs we develop a custom proof mode library, as shown in Figure 10. Our actual implementation is more generic (and therefore complicated) than what we show here, which has been simplified greatly for the sake of presentation.

We define a new type for "proof mode" calculations `[⊑] x ⇰ y` as the type of an ordered derivation starting from `x` and concluding with `y`, via a chain of equational and/or ordered reasoning. Derivations begin with `[proof-mode] do ε ∎` where `ε` is some derivation term written in Agda using the proof mode library. We write proof mode combinators inside of an Agda `abstract` block so that they are not reduced to their definitions during interactive type checking. *E.g.*, writing `[proof-mode] do ? ∎` in Agda's interactive mode will create a hole in place of `?` and display the type of the

```
begin_end : ∀ {A} → A → A
begin_end x = x
do_ : ∀ {A} → A → A
do_ x = x
abstract
  [⊑]_⇰_ : ∀ {A} → ⟪ A ⟫ → ⟪ A ⟫ → Set
  [⊑] x ⇰ y = x ⊑ y
  [proof-mode]_∎ : ∀ {A} {x y : ⟪ A ⟫} → [⊑] x ⇰ y → x ⊑ y
  [proof-mode] ε ∎ = ε
  _‣_ : ∀ {A} {x y z : ⟪ A ⟫} → [⊑] x ⇰ y → [⊑] y ⇰ z → [⊑] x ⇰ z
  ε₁ ‣ ε₂ = ε₂ ⊚[⊑] ε₁
  [[_]] : ∀ {A} (x : ⟪ A ⟫) → [⊑] x ⇰ x
  [[ x ]] = xRx[⊑]
  ⁅_⁆[≡] : ∀ {A} → {x y : ⟪ A ⟫} → x ≡ y → [⊑] x ⇰ y
  ⁅ ε ⁆[≡] = xRx[⊑/≡] ε
  ⁅_⁆[≈] : ∀ {A} → {x y : ⟪ A ⟫} → x ≈ y → [⊑] x ⇰ y
  ⁅ ε ⁆[≈] = xRx[⊑/≈] ε
  ⁅_⁆[⊑] : ∀ {A} {x y : ⟪ A ⟫} → x ⊑ y → [⊑] x ⇰ y
  ⁅ ε ⁆[⊑] = ε
  [focus-in_] : ∀ {A B} (f : ⟪ A ↗ B ⟫) {x y : ⟪ A ⟫}
    → [⊑] x ⇰ y
    → [⊑] f · x ⇰ f · y
  [focus-in f ] ε = mon[↗] f ε
  [focus-left_of_] : ∀ {A B C} (f : ⟪ A ↗ B ↗ C ⟫) {x x′ : ⟪ A ⟫} (y : ⟪ B ⟫)
    → [⊑] x ⇰ x′
    → [⊑] f · x · y ⇰ f · x′ · y
  [focus-left f of z ] ε = mon[·]L (mon[↗] fmk[⇰] ε)
  [focus-right_of_] : ∀ {A B C} (f : ⟪ A ↗ B ↗ C ⟫) (x : ⟪ A ⟫) {y y′ : ⟪ B ⟫}
    → [⊑] y ⇰ y′
    → [⊑] f · x · y ⇰ f · x · y′
  [focus-right f of z ] ε = mon[↗] (f · x) ε
```

Fig. 10. Proof Mode Library in Agda

hole to the user, which will be `[⊑] x ⇰ y` rather than its unfolding `x ⊑ y`. Ultimately, this proof mode library is just syntactic convenience for dealing with long chains of transitive and nested ordered reasoning.

The syntax `_‣_` composes two chains of reasoning and is designed with interactive use in mind. If the interactive goal is `[⊑] x ⇰ z` and the user has a sub-derivation `ε : [⊑] x ⇰ y`, they can write in the hole `ε ‣ ?` which will display the new proof state as `[⊑] y ⇰ z`. If there are unresolved meta-variables in the proof state, the user can write `[[ x ]] ‣ ?` which will succeed if the proof state can be unified with `[⊑] x ⇰ z` for some `z` (which may still contain metas). Reasoning steps may proceed by definitional equality `⁅ ε ⁆[≡]` when `ε : x ≡ y` and the proof state is `[⊑] x ⇰ z`, resulting in a new proof state `[⊑] y ⇰ v`. Analogously, they can also proceed by equivalence `⁅ ε ⁆[≈]` when `ε : x ≈ y`, and by weakening `⁅ ε ⁆[⊑]` when `ε : x ⊑ y`.

The proof library supports focusing inside the outer-term of the current proof state with `[focus-in f ] begin ε end`. This combinator is used when the current proof state is a function application `f · x` , and the user wants to proceed by ordered reasoning on the argument `x`, which is valid due to the monotonicity of `f`. Two variants are also provided for 2-ary functions, one which focuses on the first (left) argument, and another which focuses on the second (right): `[focus-left f of y ]` or `[focus-right f of x ]` when the current proof state is `f · x · y`.

It is common when using the proof mode library to begin a derivation with type `[⊑] x ⇰ y` where `x` is fully resolved—*e.g.*, as induced specification for an abstract interpreter—but where `y` is an unresolved metavariable—*e.g.*, the implementation of the abstract interpreter which will be discovered *via* the process of calculation. For example, a common setup is as follows:

```
f♯ : ⟪ A ⇗ B ⟫
f♯ = ?
calc : ∀ {x} → α · f · x ⊑ pure · f♯ · x
calc = [proof-mode]
  do [[ α · f · x ]]
   ‣ [[ (pure · η) * · (f * · (μ · x)) ]]
   ‣ ?
   ‣ [[ pure f♯ · x ]]
  ∎
```

where the hole in the definition of `calc` (written `?`) must be filled in with a derivation that calculates from the induced specification to some pure Agda function, which is guaranteed to carry algorithmic content. Once the derivation is complete, there will be some concrete term that will be unified with `f♯` within the definition of `calc`. The user can then ask Agda to automatically fill in the definition of `f♯` above using the interactive "auto" command in the Emacs frontend, which Agda will solve via unification with the derivation term `calc`.

## 7 Constructive Galois Connection Metatheory

In this section we develop the full metatheory of constructive Galois connection and prove precise claims about their relationship to classical Galois connections. We develop the metatheory of constructive Galois connections as an adjunction between posets with powerset-Kleisli adjoint functors. This is in contrast to classical Galois connections which come from an identical setup, but with the monotonic function space as adjoint functors, as shown in Figure 11. See Section 13 for a brief discussion on connections to more general category-theoretic constructions than those shown here.

We connect constructive to classical Galois connections through an isomorphism between a subset of classical to the entire space of constructive. To form this isomorphism we introduce an intermediate structure, Kleisli Galois connections, which we show are isomorphic to the classical subset, and isomorphic to constructive ones. This second isomorphism uses the constructive *theorem* of choice, as depicted in Figure 12. Both isomorphisms are themselves constructive, meaning they are suitable for use in mechanized verification with program extraction.

| *Adjunction* | classical GCs | Kleisli GCs |
|---|---|---|
| *Category* | posets | posets |
| *Adjoints* | monotonic functions | monotonic $\wp$-monadic functions |
| *Left Adjoint* | $\alpha : C \nearrow A$ | $\kappa\alpha : C \nearrow \wp(A)$ |
| *Right Adjoint* | $\gamma : A \nearrow C$ | $\kappa\gamma : A \nearrow \wp(C)$ |
| *Correspondence* | $id(x) \sqsubseteq \gamma(y) \Leftrightarrow \alpha(x) \sqsubseteq id(y)$ | $ret(x) \subseteq \kappa\gamma(y) \Leftrightarrow \kappa\alpha(x) \subseteq ret(y)$ |
| *Expansive* | $id \sqsubseteq \gamma \circ \alpha$ | $ret \sqsubseteq \kappa\gamma \circledast \kappa\alpha$ |
| *Reductive* | $\alpha \circ \gamma \sqsubseteq id$ | $\kappa\alpha \circledast \kappa\gamma \sqsubseteq ret$ |
| *Soundness* | $\alpha \circ f \circ \gamma \sqsubseteq f^\sharp$ | $\kappa\alpha \circledast f \circledast \kappa\gamma \sqsubseteq f^\sharp$ |
| *Optimality* | $\alpha \circ f \circ \gamma = f^\sharp$ | $\kappa\alpha \circledast f \circledast \kappa\gamma = f^\sharp$ |

Fig. 11. Comparison of Constructive and Classical Galois Connection Adjunctions

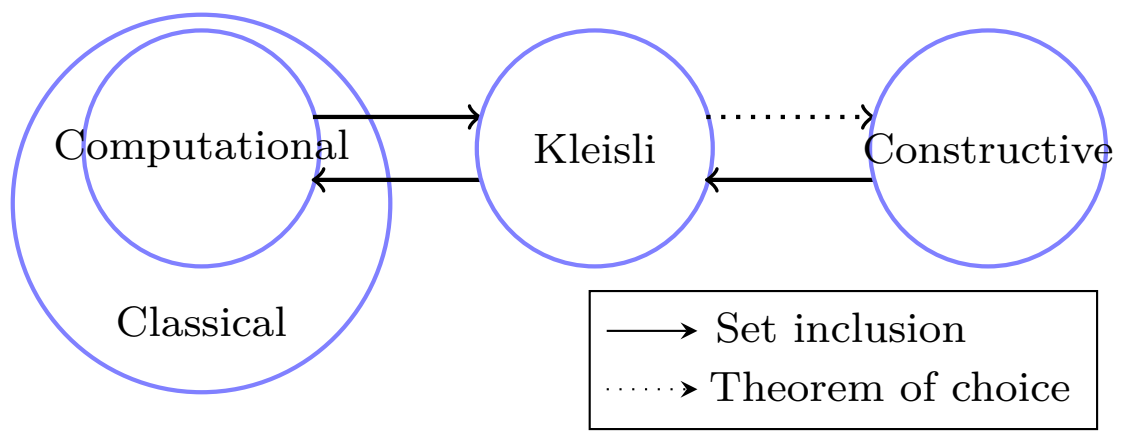

Fig. 12. Relationship Between Classical, Kleisli and Constructive Galois Connections

Kleisli Galois connections are introduced for two reasons. First, they are the "natural" structure generated by a bi-adjunction with powerset-Kleisli adjoint functors. It is therefore easier to defend Kleisli Galois connections as being a proper abstract interpretation framework because they are merely adjunctions, just like classical Galois connections. This is in contrast to constructive Galois connections which do not obviously follow an adjunction discipline. Second, we prove a surprising fact about Kleisli Galois connections, which is that they are isomorphic to constructive Galois connections. The insight gained here is that the monadic effect type on the abstraction side of the adjunction for Kleisli Galois connections ($\kappa\alpha : C \nearrow \wp(A)$) is provably benign, meaning it may as well be a pure function. Constructive Galois connections are Kleisli Galois connections where the abstraction function is written as a pure function without any loss of generality.

**Classical Galois Connections** We review classical Galois connections in Figure 11. A Galois connection between posets $C$ and $A$ contains two adjoint functors $\alpha$ and $\gamma$ which share a correspondence. An equivalent formulation of the correspondence is two unit equations called expansive and reductive. Abstract interpreters are sound by over-approximating a specification induced by $\alpha$ and $\gamma$.

**Powerset Monad** See Sections 3.1 and 3.3 for the downward-closure monotonicity property, and monad definitions and notation for the monotonic powerset monad. The monad operators obey standard monad laws. We introduce one new operator for monadic function composition: $(g \circledast f)(x) \coloneqq g^*(f(x))$.

**Kleisli Galois Connections** We summarize Kleisli Galois connections in Figure 11. Kleisli Galois connections are analogous to classical ones, but with monadic analogs to $\alpha$ and $\gamma$, and monadic identity and composition operators *ret* and $\circledast$ in place of the function space identity and composition operators *id* and $\circ$.

**Kleisli to Classical and Back** All Kleisli Galois connections $\langle \kappa\alpha, \kappa\gamma \rangle$ between $C$ and $A$ can be lifted to recover a classical Galois connection $\langle \alpha, \gamma \rangle$ between $\wp(C)$ and $\wp(A)$ through a monadic lifting operator on Kleisli Galois connections $\langle \kappa\alpha, \kappa\gamma \rangle^*$:

$$\langle \alpha, \gamma \rangle \triangleq \langle \kappa\alpha, \kappa\gamma \rangle^* \coloneqq \langle \kappa\alpha^*, \kappa\gamma^* \rangle$$

This lifting is *sound*, meaning Kleisli soundness and optimality results can be translated to classical ones.

*Theorem 1* (*KGC-Sound*)$^{AGDA\checkmark}$
For any Kleisli relationship of soundness between $f$ and $f^\sharp$, that is $\kappa\alpha \circledast f \circledast \kappa\gamma \sqsubseteq f^\sharp$, its lifting to classical is also sound, that is $\alpha \circ f^* \circ \gamma \sqsubseteq f^{\sharp *}$ where $\langle \alpha, \gamma \rangle = \langle \kappa\alpha, \kappa\gamma \rangle^*$, and likewise for optimality relationships $\alpha \circledast f \circledast \kappa y = f^\sharp$.

This lifting is also *complete*, meaning classical Galois connection soundness and optimality results can always be translated to Kleisli ones, when $\alpha$ and $\gamma$ are of lifted form.

*Theorem 2* (*KGC-Complete*)$^{AGDA\checkmark}$
For any classical relationship of soundness between $f^*$ and $f^{\sharp *}$, that is $\alpha \circ f^* \circ \gamma \sqsubseteq f^{\sharp *}$, its lowering to Kleisli is also sound when $\langle \alpha, \gamma \rangle = \langle \kappa\alpha, \kappa\gamma \rangle^*$, that is $\kappa\alpha \circledast f \circledast \kappa\gamma \sqsubseteq f^\sharp$, and likewise for optimality relationships $\alpha \circ f^* \circ \gamma = f^{\sharp *}$.

Due to soundness and completeness, one can work with the simpler setup of Kleisli Galois connections without any loss of generality. The setup is simpler in cases when the classical Galois connection is the lifting of a Kleisli Galois connection, because Kleisli Galois connection theorems only quantify over individual elements rather than elements of powersets. For example, the soundness criteria $\kappa\alpha \circledast f \circledast \kappa\gamma \sqsubseteq f^\sharp$ is proved by showing $\kappa\alpha^*(f^*(\kappa\gamma(y))) \subseteq f^\sharp(y)$ for an arbitrary element $y : A$, whereas in the classical proof (when derived from a lifted Kleisli setup) one must show $\kappa\alpha^*(f^*(\kappa\gamma^*(Y))) \subseteq f^{\sharp *}(Y)$ for arbitrary sets $Y : \wp(A)$.

**Constructive Galois Connections** Constructive Galois connections are a restriction of Kleisli Galois connections where the abstraction mapping is a pure rather than monadic function. We call the left adjoint *extraction*, notated $\eta$, and the right adjoint *interpretation*, notated $\mu$. The constructive Galois connection correspondence, alternative expansive and reductive formulation of the correspondence, and soundness and optimality criteria are identical to Kleisli Galois connections where $\langle \kappa\alpha, \kappa\gamma \rangle = \langle pure(\eta), \mu \rangle$.

**Constructive to Kleisli and Back** Our main theorem which justifies the soundness and completeness of constructive Galois connections is an isomorphism between constructive and Kleisli Galois connections. The easy direction is soundness, where a Kleisli Galois connection is formed by defining $\langle \kappa\alpha, \kappa\gamma\rangle = \langle pure(\eta), \mu\rangle$. Soundness and optimality theorems are then lifted from constructive to Kleisli without modification.

*Theorem 3* (*CGC-Sound*)$^{\mathsf{AGDA}\checkmark}$
For any constructive relationship of soundness between $f$ and $f^\sharp$, that is $pure(\eta) \circledast f \circledast \mu \sqsubseteq f^\sharp$, its lifting to Kleisli is sound, that is $\kappa\alpha \circledast f \circledast \kappa\gamma \sqsubseteq f^\sharp$ where $\langle \kappa\alpha, \kappa\gamma\rangle = \langle pure(\eta), \mu\rangle$, and likewise for optimality relationships $pure(\eta) \circledast f \circledast \mu = f^\sharp$.

The other direction, completeness, is much more surprising. First we establish a lowering for Kleisli Galois connections.

*Lemma 1* (*CGC-Induce*)$^{\mathsf{AGDA}\checkmark}$
For every Kleisli Galois connection $\langle \kappa\alpha, \kappa\gamma\rangle$, there exists a constructive Galois connection $\langle \eta, \mu\rangle$ where $\langle pure(\eta), \mu\rangle = \langle \kappa\alpha, \kappa\gamma\rangle$.

*Proof*
Because the mapping from Kleisli to constructive is interesting we provide a proof, which to our knowledge is novel. The proof builds a constructive Galois connection $\langle \eta, \mu\rangle$ from a Kleisli $\langle \kappa\alpha, \kappa\gamma\rangle$ by exploiting the Kleisli correspondence and making use of the constructive theorem of choice.

To turn an arbitrary Kleisli Galois connection into a constructive one, we show that the effect on $\kappa\alpha : C \nearrow \wp(A)$ is benign, or in other words, that there exists some $\eta$ such that $\kappa\alpha = pure(\eta)$. We prove this using two ingredients: a constructive interpretation of the Kleisli expansive law, and the constructive *theorem* of choice.

We first expand the Kleisli expansive property, unfolding definitions of $\circledast$ and *ret*, to get an equivalent logical statement:

$$\forall x. \exists y. y \in \kappa\alpha(x) \wedge x \in \kappa\gamma(y) \tag{KGC-Exp}$$

Statements of this form can be used in conjunction with an axiom of choice in classical mathematics, which is:

$$(\forall x. \exists y. R(x,y)) \implies \exists f. \forall x. R(x, f(x)) \tag{AxChoice}$$

This theorem is admitted as an *axiom* in classical mathematics, but in constructive logic—the setting used for extracting verified algorithms–(AxChoice) is definable as a *theorem*, due to the computational interpretation of logical connectives $\forall$ and $\exists$. (The formula AxChoice technically changes meaning when embedded in constructive logic, and is no longer equivalent to the classical axiom of choice once interpreted constructively.) We define (AxChoice) as a theorem in Agda without trouble:

```
choice : ∀ {A B} {R : A → B → Set}
  → (∀ x → ∃ y st R x y)
  → (∃ f st ∀ x → R x (f x))
choice P = ⟨∃ (λ x → π₁ (P x)) , (λ x → π₂ (P x)) ⟩
```

Applying (AxChoice) to (KGC-Exp) then gives:

$$\exists \eta . \forall x . \eta(x) \in \kappa\alpha(x) \wedge x \in \kappa\gamma(\eta(x)) \quad \text{(ExpChoice)}$$

which proves the existence of a pure function $\eta : C \nearrow A$.

In order to form a constructive Galois connection $\eta$ and $\mu$ must satisfy the correspondence, which we prove in split form:

$$x \in \mu(\eta(x)) \quad \text{(CGC-Exp)}$$

$$x \in \mu(y) \implies \eta(x) \sqsubseteq y \quad \text{(CGC-Red)}$$

The expansive property is immediate from the second conjunct in (ExpChoice). The reductive property follows from the Kleisli reductive property:

$$x \in \kappa\gamma(y) \wedge y' \in \kappa\alpha(x) \implies y' \sqsubseteq y \quad \text{(KGC-Red)}$$

The constructive variant of reductive is proved by satisfying the first two premises of (KGC-Red), where $x \in \kappa\gamma(y)$ is by assumption and $y' \in \kappa\alpha(x)$ is by the first conjunct in (ExpChoice).

So far we have shown that for a Kleisli Galois connection $\langle \kappa\alpha, \kappa\gamma \rangle$, there exists a constructive Galois connection $\langle \eta, \mu \rangle$ where $\mu = \kappa\gamma$. However, we have yet to show $pure(\eta) = \kappa\alpha$. To show this, we prove an analog of a standard result for classical Galois connections: that $\alpha$ and $\gamma$ uniquely determine each other.

*Lemma 2 (Unique Abstraction)*$^{\text{AGDA}\checkmark}$
For any two Kleisli Galois connections $\langle \kappa\alpha_1, \kappa\gamma_1 \rangle$ and $\langle \kappa\alpha_2, \kappa\gamma_2 \rangle$, $\kappa\alpha_1 = \kappa\alpha_2$ *iff* $\kappa\gamma_1 = \kappa\gamma_2$

We then conclude $pure(\eta) = \kappa\alpha$ as a consequence of the above lemma and the fact that $\mu = \kappa\gamma$.
□

Given the above mapping from Kleisli Galois connections to constructive ones, we prove the completeness of this mapping.

*Theorem 4 (CGC-Complete)*$^{\text{AGDA}\checkmark}$
For any Kleisli relationship of soundness between $f$ and $f^\sharp$, that is $\kappa\alpha \circledast f \circledast \kappa\gamma \sqsubseteq f^\sharp$, its lowering to constructive is also sound, that is $pure(\eta) \circledast f \circledast \mu \sqsubseteq f^\sharp$ where $\langle \eta, \mu \rangle$ is induced, and likewise for optimality relationships $\kappa\alpha \circledast f \circledast \kappa\gamma = f^\sharp$.

**Wrapping Up** In this section we showed that constructive Galois connections are sound w.r.t. classical Galois connections, and complete w.r.t. the subset of classical Galois connections recovered by lifting constructive ones. We showed this by introducing Kleisli Galois connections, and by establishing two isomorphisms: (1) between a subset of classical and Kleisli, and (2) between Kleisli and constructive. The proof of isomorphism between constructive and Kleisli yielded an interesting proof which applies the constructive theorem of choice to the Kleisli Galois connection correspondence laws.

## 8 Constructing Constructive Galois Connections

The classical Galois connection framework comes with a library of connectives which are used to build larger Galois connections out of smaller, primitive ones (Cousot & Cousot, 1994). For example, it is common to create a Galois connection for Cartesian products $(A \times B)$ as the product abstraction of two Galois connections, one for each side ($A$ and $B$).

In this section we define the constructive analog of many classical Galois connection connectives and primitives. Each constructive Galois connection we define is uniquely determined by just $\eta$, since $\mu$ is always derivable as its inverse image $\mu(y) := \{x \mid \eta(x) \sqsubseteq y\}$. However, we provide the canonical $\mu$ with a more direct definition. In later sections we will highlight similarities and differences between constructive and classical calculations (§ 9), how derivations of optimal abstract interpreters varies between the two settings (§ 10), and how multivalued computations are supported in the constructive setting (§ 11). Each section will make use of the connectives and primitives defined in this section without explicit introduction.

By convention we notate *classical* Galois connections $A \xleftrightarrow[\alpha]{\gamma} B$, that is with $\alpha$ and $\gamma$ below and above the arrows, and constructive Galois connections $A \xleftrightarrow[\eta]{\mu} B$, that is with $\eta$ and $\mu$ below and above the arrows. In the case of classical Galois connections, the domain and codomain of $\alpha$ and $\gamma$ are immediate from the notation, that is, $\alpha : A \nearrow B$ and $\gamma : B \nearrow A$. However for constructive Galois connections, the domain and codomain is only immediate from the notation for $\eta$ but not $\mu$ because it maps to a powerset in the codomain, that is $\eta : A \nearrow B$ but $\mu : B \nearrow \wp(A)$. We notate *pure*$(x)$ compactly as $\lfloor x \rfloor$ and assume all powersets are downward closed. In this section we abandon the convention of writing $C$ for the concrete set and $A$ for the abstract set. Instead, we write $A$, $B$, etc. for arbitrary posets, and annotate with a sharp sign for abstractions, *e.g.*, $A^\sharp$ as the abstraction of $A$.

### *8.1 Strictly Classical Galois Connections*

**Independent Attributes Abstraction** The independent attributes abstraction is defined for relations $(\wp(A \times B))$, and constructs the classical Galois connection:

$$\wp(A \times B) \xleftrightarrow[\overset{IA}{\alpha}]{\overset{IA}{\gamma}} \wp(A) \times \wp(B) \qquad \begin{aligned} \overset{IA}{\alpha} &: \wp(A \times B) \nearrow \wp(A) \times \wp(B) \\ \overset{IA}{\gamma} &: \wp(A) \times \wp(B) \nearrow \wp(A \times B) \end{aligned}$$

$$\begin{aligned} \overset{IA}{\alpha}(XY) &:= \langle \{x \mid \exists y.\langle x,y\rangle \in XY\}, \{y \mid \exists x.\langle x,y\rangle \in XY\}\rangle \\ \overset{IA}{\gamma}(X,Y) &:= \{\langle x,y\rangle \mid x \in X \wedge y \in Y\} \end{aligned}$$

### 8.2 Primitive Galois Connections—Classical and Constructive

**Identity Abstraction** The classical identity abstraction is defined for partially ordered sets $A$, and constructs the classical Galois connection:

$$A \underset{\alpha^{ID}}{\overset{\gamma^{ID}}{\leftrightarrows}} A \qquad \begin{array}{rl} \alpha^{ID} & : A \nearrow A \\ \gamma^{ID} & : A \nearrow A \end{array} \qquad \begin{array}{rl} \alpha^{ID}(x) & := x \\ \gamma^{ID}(x) & := x \end{array}$$

The constructive analog is defined for partially ordered sets $A$, and constructs the constructive Galois connection:

$$A \underset{\eta^{ID}}{\overset{\mu^{ID}}{\leftrightarrows}} A \qquad \begin{array}{rl} \eta^{ID} & : A \nearrow A \\ \mu^{ID} & : A \nearrow \wp(A) \end{array} \qquad \begin{array}{rl} \eta^{ID}(x) & := x \\ \mu^{ID}(x) & := \{x\} \end{array}$$

*Fact 1* (*Identity Abstraction Correspondence*)
The classical identity abstraction instantiated to $\wp(B)$ is equal to the classical lifting of the constructive identity abstraction, that is: $\alpha^{ID} = \lfloor \eta^{ID} \rfloor^{*}$ and $\gamma^{ID} = \mu^{ID*}$.

**Elementwise Abstraction** The elementwise abstraction (generalized to posets) is defined given a monotonic function $f : A \nearrow B$, and constructs the classical Galois connection:

$$\wp(A) \underset{\overset{[f]}{\alpha}}{\overset{\overset{[f]}{\gamma}}{\leftrightarrows}} \wp(B) \qquad \begin{array}{rl} \overset{[f]}{\alpha} & : \wp(A) \nearrow \wp(B) \\ \overset{[f]}{\gamma} & : \wp(B) \nearrow \wp(A) \end{array} \qquad \begin{array}{rl} \overset{[f]}{\alpha}(X) & := \{f(x) \mid x \in X\} \\ \overset{[f]}{\gamma}(Y) & := \{x \mid f(x) \in Y\} \end{array}$$

The constructive analog is defined given a monotonic function $f : A \nearrow B$ and constructs a constructive Galois connection $A \underset{\eta}{\overset{\mu}{\leftrightarrows}} B$ where:

$$\begin{array}{rl} \overset{[f]}{\eta} & : A \nearrow B \\ \overset{[f]}{\mu} & : B \nearrow \wp(A) \end{array} \qquad \begin{array}{rl} \overset{[f]}{\eta}(x) & := f(x) \\ \overset{[f]}{\mu}(y) & := \{x \mid f(x) \sqsubseteq y\} \end{array}$$

*Fact 2* (*Elementwise Abstraction Correspondence*)
The classical elementwise abstraction is equal to the classical lifting of the constructive elementwise abstraction, that is: $\overset{[f]}{\alpha} = \lfloor \overset{[f]}{\eta} \rfloor^{*}$ and $\overset{[f]}{\gamma} = \overset{[f]}{\mu}^{*}$.

**Least-upper-bound Abstraction** The least-upper-bound abstraction is defined for join-semilattices $A$, and constructs the classical Galois connection:

$$\wp(A) \underset{\overset{\sqcup}{\alpha}}{\overset{\overset{\sqcup}{\gamma}}{\leftrightarrows}} A \qquad \begin{array}{rl} \overset{\sqcup}{\alpha} & : \wp(A) \nearrow A \\ \overset{\sqcup}{\gamma} & : A \nearrow \wp(A) \end{array} \qquad \begin{array}{rl} \overset{\sqcup}{\alpha}(X) & := \bigsqcup_{x \in X} x \\ \overset{\sqcup}{\gamma}(x) & := \{x\} \end{array}$$

The constructive analog is defined for join-semilattices $A$, and constructs the *classical* Galois connection:

$$\wp(A) \underset{\overset{\sqcup\wp}{\alpha}}{\overset{\overset{\sqcup\wp}{\gamma}}{\leftrightarrows}} \wp^{1}(A) \qquad \begin{array}{rl} \overset{\sqcup\wp}{\alpha} & : \wp(A) \nearrow \wp^{1}(A) \\ \overset{\sqcup\wp}{\gamma} & : \wp^{1}(A) \nearrow \wp(A) \end{array} \qquad \begin{array}{rl} \overset{\sqcup\wp}{\alpha}(X) & := \{x \mid x \sqsubseteq \bigsqcup_{x \in X} x\} \\ \overset{\sqcup\wp}{\gamma}(X) & := \{x \mid x \in X\} \end{array}$$

We notate singleton (downward closed) powersets $\wp^1(_)$, which classically are isomorphic to the carrier set ($\wp^1(A) \leftrightarrows A$), but not constructively.

### 8.3 Composing Galois Connections—Classical and Constructive

**Abstraction Composition** The composition of two classical abstractions is defined given abstractions $B \underset{\alpha_1}{\overset{\gamma_1}{\leftrightarrows}} C$ and $A \underset{\alpha_2}{\overset{\gamma_2}{\leftrightarrows}} B$, and constructs the classical Galois connection:

$$A \underset{\overset{1\circ2}{\alpha}}{\overset{\overset{1\circ2}{\gamma}}{\leftrightarrows}} C \qquad \begin{aligned} \overset{1\circ2}{\alpha} &: A \nearrow C \\ \overset{1\circ2}{\gamma} &: C \nearrow A \end{aligned} \qquad \begin{aligned} \overset{1\circ2}{\alpha}(x) &:= \alpha_1(\alpha_2(x)) \\ \overset{1\circ2}{\gamma}(z) &:= \gamma_2(\gamma_1(z)) \end{aligned}$$

The constructive analog is defined given abstractions $B \underset{\eta_1}{\overset{\mu_1}{\leftrightarrows}} C$ and $A \underset{\eta_2}{\overset{\mu_2}{\leftrightarrows}} B$, and constructs the constructive Galois connection:

$$A \underset{\overset{1\circ2}{\eta}}{\overset{\overset{1\circ2}{\mu}}{\leftrightarrows}} C \qquad \begin{aligned} \overset{1\circ2}{\eta} &: A \nearrow C \\ \overset{1\circ2}{\mu} &: C \nearrow \wp(A) \end{aligned} \qquad \begin{aligned} \overset{1\circ2}{\eta}(x) &:= \eta_1(\eta_2(x)) \\ \overset{1\circ2}{\mu}(z) &:= \mu_2^*(\mu_1(z)) \end{aligned}$$

*Fact 3* (*Abstraction Composition Correspondence*)
The classical composition of lifted constructive abstractions is equal to the lifting of the constructive composition of those abstractions, that is: $\lfloor\eta_1\rfloor^* \circ \lfloor\eta_2\rfloor^* = (\lfloor\eta_1\rfloor \circledast \lfloor\eta_2\rfloor)^*$ and $\mu_2^* \circ \mu_1^* = (\mu_2 \circledast \mu_1)^*$.

**Product Abstraction** The classical product abstraction is defined given abstractions $A \underset{\alpha^A}{\overset{\gamma^A}{\leftrightarrows}} A^\sharp$ and $B \underset{\alpha^B}{\overset{\gamma^B}{\leftrightarrows}} B^\sharp$, and constructs the classical Galois connection:

$$A \times B \underset{\overset{A\times B}{\alpha}}{\overset{\overset{A\times B}{\gamma}}{\leftrightarrows}} A^\sharp \times B^\sharp \qquad \begin{aligned} \overset{A\times B}{\alpha} &: A \times B \nearrow A^\sharp \times B^\sharp \\ \overset{A\times B}{\gamma} &: A^\sharp \times B^\sharp \nearrow A \times B \end{aligned} \qquad \begin{aligned} \overset{A\times B}{\alpha}(x,y) &:= \langle \alpha^A(x), \alpha^B(y) \rangle \\ \overset{A\times B}{\gamma}(x^\sharp,y^\sharp) &:= \langle \gamma^A(x^\sharp), \gamma^B(y^\sharp) \rangle \end{aligned}$$

The constructive analog is defined given abstractions $A \underset{\eta^A}{\overset{\mu^A}{\leftrightarrows}} A^\sharp$ and $B \underset{\eta^B}{\overset{\mu^B}{\leftrightarrows}} B^\sharp$, and constructs the constructive Galois connection:

$$A \times B \underset{\overset{A\times B}{\eta}}{\overset{\overset{A\times B}{\mu}}{\leftrightarrows}} A^\sharp \times B^\sharp \qquad \begin{aligned} \overset{A\times B}{\eta} &: A \times B \nearrow A^\sharp \times B^\sharp \\ \overset{A\times B}{\mu} &: A^\sharp \times B^\sharp \nearrow \wp(A \times B) \end{aligned}$$

$$\begin{aligned} \overset{A\times B}{\eta}(x,y) &:= \langle \eta^A(x), \eta^B(y) \rangle \\ \overset{A\times B}{\mu}(x^\sharp,y^\sharp) &:= \{ \langle x,y \rangle \mid x \in \mu^A(x^\sharp) \wedge y \in \mu^B(y) \} \end{aligned}$$

*Fact 4* (*Product Abstraction Correspondence*)
The classical product abstraction instantiated to powersets is equal to the lifted constructive product abstraction composed with the independent attributes abstraction *when applied to non-empty powersets*, that is: $\alpha^{A\times B}(X,Y) = (\alpha^{IA} \circ \lfloor\eta^{A\times B}\rfloor^* \circ \gamma^{IA})(X,Y)$ and $\gamma^{A\times B}(X^\sharp,Y^\sharp) = (\alpha^{IA} \circ \mu^{A\times B*} \circ \gamma^{IA})(X^\sharp,Y^\sharp)$ when $X$,$Y$,$X^\sharp$ and $Y^\sharp$ are non-empty. (In the case that $X$–$Y^\sharp$ could be empty, the classical product is equal-to *or larger* ($\sqsupseteq$) than the lifted constructive product composed with IA.)

**Functional Abstraction** The classical functional abstraction is defined given abstractions $A \underset{\alpha^A}{\overset{\gamma^A}{\leftrightarrows}} A^\sharp$ and $B \underset{\alpha^B}{\overset{\gamma^B}{\leftrightarrows}} B^\sharp$, and constructs the classical Galois connection:

$$A \nearrow B \underset{\overset{A\mapsto B}{\alpha}}{\overset{\overset{A\mapsto B}{\gamma}}{\leftrightarrows}} A^\sharp \nearrow B^\sharp \qquad \begin{aligned} \overset{A\mapsto B}{\alpha} &: (A \nearrow B) \nearrow A^\sharp \nearrow B^\sharp \\ \overset{A\mapsto B}{\gamma} &: (A^\sharp \nearrow B^\sharp) \nearrow A \nearrow B \end{aligned}$$

$$\begin{aligned} \overset{A\mapsto B}{\alpha}(f)(x^\sharp) &:= \alpha^B(f(\gamma^A(x^\sharp))) \\ \overset{A\mapsto B}{\gamma}(f^\sharp)(X) &:= \gamma^B(f^\sharp(\alpha^A(X))) \end{aligned}$$

The constructive analog is defined given constructive abstractions $A \underset{\eta^A}{\overset{\mu^A}{\leftrightarrows}} A^\sharp$ and $B \underset{\eta^B}{\overset{\mu^B}{\leftrightarrows}} B^\sharp$, and constructs the *classical* Galois connection:

$$A \nearrow \wp(B) \underset{\overset{A\overset{\wp}{\mapsto}B}{\alpha}}{\overset{\overset{A\overset{\wp}{\mapsto}B}{\gamma}}{\leftrightarrows}} A^\sharp \nearrow \wp(B^\sharp) \qquad \begin{aligned} \overset{A\overset{\wp}{\mapsto}B}{\alpha} &: (A \nearrow \wp(B)) \nearrow A^\sharp \nearrow \wp(B^\sharp) \\ \overset{A\overset{\wp}{\mapsto}B}{\gamma} &: (A^\sharp \nearrow \wp(B^\sharp)) \nearrow A \nearrow \wp(B) \end{aligned}$$

$$\begin{aligned} \overset{A\overset{\wp}{\mapsto}B}{\alpha}(f)(x^\sharp) &:= \lfloor \eta^B \rfloor^*(f^*(\gamma^A(x^\sharp))) \\ \overset{A\overset{\wp}{\mapsto}B}{\gamma}(f^\sharp)(x) &:= \mu^{B*}(f^\sharp(\eta^A(x))) \end{aligned}$$

*Fact 5* (*Functional Abstraction Correspondence*)
The classical functional abstraction instantiated to powersets $\wp(A)$, $\wp(B)$, $\wp(A^\sharp)$ and $\wp(B^\sharp)$ is equal to the lifted constructive analog, that is: $\overset{\wp(A)\mapsto\wp(B)}{\alpha} = (\overset{A\overset{\wp}{\mapsto}B}{\alpha})^*$ and $\overset{\wp(A)\mapsto\wp(B)}{\gamma} = (\overset{A\overset{\wp}{\mapsto}B}{\gamma})^*$.

## 9 Comparing Classical and Constructive Approaches

In this section we aim to further clarify to what extent classical Galois connection calculations, which have been used successfully for decades, are related and/or interderivable with constructive Galois connection calculations. We will demonstrate this relationship an extended example drawn from our first case study.

In Section 4 we showed calculations for the random number expression (`rand`) and variable reference ($x$). The inductive case for binary operators ($ae \oplus ae$) was omitted for brevity, however its calculation is particularly interesting because it involves interacting with a classical Galois connection during the calculation (in both constructive and classical settings). In this section we will work through this calculation in detail to demonstrate the differences and similarities between classical and constructive approaches, as well as to demonstrate the effectiveness of constructive Galois connections used in conjunction with classical ones.

**Setup** To set the stage, we review in Figure 13 the types for the arithmetic operator denotation ($[\![_]\!]^a$), its abstraction ($[\![_]\!]^{a\sharp}$), the arithmetic expression relational

$$
\begin{array}{rll}
[\![_]\!]^{a} & : & \mathbb{Z} \times \mathbb{Z} \rightharpoonup \mathbb{Z} \\
[\![_]\!]^{a\sharp} & : & \mathbb{Z}^{\sharp} \times \mathbb{Z}^{\sharp} \to \mathbb{Z}^{\sharp} \\
_\vdash_\Downarrow^{a}_ & : & \wp(\mathsf{env} \times \mathsf{aexp} \times \mathbb{Z}) \\
\mathcal{A}[_] & : & \mathsf{aexp} \to \mathsf{env} \to \wp(\mathbb{Z}) \\
\mathcal{A}_{\wp}[_] & : & \mathsf{aexp} \to \wp(\mathsf{env}) \nearrow \wp(\mathbb{Z}) \\
\mathcal{A}^{\sharp}[_] & : & \mathsf{aexp} \to \mathsf{env}^{\sharp} \nearrow \mathbb{Z}^{\sharp}
\end{array}
\quad
\begin{array}{rll}
\eta^{z} & : & \mathbb{Z} \to \mathbb{Z}^{\sharp} \\
\alpha^{z} & : & \wp(\mathbb{Z}) \nearrow \mathbb{Z}^{\sharp} \\
\eta^{r} & : & \mathsf{env} \to \mathsf{env}^{\sharp} \\
\alpha^{r} & : & \wp(\mathsf{env}) \nearrow \mathsf{env}^{\sharp}
\end{array}
\quad
\begin{array}{rll}
\mu^{z} & : & \mathbb{Z}^{\sharp} \nearrow \wp(\mathbb{Z}) \\
\gamma^{z} & : & \mathbb{Z}^{\sharp} \nearrow \wp(\mathbb{Z}) \\
\mu^{r} & : & \mathsf{env}^{\sharp} \nearrow \wp(\mathsf{env}) \\
\gamma^{r} & : & \mathsf{env}^{\sharp} \nearrow \wp(\mathsf{env})
\end{array}
$$

Fig. 13. Review: Calculational Derivation for Binary Arithmetic Operator Expressions

semantics ($_\vdash_\Downarrow^{a}_$), its functional variant ($\mathcal{A}[_]$) and collecting semantics ($\mathcal{A}_{\wp}[_]$), its abstraction ($\mathcal{A}^{\sharp}[_]$), as well as classical and constructive Galois connections for integers ($\mathbb{Z} \underset{\alpha^{z}}{\overset{\gamma^{z}}{\leftrightarrows}} \mathbb{Z}^{\sharp}$ and $\mathbb{Z} \underset{\eta^{z}}{\overset{\mu^{z}}{\leftrightarrows}} \mathbb{Z}^{\sharp}$) and environments ($\mathsf{env} \underset{\alpha^{r}}{\overset{\gamma^{r}}{\leftrightarrows}} \mathsf{env}^{\sharp}$ and $\mathsf{env} \underset{\eta^{r}}{\overset{\mu^{r}}{\leftrightarrows}} \mathsf{env}^{\sharp}$).

First we will show the original classical calculation for binary arithmetic operator expressions which does not make explicit use of the independent attributes abstraction (§ 9.1). We will then make independent attributes explicit in the classical calculation (§ 9.2), and then show the constructive analog with explicit use of independent attributes (§ 9.3).

### *9.1 Review: Cousot's Original Classical Calculation*

In the classical Galois connection framework, the abstraction ($\mathcal{A}^{\sharp}[_]$) for the arithmetic relational semantics ($_\vdash_\Downarrow^{a}_$) is calculated by first defining the collecting semantics ($\mathcal{A}_{\wp}[_] \ : \ \mathsf{aexp} \to \wp(\mathsf{env}) \nearrow \wp(\mathbb{Z})$), and then relating the collecting semantics to the abstract semantics through a functional abstraction, that is:

$$\overset{r\mapsto z}{\alpha}(\mathcal{A}_{\wp}[ae])(\rho^{\sharp}) = \alpha^{z}(\mathcal{A}_{\wp}[ae](\gamma^{r}(\rho^{\sharp}))) \sqsubseteq \ldots \triangleq \mathcal{A}^{\sharp}[ae](\rho^{\sharp})$$

Cousot's original calculation proceeds by induction on the syntax for arithmetic expressions, so for arithmetic operator expressions, the calculation goal is:

$$\alpha^{z}(\mathcal{A}_{\wp}[ae_1 \oplus ae_2](\gamma^{r}(\rho^{\sharp}))) \sqsubseteq \ldots \triangleq \mathcal{A}^{\sharp}[ae_1 \oplus ae_2](\rho^{\sharp})$$

along with an assumed inductive hypothesis for subexpressions $ae_1$ and $ae_2$. The calculation is shown in Figure 14. Steps 1–3 unfold semantic function and relation definitions; at Step 4 the specification is weakened explicitly to break the equality relationship between the environment used to evaluate $ae_1$ and $ae_2$; Step 5 rewrites the goal in terms of collecting semantics operations; Step 6 applies the inductive hypothesis; Step 7 applies a sound abstract interpreter for binary operators (a parameter to the calculation); Step 8 collapses neighboring abstraction and concretization functions; and Step 9 declares the final state of the calculation to be the definition of the algorithm.

$$
\begin{array}{ll}
 & \alpha^{z}(\mathcal{A}_{\wp}[ae_1 \oplus ae_2](\gamma^{r}(\rho^{\sharp}))) \\
(1) & = \wr \text{ defn. of } \mathcal{A}_{\wp}[ae_1 \oplus ae_2] \int \\
 & \alpha^{z}(\bigcup\limits_{\rho \in \gamma^{r}(\rho^{\sharp})} \mathcal{A}[ae_1 \oplus ae_2](\rho)) \\
(2) & = \wr \text{ defn. of } \mathcal{A}[ae_1 \oplus ae_2] \int \\
 & \alpha^{z}(\bigcup\limits_{\rho \in \gamma^{r}(\rho^{\sharp})} \{[\![\oplus]\!]^{a}(i_1, i_2) \mid \rho \vdash ae_1 \Downarrow^{a} i_1 \wedge \rho \vdash ae_2 \Downarrow^{a} i_2\}) \\
(3) & = \wr \text{ defn. of } \mathcal{A}[ae_1] \text{ and } \mathcal{A}[ae_2] \int \\
 & \alpha^{z}(\bigcup\limits_{\rho \in \gamma^{r}(\rho^{\sharp})} \{[\![\oplus]\!]^{a}(i_1, i_2) \mid i_1 \in \mathcal{A}[ae_1](\rho) \wedge i_2 \in \mathcal{A}[ae_2](\rho)\}) \\
(4) & \sqsubseteq \wr \text{ monotonicity of } \alpha^{z} \int \\
 & \alpha^{z}(\bigcup\limits_{\rho_1 \in \gamma^{r}(\rho^{\sharp})} \bigcup\limits_{\rho_2 \in \gamma^{r}(\rho^{\sharp})} \{[\![\oplus]\!]^{a}(i_1, i_2) \mid i_1 \in \mathcal{A}[ae_1](\rho_1) \wedge i_2 \in \mathcal{A}[ae_2](\rho_2)\}) \\
(5) & = \wr \text{ set equality and defn. of } \mathcal{A}_{\wp} \int \\
 & \alpha^{z}(\{[\![\oplus]\!]^{a}(i_1, i_2) \mid i_1 \in \mathcal{A}_{\wp}[ae_1](\gamma^{r}(\rho^{\sharp})) \wedge i_2 \in \mathcal{A}_{\wp}[ae_2](\gamma^{r}(\rho^{\sharp}))\}) \\
(6) & \sqsubseteq \wr \text{ inductive hypothesis } (\mathcal{A}_{\wp}[ae] \circ \gamma^{r} \sqsubseteq \gamma^{z} \circ \mathcal{A}^{\sharp}[ae]) \int \\
 & \alpha^{z}(\{[\![\oplus]\!]^{a}(i_1, i_2) \mid i_1 \in \gamma^{z}(\mathcal{A}^{\sharp}[ae_1](\rho^{\sharp})) \wedge i_2 \in \gamma^{z}(\mathcal{A}^{\sharp}[ae_2](\rho^{\sharp}))\}) \\
(7) & \sqsubseteq \wr [\![\oplus]\!]^{a\sharp} \text{ sound } ([\![\oplus]\!]^{a}_{\wp} \circ \overset{z\times z}{\gamma} \sqsubseteq \gamma^{z} \circ [\![\oplus]\!]^{a\sharp}) \int \\
 & \alpha^{z}(\gamma^{z}([\![\oplus]\!]^{a\sharp}(\mathcal{A}^{\sharp}[ae_1](\rho^{\sharp}), \mathcal{A}^{\sharp}[ae_2](\rho^{\sharp})))) \\
(8) & \sqsubseteq \wr \alpha^{z} \circ \gamma^{z} \text{ reductive } (\alpha^{z} \circ \gamma^{z} \sqsubseteq id) \int \\
 & [\![\oplus]\!]^{a\sharp}(\mathcal{A}^{\sharp}[ae_1](\rho^{\sharp}), \mathcal{A}^{\sharp}[ae_2](\rho^{\sharp})) \\
(9) & \triangleq \wr \text{ by defining } \mathcal{A}^{\sharp}[ae_1 \oplus ae_2](\rho^{\sharp}) \coloneqq [\![\oplus]\!]^{a\sharp}(\mathcal{A}^{\sharp}[ae_1](\rho^{\sharp}), \mathcal{A}^{\sharp}[ae_2](\rho^{\sharp})) \int \\
 & \mathcal{A}^{\sharp}[ae_1 \oplus ae_2](\rho^{\sharp}) \quad \blacksquare
\end{array}
$$

Fig. 14. Classical Calculation for Binary Arithmetic Operator Expressions

Although there was no mention of the independent attributes abstraction in this calculation, its effects are there implicitly. In particular, Step 4, which breaks the equality relationship between environments, is implicitly performing the function of the independent attributes abstraction: to break relationships between elements of concrete sets of pairs. Step 4 is also the only step in the derivation which loses precision (uses $\sqsubseteq$ instead of $=$) unnecessarily, whereas the other losses of precision are unavoidable (inductive hypothesis, abstraction for binary operators, and collapsing abstraction and concretization function). In the next subsection, we will make explicit use of the independent attributes abstraction, rather than through the ad-hoc line of reasoning contained in Step 4.

### *9.2 Using Independent Attributes Explicitly*

In this section we recreate the calculation for binary arithmetic operator expressions from last section, but in a way that makes explicit use of the independent attributes abstraction.

The calculation is shown in Figure 15. The beginning of the derivation is as before (Steps 1–3); Step 4.1 rewrites the calculation into a form that mentions independent

$$
\begin{array}{ll}
 & \ldots \textit{ initial calculation as before (Steps 1–3)} \\
 & \alpha^z(\bigcup_{\rho \in \gamma^r(\rho^\sharp)} \{[\![\oplus]\!]^a(i_1,i_2) \mid i_1 \in \mathcal{A}[ae_1](\rho) \wedge i_2 \in \mathcal{A}[ae_2](\rho)\}) \\
\textit{(4.1)} & = \wr \text{ defn. of } \overset{IA}{\gamma} \text{ and } [\![\oplus]\!]^a_{\wp} \,\int \\
 & \alpha^z(\bigcup_{\rho \in \gamma^r(\rho^\sharp)} [\![\oplus]\!]^a_{\wp}(\overset{IA}{\gamma}(\mathcal{A}[ae_1](\rho), \mathcal{A}[ae_2](\rho)))) \\
\textit{(4.2)} & = \wr \text{ set equality } \int \\
 & \alpha^z([\![\oplus]\!]^a_{\wp}(\bigcup_{\rho \in \gamma^r(\rho^\sharp)} \overset{IA}{\gamma}(\mathcal{A}[ae_1](\rho), \mathcal{A}[ae_2](\rho)))) \\
\textit{(5.1)} & \sqsubseteq \wr \overset{IA}{\gamma} \circ \overset{IA}{\alpha} \text{ expansive } (id \sqsubseteq \overset{IA}{\gamma} \circ \overset{IA}{\alpha}) \int \\
 & \alpha^z([\![\oplus]\!]^a_{\wp}(\overset{IA}{\gamma}(\overset{IA}{\alpha}(\bigcup_{\rho \in \gamma^r(\rho^\sharp)} \overset{IA}{\gamma}(\mathcal{A}[ae_1](\rho), \mathcal{A}[ae_2](\rho)))))) \\
\textit{(5.2)} & = \wr \text{ set equality (see IA-Split below) } \int \\
 & \alpha^z([\![\oplus]\!]^a_{\wp}(\overset{IA}{\gamma}(\mathcal{A}_{\wp}[ae_1](\gamma^r(\rho^\sharp)), \mathcal{A}_{\wp}[ae_2](\gamma^r(\rho^\sharp))))) \\
\textit{(5.3)} & \sqsubseteq \wr \text{ defn. of } \overset{IA}{\gamma} \text{ and } [\![\oplus]\!]^a_{\wp} \,\int \\
 & \alpha^z(\{[\![\oplus]\!]^a(i_1,i_2) \mid i_1 \in \mathcal{A}_{\wp}[ae_1](\gamma^r(\rho^\sharp)) \wedge i_2 \in \mathcal{A}_{\wp}[ae_2](\gamma^r(\rho^\sharp))\}) \\
 & \ldots \textit{ final calculation as before (Steps 6–9)}
\end{array}
$$

Fig. 15. Classical Calculation for Binary Arithmetic Operator Expressions Using Independent Attributes

attributes concretization; Step 4.2 pulls the collecting semantics for binary operators out of the union operation; Step 5.1 introduces the explicit independent attributes abstraction; Step 5.2 collapses the union operation between independent attributes abstraction and concretization based on a key observation (see below); Step 5.3 unfolds the definition of independent attributes concretization; and the rest of the derivation is as before (Steps 6–9).

The key observation in this derivation is the fact that the independent attributes abstraction is transparent w.r.t. element-wise relationships, that is pairing ($\overset{IA}{\gamma}$) and splitting ($\overset{IA}{\alpha}$) two functions over related elements ($f(x_1)$ and $g(x_2)$ for $x_1 = x_2 \in X$), is equivalent to pairing each functions applied to unrelated elements ($f^*(X)$ and $g^*(X)$):

*Fact 6* (*Independent Attributes Split Equality*)

$$\overset{IA}{\alpha}(\bigcup_{x \in X} \overset{IA}{\gamma}(f(x), g(x))) = \langle f^*(X), g^*(X) \rangle \tag{IA-Split}$$

This observation captures locally the fact that if relational information is eventually going to be explicitly removed, then nothing is lost by splitting the equality relationship between arguments to each function.

One of the benefits of the calculational approach to abstract interpretation is that any loss of precision w.r.t. the induced specification is made explicit. In this derivation, the only non-essential loss in precision came from an explicit introduction of the independent attributes abstraction, which in turn makes explicit the fact that the resulting analysis may not be relational. If a relational analyzer (Cousot & Halbwachs, 1978) was desired, one could point exactly where in the calculation this information was lost *via* the independent attributes abstraction, and correct it locally. *E.g.*, recent results in information flow analysis show how to obtain more precise analyzers in exactly this way: by pinpointing and correcting the loss of precision after deriving the analysis using the calculational method (Assaf *et al.*, 2017).

### 9.3 Calculating with Constructive Galois Connections

In the constructive framework, the abstract interpretation of binary arithmetic operator expressions ($\mathcal{A}^\sharp[ae_1 \oplus ae_2]$) is derived in a similar way, and also has the option of explicitly using the classical independent attributes abstraction along the way. The constructive calculation proceeds from the induced specification:

$$\overset{r\overset{\wp}{\to}z}{\alpha}(\mathcal{A}[ae])(\rho^\sharp) = \lfloor \eta^z \rfloor^*(\mathcal{A}[ae]^*(\mu^r(\rho^\sharp))) \sqsubseteq \ldots \triangleq \lfloor \mathcal{A}^\sharp[ae] \rfloor(\rho^\sharp)$$

Two notable difference in the constructive calculation setup are:

1. The codomain type for both sides is $\wp(\mathbb{Z}^\sharp)$, not $\mathbb{Z}^\sharp$. This powerset modality makes explicit the transition from "specification" to "algorithm."
2. The specification on the left-hand-side is *stronger* than the classical one, because it does not collapse the set of abstract integers $I^\sharp : \wp(\mathbb{Z}^\sharp)$ into a single least-upper-bound abstract integer $i^\sharp = \bigsqcup_{i^{\sharp\prime} \in I^\sharp} i^{\sharp\prime}$.

The original classical equation is recovered (in a constructive setting) by composing with the constructive least-upper-bound-abstraction ($\overset{\sqcup_\wp}{\alpha} : \wp(\mathbb{Z}^\sharp) \nearrow \wp^1(\mathbb{Z}^\sharp)$):

$$\overset{\sqcup_\wp}{\alpha}(\lfloor \eta^z \rfloor^*(\mathcal{A}[ae]^*(\mu^r(\rho^\sharp)))) \sqsubseteq \ldots \triangleq \lfloor \mathcal{A}^\sharp[ae] \rfloor(\rho^\sharp)$$

However, we will continue our demonstration with the original induced equation, where the constructive least-upper-bound-abstraction is not present.

The constructive calculation for the binary expression case proceeds in a similar fashion to Cousot's classical derivation. To mimic the classical derivation, the independent attributes abstraction is introduced to weaken the specification to discard the equality relationship between evaluation environments used to evaluate $ae_1$ and $ae_2$.

The calculation is shown in Figure 16. Steps 1–4 unfold semantic function and relation definitions; Step 5 explicitly weakens the specification using independent attributes; Step 6 applies the key independent attributes observation; Step 7 applies the inductive hypothesis; Step 8 combines concretization for independent attributes and the abstraction for integers; Step 9 applies a sound abstract interpreter for binary arithmetic operators (a parameter to the calculation); Step 10 collapses neighboring abstraction and concretization functions; and Step 11 declares the final state of the calculation to be the definition of the algorithm.

$$\lfloor\eta^z\rfloor^*(\mathcal{A}[ae_1 \oplus ae_2]^*(\mu^r(\rho^\sharp)))$$

*(1)* $= \wr$ defn. of $\mathcal{A}[ae_1 \oplus ae_2]$ $\int$

$$\lfloor\eta^z\rfloor^*(\bigcup_{\rho\in\mu^r(\rho^\sharp)} \{[\![\oplus]\!]^a(i_1,i_2) \mid \rho \vdash ae_1 \Downarrow^a i_1 \wedge \rho \vdash ae_2 \Downarrow^a i_2\})$$

*(2)* $= \wr$ defn. of $\mathcal{A}[ae_1]$ and $\mathcal{A}[ae_2]$ $\int$

$$\lfloor\eta^z\rfloor^*(\bigcup_{\rho\in\mu^r(\rho^\sharp)} \{[\![\oplus]\!]^a(i_1,i_2) \mid i_1 \in \mathcal{A}[ae_1](\rho) \wedge i_2 \in \mathcal{A}[ae_2](\rho)\})$$

*(3)* $= \wr$ defn. of $\overset{IA}{\gamma}$ $\int$

$$\lfloor\eta^z\rfloor^*(\bigcup_{\rho\in\mu^r(\rho^\sharp)} \lfloor[\![\oplus]\!]^a\rfloor^*(\overset{IA}{\gamma}(\mathcal{A}[ae_1](\rho),\mathcal{A}[ae_2](\rho))))$$

*(4)* $= \wr$ set equality $\int$

$$\lfloor\eta^z\rfloor^*(\lfloor[\![\oplus]\!]^a\rfloor^*(\bigcup_{\rho\in\mu^r(\rho^\sharp)} \overset{IA}{\gamma}(\mathcal{A}[ae_1](\rho),\mathcal{A}[ae_2](\rho))))$$

*(5)* $\sqsubseteq \wr$ $\overset{IA}{\gamma} \circ \overset{IA}{\alpha}$ expansive ($id \sqsubseteq \overset{IA}{\gamma} \circ \overset{IA}{\alpha}$) $\int$

$$\lfloor\eta^z\rfloor^*(\lfloor[\![\oplus]\!]^a\rfloor^*(\overset{IA}{\gamma}(\overset{IA}{\alpha}(\bigcup_{\rho\in\mu^r(\rho^\sharp)} \overset{IA}{\gamma}(\mathcal{A}[ae_1](\rho),\mathcal{A}[ae_2](\rho))))))$$

*(6)* $= \wr$ set equality (see IA-Split above) $\int$

$$\lfloor\eta^z\rfloor^*(\lfloor[\![\oplus]\!]^a\rfloor^*(\overset{IA}{\gamma}(\mathcal{A}[ae_1]^*(\mu^r(\rho^\sharp)),\mathcal{A}[ae_2]^*(\mu^r(\rho^\sharp)))))$$

*(7)* $\sqsubseteq \wr$ inductive hypothesis ($\mathcal{A}[ae] \circledast \mu^r \sqsubseteq \mu^z \circledast \lfloor\mathcal{A}^\sharp[ae]\rfloor$) $\int$

$$\lfloor\eta^z\rfloor^*(\lfloor[\![\oplus]\!]^a\rfloor^*(\overset{IA}{\gamma}(\mu^z(\mathcal{A}^\sharp[ae_1](\rho^\sharp)),\mu^z(\mathcal{A}^\sharp[ae_2](\rho^\sharp)))))$$

*(8)* $= \wr$ defn. of $\overset{IA}{\gamma}$ and $\overset{z\times z}{\mu}$ $\int$

$$\lfloor\eta^z\rfloor^*(\lfloor[\![\oplus]\!]^a\rfloor^*(\overset{z\times z}{\mu}(\mathcal{A}^\sharp[ae_1](\rho^\sharp),\mathcal{A}^\sharp[ae_2](\rho^\sharp))))$$

*(9)* $\sqsubseteq \wr$ $[\![\oplus]\!]^{a\sharp}$ sound ($\lfloor[\![\oplus]\!]^a\rfloor \circledast \overset{z\times z}{\mu} \sqsubseteq \mu^z \circledast \lfloor[\![\oplus]\!]^{a\sharp}\rfloor$) $\int$

$$\lfloor\eta^z\rfloor^*(\mu^z([\![\oplus]\!]^{a\sharp}(\mathcal{A}^\sharp[ae_1](\rho^\sharp),\mathcal{A}^\sharp[ae_2](\rho^\sharp))))$$

*(10)* $\sqsubseteq \wr$ $\lfloor\eta^z\rfloor \circledast \mu^z$ reductive ($\lfloor\eta^z\rfloor \circledast \mu^z \sqsubseteq ret$) $\int$

$$\{[\![\oplus]\!]^{a\sharp}(\mathcal{A}^\sharp[ae_1](\rho^\sharp),\mathcal{A}^\sharp[ae_2](\rho^\sharp))\}$$

*(11)* $\triangleq \wr$ by defining $\mathcal{A}^\sharp[ae_1 \oplus ae_2](\rho^\sharp) := [\![\oplus]\!]^{a\sharp}(\mathcal{A}^\sharp[ae_1](\rho^\sharp),\mathcal{A}^\sharp[ae_2](\rho^\sharp))$ $\int$

$$\lfloor\mathcal{A}^\sharp[ae_1 \oplus ae_2]\rfloor(\rho^\sharp) \quad \blacksquare$$

Fig. 16. Constructive Calculation for Binary Arithmetic Operator Expressions

What this calculation shows is that constructive Galois connections are able to work in tandem with classical Galois connections, as this constructive calculation made use of the classical independent attributes abstraction.

## 10 Optimal Calculations—Constructive and Classical

All of the derivations shown in the previous section follow a $\gamma$-directed approach to calculation. In this style, the next step of the calculation pushes concretization ($\gamma$) through the concrete semantics, from right to left, until it meets abstraction ($\alpha$) on the far left-hand-side, at which point they collapse. In this section we explore the alternative approach of going the other direction: push abstraction from left-to-right until it meets concretization.

In the classical Galois connection framework, both $\gamma$-directed and $\alpha$-directed approaches are similar, and the choice to use one or the other appears at first to be cosmetic. However, in the constructive framework, abstraction ($\eta$) is of a different nature than concretization ($\mu$): it is a pure function with algorithmic content, rather than a relation. This means abstraction is easier to push through the concrete semantics, and therefore $\eta$-directed derivations can be simpler than $\mu$-directed ones.

Because constructive and classical Galois connections are so tightly connected, we show how this insight of $\eta$-directed calculations can be translated back to the world of classical Galois connections. To do this, we (1) recall a fact about all classical Galois connections, and then (2) introduce a restriction on collecting semantics which often holds in practice:

1. Fact: All abstraction functions ($\alpha : \wp(A) \nearrow A^\sharp$) are *complete join morphisms*, that is:

$$\alpha(\bigcup_{i \in I} X_i) = \bigsqcup_{i \in I}(\alpha(X_i))$$

   for all indexed families $X_ : I \to A^\sharp$
2. Restriction: The predicate transformer ($t : \wp(A) \to \wp(B)$) must be a *complete union morphism*, that is:

$$f(\bigcup_{i \in I} X_i) = \bigcup_{i \in I}(f(X_i))$$

   for all indexed families $X_ : I \to \wp(A)$

The restriction (2) is equivalent to the existence of a monadic semantics relation, or $f : A \to \wp(B)$, where:

$$t(X) = \bigcup_{x \in X} f(x) \qquad \textit{and} \qquad f(x) = t(\{x\})$$

It follows that, in any setting where classical Galois connections are used where the collecting semantics $t : \wp(A) \nearrow \wp(B)$ is a complete union morphism, it suffices to work purely with constructive Galois connections without any loss of generality. These generality results coincide with the completeness theorems for Kleisli and constructive Galois connections described in Section 7 (KGC-Complete and CGC-Induce).

As a consequence of this, our observation above about $\eta$-directed calculations being easier to "push through" the calculation for constructive Galois connections also holds for $\alpha$-directed classical calculations when the collecting semantics is a complete union morphism.

The $\eta$-directed calculation of an abstract interpreter for binary arithmetic operator expressions is shown in Figure 17. The beginning of the calculation is as before (Steps 1–2); Step 3 pushes the abstraction function through the union operation; Step 4 applies a sound abstract interpretation for binary operators (a

$$
\begin{array}{rll}
& & \textit{... initial calculation as before (Steps 1–2)} \\
& & \lfloor \eta^z \rfloor^*\Big( \bigcup\limits_{\rho \in \mu^r(\rho^\sharp)} \{ \llbracket \oplus \rrbracket^a(i_1, i_2) \mid i_1 \in \mathcal{A}[ae_1](\rho) \wedge i_2 \in \mathcal{A}[ae_2](\rho) \} \Big) \\
\textit{(3)} & = & \wr \text{ set equality } \int \\
& & \bigcup\limits_{\rho \in \mu^r(\rho^\sharp)} \{ \eta^z(\llbracket \oplus \rrbracket^a(i_1, i_2)) \mid i_1 \in \mathcal{A}[ae_1](\rho) \wedge i_2 \in \mathcal{A}[ae_2](\rho) \} \\
\textit{(4)} & \sqsubseteq & \wr\ \llbracket \oplus \rrbracket^{a\sharp} \text{ sound } (\eta^z \circ \llbracket \oplus \rrbracket^a \sqsubseteq \llbracket \oplus \rrbracket^{a\sharp} \circ \overset{z \times z}{\eta}) \int \\
& & \bigcup\limits_{\rho \in \mu^r(\rho^\sharp)} \{ \llbracket \oplus \rrbracket^{a\sharp}(\eta^z(i_1), \eta^z(i_2)) \mid i_1 \in \mathcal{A}[ae_1](\rho) \wedge i_2 \in \mathcal{A}[ae_2](\rho) \} \\
\textit{(5)} & = & \wr \text{ set equality } \int \\
& & \bigcup\limits_{\rho \in \mu^r(\rho^\sharp)} \{ \llbracket \oplus \rrbracket^{a\sharp}(i_1^\sharp, i_2^\sharp) \mid i_1^\sharp \in \lfloor \eta^z \rfloor^*(\mathcal{A}[ae_1](\rho)) \wedge i_2^\sharp \in \lfloor \eta^z \rfloor^*(\mathcal{A}[ae_2](\rho)) \} \\
\textit{(6)} & \sqsubseteq & \wr \text{ inductive hypothesis } (\lfloor \eta^z \rfloor \circledast \mathcal{A}[ae] \sqsubseteq \lfloor \mathcal{A}^\sharp[ae] \rfloor \circledast \lfloor \eta^r \rfloor) \int \\
& & \bigcup\limits_{\rho \in \mu^r(\rho^\sharp)} \{ \llbracket \oplus \rrbracket^{a\sharp}(i_1^\sharp, i_2^\sharp) \mid i_1^\sharp \sqsubseteq \mathcal{A}^\sharp[ae_1](\eta^r(\rho)) \wedge i_2^\sharp \sqsubseteq \mathcal{A}^\sharp[ae_1](\eta^r(\rho)) \} \\
\textit{(7)} & = & \wr \text{ powerset downward-closed } \int \\
& & \bigcup\limits_{\rho \in \mu^r(\rho^\sharp)} \{ \llbracket \oplus \rrbracket^{a\sharp}(\mathcal{A}^\sharp[ae_1](\eta^r(\rho)), \mathcal{A}^\sharp[ae_2](\eta^r(\rho))) \} \\
\textit{(8)} & = & \wr \text{ powerset equality } \int \\
& & \bigcup\limits_{\rho^{\sharp\prime} \in \lfloor \eta^r \rfloor^*(\mu^r(\rho^\sharp))} \{ \llbracket \oplus \rrbracket^{a\sharp}(\mathcal{A}^\sharp[ae_1](\rho^{\sharp\prime}), \mathcal{A}^\sharp[ae_2](\rho^{\sharp\prime})) \} \\
\textit{(9)} & \sqsubseteq & \wr\ \lfloor \eta^r \rfloor \circledast \mu^r \text{ reductive } (\lfloor \eta^r \rfloor \circledast \mu^r \sqsubseteq \mathit{ret}) \int \\
& & \{ \llbracket \oplus \rrbracket^{a\sharp}(\mathcal{A}^\sharp[ae_1](\rho^\sharp), \mathcal{A}^\sharp[ae_2](\rho^\sharp)) \} \\
\textit{(10)} & \triangleq & \wr \text{ by defining } \mathcal{A}^\sharp[ae_1 \oplus ae_2](\rho^\sharp) := \llbracket \oplus \rrbracket^{a\sharp}(\mathcal{A}^\sharp[ae_1](\rho^\sharp), \mathcal{A}^\sharp[ae_2](\rho^\sharp)) \int \\
& & \lfloor \mathcal{A}^\sharp[ae_1 \oplus ae_2] \rfloor(\rho^\sharp) \quad \blacksquare
\end{array}
$$

Fig. 17. Constructive Calculation for Binary Arithmetic Operator Expressions—Optimal and $\eta$-directed

parameter to the calculation); Step 5 pushes the abstraction function through the set comprehension; Step 6 applies the inductive hypothesis; Step 7 applies the fact that the abstract denotation for binary operators is monotonic, and that powerset are downward closed; Step 8 pushes abstraction again through the set comprehension; Step 9 collapses the neighboring abstraction and concretization functions; and Step 10 declares the final state of the calculation to be the definition of the algorithm.

This abstraction-directed calculation is not only simpler due to how easily the abstraction function distributes through powerset operations, but it is also optimal. Unlike the classical calculation (and the constructive $\mu$-directed calculation), no loss in precision is explicitly introduced, and no use of independent attributes is made, explicitly or implicitly. This does not mean that the prior derivations resulting in a less-precise algorithm. (The resulting algorithm is the same as before.) Rather, it means that before there was no guarantee *via* the derivation process that the result was optimal, whereas now we have such a guarantee. The prior derivations *were*

*... initial calculation as before (Steps 1–3)*

$$
\begin{array}{ll}
 & \alpha^z(\bigcup_{\rho \in \gamma^r(\rho^\sharp)} \{[\![\oplus]\!]^a(i_1,i_2) \mid i_1 \in \mathcal{A}[ae_1](\rho) \wedge i_2 \in \mathcal{A}[ae_2](\rho)\}) \\
\textit{(4)} & \quad \wr\ \alpha^z \text{ complete join morphism } \int \\
 & \bigsqcup_{\rho \in \gamma^r(\rho^\sharp)} \alpha^z(\{[\![\oplus]\!]^a(i_1,i_2) \mid i_1 \in \mathcal{A}[ae_1](\rho) \wedge i_2 \in \mathcal{A}[ae_2](\rho)\}) \\
\textit{(5)} & \sqsubseteq\ \wr\ [\![\oplus]\!]^{a\sharp} \text{ sound } (\alpha^z \circ [\![\oplus]\!]^a_{\wp} \sqsubseteq [\![\oplus]\!]^{a\sharp} \circ \overset{z\times z}{\alpha})\ \int \\
 & \bigsqcup_{\rho \in \gamma^r(\rho^\sharp)} [\![\oplus]\!]^{a\sharp}(\alpha^z(\mathcal{A}[ae_1](\rho)), \alpha^z(\mathcal{A}[ae_2](\rho))) \\
\textit{(6)} & \sqsubseteq\ \wr\ \text{inductive hypothesis } (\alpha^z \circ \mathcal{A}_{\wp}[ae] = \mathcal{A}_{\wp}[ae] \circ \alpha^r)\ \int \\
 & \bigsqcup_{\rho \in \gamma^r(\rho^\sharp)} [\![\oplus]\!]^{a\sharp}(\mathcal{A}^\sharp[ae_1](\alpha^r(\{\rho\})), \mathcal{A}^\sharp[ae_2](\alpha^r(\{\rho\}))) \\
\textit{(7)} & =\ \wr\ \text{set equality } \int \\
 & \bigsqcup_{\rho^{\sharp\prime} \in \{\alpha^r(\{\rho\}) \mid \rho \in \gamma^r(\rho^\sharp)\}} [\![\oplus]\!]^{a\sharp}(\mathcal{A}^\sharp[ae_1](\rho^{\sharp\prime}), \mathcal{A}^\sharp[ae_2](\rho^{\sharp\prime})) \\
\textit{(8)} & =\ \wr\ \alpha^r \text{ complete join morphism } \int \\
 & \bigsqcup_{\rho^{\sharp\prime} \in \{\alpha^r(\gamma^r(\rho^\sharp))\}} [\![\oplus]\!]^{a\sharp}(\mathcal{A}^\sharp[ae_1](\rho^{\sharp\prime}), \mathcal{A}^\sharp[ae_2](\rho^{\sharp\prime})) \\
\textit{(9)} & \sqsubseteq\ \wr\ \alpha^r \circ \gamma^r \text{ reductive } (\alpha^r \circ \gamma^r \sqsubseteq id)\ \int \\
 & [\![\oplus]\!]^{a\sharp}(\mathcal{A}^\sharp[ae_1](\rho^\sharp), \mathcal{A}^\sharp[ae_2](\rho^\sharp)) \\
\textit{(10)} & \triangleq\ \wr\ \text{by defining } \mathcal{A}^\sharp[ae_1 \oplus ae_2](\rho^\sharp) := [\![\oplus]\!]^{a\sharp}(\mathcal{A}^\sharp[ae_1](\rho^\sharp), \mathcal{A}^\sharp[ae_2](\rho^\sharp))\ \int \\
 & \mathcal{A}^\sharp[ae_1 \oplus ae_2](\rho^\sharp) \quad \blacksquare
\end{array}
$$

Fig. 18. Classical Calculation for Binary Arithmetic Operator Expressions—Optimal and $\alpha$-directed

optimal, but this fact was not made manifest in the calculation. Next, we show how to port this optimal calculation back to the classical Galois connection framework.

**Porting the Optimal Derivation Back to Classical** In this $\eta$-directed constructive calculation, no steps lose precision unnecessarily. However, the classical calculation seemed to require an explicit loss of precision through the independent attributes abstraction. How can this be? To shed light on this question, we show that the constructive abstraction-directed calculation can be back-ported to a classical calculation, leveraging the fact that the abstraction side of Galois connections are always complete join morphisms, that is:

$$\alpha^z(\bigcup_{i \in I} X_i) = \bigsqcup_{i \in I} (\alpha^z(X_i))$$

With this observation, a classical derivation is possible which doesn't need to interact with independent attributes to induce a final algorithm.

The classical calculation of binary arithmetic operator expressions is shown in Figure 18. The beginning of the calculation is as before (Steps 1–3); Step 4 pushes abstraction through the union operation, due to being a complete join morphism; Step 5 applies a sound abstraction for binary operators; Step 6 applies the inductive

$$
\begin{array}{rcl@{\qquad}rcl@{\qquad}rcl}
\varsigma \in \Sigma & \coloneqq & \mathsf{env} \times \mathsf{cexp} & & & & & & \\
\varsigma^\sharp \in \Sigma^\sharp & \coloneqq & \mathsf{env}^\sharp \times \wp(\mathsf{cexp}) & \eta^z & : & \mathbb{Z} \rightarrow \mathbb{Z}^\sharp & \mu^z & : & \mathbb{Z}^\sharp \nearrow \wp(\mathbb{Z}) \\
_ \mapsto^c _ & : & \wp(\Sigma \times \Sigma) & \alpha^z & : & \wp(\mathbb{Z}) \nearrow \mathbb{Z}^\sharp & \gamma^z & : & \mathbb{Z}^\sharp \nearrow \wp(\mathbb{Z}) \\
\mathcal{C}[_] & : & \mathsf{cexp} \rightarrow \Sigma \nearrow \wp(\Sigma) & \eta^r & : & \mathsf{env} \rightarrow \mathsf{env}^\sharp & \mu^r & : & \mathsf{env}^\sharp \nearrow \wp(\mathsf{env}) \\
\mathcal{C}_\wp[_] & : & \mathsf{cexp} \rightarrow \wp(\Sigma) \nearrow \wp(\Sigma) & \alpha^r & : & \wp(\mathsf{env}) \nearrow \mathsf{env}^\sharp & \gamma^r & : & \mathsf{env}^\sharp \nearrow \wp(\mathsf{env}) \\
\mathcal{C}^\sharp[_] & : & \mathsf{cexp} \rightarrow \Sigma^\sharp \nearrow \Sigma^\sharp & & & & & &
\end{array}
$$

Fig. 19. Review: calculating abstraction for conditional expressions

hypothesis; Step 7 pulls abstraction out of the set comprehension; Step 8 pushes abstraction through the set comprehension, due to being a complete join morphism; Step 9 collapses adjacent abstraction and concretization functions; and Step 10 declares the final state of the calculation to be the definition of the algorithm.

In this section we have shown two new calculations which are equivalent to Cousot's original derivation, but which are also guaranteed to be optimal by construction. The insight for optimality came from the constructive Galois connection framework, where the extraction function ($\eta$) is algorithmic, and therefore easier to "push through" the induced specification towards the definition of an algorithm. This insight was then ported to the classical setting, where it took the form of exploiting the complete-join-morphism property of abstraction functions ($\alpha$).

## 11 Multivalued Constructive Galois Connections

In this section we argue that constructive Galois connections support multivalued Galois connections, concrete semantics, and abstract interpreters, while maintaining their ability to be mechanized effectively.

To explore multivalued constructive Galois connections, we again work through an extended example based on the first case study, but this time deriving an abstract interpreter for conditional expressions (`if` *be* `then` *ce* `else` *ce*) in the command language (`cexp`) rather than arithmetic expressions (`aexp`).

**Setup** To set the stage, we review in Figure 19 the types for the command expression relational semantics ($_ \mapsto^c _$), its functional variant ($\mathcal{C}[_]$) and collecting semantics ($\mathcal{C}_\wp[_]$), its abstraction ($\mathcal{C}^\sharp[_]$), as well as classical and constructive Galois connections for integers ($\mathbb{Z} \underset{\eta^z}{\overset{\mu^z}{\leftrightarrows}} \mathbb{Z}^\sharp$ and $\mathbb{Z} \underset{\alpha^z}{\overset{\gamma^z}{\leftrightarrows}} \mathbb{Z}^\sharp$) and environments ($\mathsf{env} \underset{\eta^r}{\overset{\mu^r}{\leftrightarrows}} \mathsf{env}^\sharp$ and $\mathsf{env} \underset{\alpha^r}{\overset{\gamma^r}{\leftrightarrows}} \mathsf{env}^\sharp$).

### *11.1 Review: Cousot's Original Classical Calculation*

In the classical Galois connection framework, the abstraction ($\mathcal{C}^\sharp[_]$) for the command small-step relational semantics ($_ \mapsto^c _$) is calculated first by constructing the collecting semantics ($\mathcal{C}_{\wp}[_]$), and then relating the collecting semantics to the abstract semantics through a functional abstraction, that is:

$$\overset{\Sigma\mapsto\Sigma}{\alpha}(\mathcal{C}_{\wp}[ce])(\Sigma^{\sharp}) \triangleq \alpha^{\Sigma}(\mathcal{C}_{\wp}[ce](\gamma^{\Sigma}(\Sigma^{\sharp}))) \sqsubseteq \ldots \triangleq \mathcal{C}^{\sharp}[ce](\Sigma^{\sharp})$$

where configurations ($\varsigma \in \Sigma$) are abstracted through a composition of independent attributes and a product abstraction over environments:

$$\wp(\Sigma) \underset{\overset{IA}{\alpha}}{\overset{\overset{IA}{\gamma}}{\leftrightarrows}} \wp(\mathsf{env}) \times \wp(\mathsf{cexp}) \underset{\overset{r\times id}{\alpha}}{\overset{\overset{r\times id}{\gamma}}{\leftrightarrows}} \Sigma^{\sharp} \qquad \begin{array}{l} \alpha^{\Sigma} : \wp(\Sigma) \nearrow \Sigma^{\sharp} \\ \gamma^{\Sigma} : \Sigma^{\sharp} \nearrow \wp(\Sigma) \end{array} \qquad \begin{array}{l} \alpha^{\Sigma} := \overset{r\times id}{\alpha} \circ \overset{IA}{\alpha} \\ \gamma^{\Sigma} := \overset{IA}{\gamma} \circ \overset{r\times id}{\gamma} \end{array}$$

In Cousot's original derivation, the abstract interpreter is derived for the reflexive transitive closure of the small step relation directly. We will instead present the abstract interpreter for just the small step relation, factored out from the reflexive transitive closure.

The classical calculation begins by case analysis on the syntax for command expressions, so for conditional expressions the calculation is:

$$\alpha^{\Sigma}(\mathcal{C}_{\wp}[\mathtt{if}\ be\ \mathtt{then}\ ce_1\ \mathtt{else}\ ce_2](\gamma^{r}(\rho^{\sharp}))) \sqsubseteq \ldots \triangleq \mathcal{C}^{\sharp}[\mathtt{if}\ be\ \mathtt{then}\ ce_1\ \mathtt{else}\ ce_2](\rho^{\sharp})$$

The calculation is shown in Figure 20. Steps 1–4 unfold semantic function and relation definitions; Step 5 weakens the specification through an (implicit) independent attributes abstraction; Step 6 applies a sound abstract interpreter for boolean expressions (a parameter to the calculation); Step 7 weakens the case when neither branch is valid, which would result in the returned abstract environment being bottom ($\bot$), or the empty map ($\varnothing$); Step 8 collapses adjacent abstraction and concretization functions; and Step 9 declares the final state of the calculation as the definition of the algorithm.

### *11.2 The Constructive Calculation*

The goal is now to recreate this calculation using constructive Galois connections. Up until this point, the use of powersets has been entirely restricted to describing classical specifications. However, in this classical derivation, *finite* powersets appear in the resulting algorithm. Thus, powersets served double-duty: both for classical specification and for multivalued algorithmic results. When porting to constructive Galois connections, this distinction must be made explicit in order to support extraction of a verified algorithm.

**Constructed Finite Sets** We introduce new notation to distinguish between classical powersets and constructed finite sets. We will continue to notate classical powersets as $\wp(A)$, which are modeled as downward-closed $A \searrow prop$, and introduce new notation for constructed finite sets as $\mathfrak{p}(A)$, which must be representable using a data structure such as a sorted list, binary tree, or hashed dictionary. We will continue to notate elements of powersets of posets $X : \wp(A)$ as $\{x \mid P(x)\}$, which is valid for any downward-closed proposition $P : A \searrow prop$, and introduce notation for elements of constructed finite sets ($\mathfrak{X} : \mathfrak{p}(A)$) as $\{\!\{x \mid P(x)\}\!\}$, which is valid for any *decidable* downward-closed proposition $P : A \searrow \mathbb{B}$ with finite support ($A$ finite).

$$
\begin{array}{ll}
 & \alpha^{\Sigma}(\mathcal{C}_{\wp}[\texttt{if}\ be\ \texttt{then}\ ce_1\ \texttt{else}\ ce_2](\gamma^{r}(\rho^{\sharp}))) \\
(1) & = \wr\ \text{defn. of } \mathcal{C}_{\wp}[\texttt{if}\ be\ \texttt{then}\ ce_1\ \texttt{else}\ ce_2]\ \int \\
 & \alpha^{\Sigma}(\bigcup\limits_{\rho\in\gamma^{r}(\rho^{\sharp})} \{\langle\rho',ce\rangle \mid \langle\rho,\texttt{if}\ be\ \texttt{then}\ ce_1\ \texttt{else}\ ce_2\rangle \mapsto^{c} \langle\rho',ce\rangle\}) \\
(2) & = \wr\ \text{defn. of } \langle\rho,\texttt{if}\ be\ \texttt{then}\ ce_1\ \texttt{else}\ ce_2\rangle \mapsto^{c} \langle\rho',ce'\rangle\ \int \\
 & \alpha^{\Sigma}(\bigcup\limits_{\rho\in\gamma^{r}(\rho^{\sharp})} \{\langle\rho,ce_1\rangle \mid \rho\vdash be \Downarrow^{b} true\} \cup \{\langle\rho,ce_2\rangle \mid \rho\vdash be \Downarrow^{b} false\}) \\
(3) & = \wr\ \text{defn. of } \rho\vdash be\Downarrow^{b} b\ \int \\
 & \alpha^{\Sigma}(\bigcup\limits_{\rho\in\gamma^{r}(\rho^{\sharp})} \{\langle\rho,ce_1\rangle \mid true = \mathcal{B}[be](\rho)\} \cup \{\langle\rho,ce_2\rangle \mid false = \mathcal{B}[be](\rho)\}) \\
(4) & = \wr\ \text{set equality (union commutativity)}\ \int \\
 & \alpha^{\Sigma}\left(\bigcup\left\{\begin{array}{l} \bigcup\limits_{\rho\in\gamma^{r}(\rho^{\sharp})} \{\langle\rho,ce_1\rangle \mid true = \mathcal{B}[be](\rho)\} \\ \bigcup\limits_{\rho\in\gamma^{r}(\rho^{\sharp})} \{\langle\rho,ce_2\rangle \mid false = \mathcal{B}[be](\rho)\} \end{array}\right.\right) \\
(5) & \sqsubseteq \wr\ \text{monotonicity (independent attributes)}\ \int \\
 & \alpha^{\Sigma}\left(\bigcup\left\{\begin{array}{l} \{\langle\rho,ce_1\rangle \mid \rho\in\gamma^{r}(\rho^{\sharp}) \wedge \exists\rho'.true = \mathcal{B}[be](\rho')\} \\ \{\langle\rho,ce_2\rangle \mid \rho\in\gamma^{r}(\rho^{\sharp}) \wedge \exists\rho'.false = \mathcal{B}[be](\rho')\} \end{array}\right.\right) \\
(6) & \sqsubseteq \wr\ \mathcal{B}^{\sharp}[be]\ \text{sound}\ (\mathcal{B}_{\wp}[be]\circ\gamma^{r} \sqsubseteq \gamma^{b}\circ\mathcal{B}^{\sharp}[be])\ \int \\
 & \alpha^{\Sigma}\left(\bigcup\left\{\begin{array}{lll} \{\langle\rho,ce_1\rangle \mid \rho\in\gamma^{r}(\rho^{\sharp})\} & \mathit{if} & true \sqsubseteq \mathcal{B}^{\sharp}[be](\rho^{\sharp}) \\ \{\langle\rho,ce_2\rangle \mid \rho\in\gamma^{r}(\rho^{\sharp})\} & \mathit{if} & false \sqsubseteq \mathcal{B}^{\sharp}[be](\rho^{\sharp}) \end{array}\right.\right) \\
(7) & \sqsubseteq \wr\ \text{ignore case } \neg(true \sqsubseteq \mathcal{B}^{\sharp}[be](\rho^{\sharp}) \vee false \sqsubseteq \mathcal{B}^{\sharp}[be](\rho^{\sharp}))\ \int \\
 & \left\langle \alpha^{r}(\gamma^{r}(\rho^{\sharp})), \bigcup\left\{\begin{array}{lll} \{ce_1\} & \mathit{if} & true \sqsubseteq \mathcal{B}^{\sharp}[be](\rho^{\sharp}) \\ \{ce_2\} & \mathit{if} & false \sqsubseteq \mathcal{B}^{\sharp}[be](\rho^{\sharp}) \end{array}\right.\right\rangle \\
(8) & \sqsubseteq \wr\ \alpha^{r}\circ\gamma^{r}\ \text{reductive}\ (\alpha^{r}\circ\gamma^{r} \sqsubseteq id)\ \int \\
 & \left\langle \rho^{\sharp}, \bigcup\left\{\begin{array}{lll} \{ce_1\} & \mathit{if} & true \sqsubseteq \mathcal{B}^{\sharp}[be](\rho^{\sharp}) \\ \{ce_2\} & \mathit{if} & false \sqsubseteq \mathcal{B}^{\sharp}[be](\rho^{\sharp}) \end{array}\right.\right\rangle \\
(9) & \triangleq \wr\ \text{by defining } \mathcal{C}^{\sharp}[\texttt{if}\ be\ \texttt{then}\ ce_1\ \texttt{else}\ ce_2](\rho^{\sharp}) := \left\langle \rho^{\sharp}, \bigcup\left\{\begin{array}{lll} \{ce_1\} & \mathit{if} & true \sqsubseteq \mathcal{B}^{\sharp}[be](\rho^{\sharp}) \\ \{ce_2\} & \mathit{if} & false \sqsubseteq \mathcal{B}^{\sharp}[be](\rho^{\sharp}) \end{array}\right.\right\rangle\ \int \\
 & \mathcal{C}^{\sharp}[\texttt{if}\ be\ \texttt{then}\ ce_1\ \texttt{else}\ ce_2](\rho^{\sharp}) \quad \blacksquare
\end{array}
$$

Fig. 20. Classical Calculation for Conditional Command Expressions

We relate classical powersets ($\wp(A)$) to constructive finite sets ($\mathfrak{p}(A)$) using a constructive Galois connection:

$$
\mathfrak{p}(A) \underset{\overset{\mathfrak{p}}{\eta}}{\overset{\overset{\mathfrak{p}}{\mu}}{\leftrightarrows}} \wp(A)
\qquad
\begin{array}{lll}
\overset{\mathfrak{p}}{\eta} : \mathfrak{p}(A) \nearrow \wp(A) & & \overset{\mathfrak{p}}{\eta}(\mathfrak{X}) := \{x \mid x \in X\} \\
\overset{\mathfrak{p}}{\mu} : \wp(A) \nearrow \wp(\mathfrak{p}(A)) & & \overset{\mathfrak{p}}{\mu}(X) := \{\mathfrak{X} \mid \forall x.x \in X \Leftrightarrow x \in \mathfrak{X}\}
\end{array}
$$

and define a singleton abstraction for constructive finite sets:

$$
A \underset{\overset{1\mathfrak{p}}{\eta}}{\overset{\overset{1\mathfrak{p}}{\mu}}{\leftrightarrows}} \mathfrak{p}(A)
\qquad
\begin{array}{lll}
\overset{1\mathfrak{p}}{\eta} : A \nearrow \mathfrak{p}(A) & & \overset{1\mathfrak{p}}{\eta}(x) := \{\!\{x\}\!\} \\
\overset{1\mathfrak{p}}{\mu} : \mathfrak{p}(A) \nearrow \wp(A) & & \overset{1\mathfrak{p}}{\mu}(\mathfrak{X}) := \{x \mid x \in \mathfrak{X}\}
\end{array}
$$

Finally, we redefine abstract configurations ($\varsigma^\sharp \in \Sigma^\sharp$) to use constructive finite sets:

$$\varsigma^\sharp \in \Sigma^\sharp \coloneqq \mathsf{env}^\sharp \times \mathfrak{p}(\mathsf{cexp})$$

In this new setting for abstract configurations, the constructive Galois connection for concrete configurations ($\varsigma \in \Sigma$) is:

$$\Sigma \underset{\overset{r\times 1\mathfrak{p}}{\eta}}{\overset{\overset{r\times 1}{\mu}}{\leftrightarrows}} \Sigma^\sharp \qquad \begin{aligned} \overset{r\times 1\mathfrak{p}}{\eta} &: \Sigma \rightarrow \Sigma^\sharp \\ \overset{r\times 1\mathfrak{p}}{\mu} &: \Sigma^\sharp \nearrow \wp(\Sigma) \end{aligned}$$

$$\begin{aligned} \overset{r\times 1\mathfrak{p}}{\eta}(\rho, ce) &\coloneqq \langle \eta^r(\rho), \{\!\{ce\}\!\}\rangle \\ \overset{r\times 1\mathfrak{p}}{\mu}(\rho^\sharp, CE) &\coloneqq \{\langle \rho, ce\rangle \mid \rho \in \mu^r(\rho^\sharp) \wedge cd \in CE\} \end{aligned}$$

Using constructive finite sets and this new definition for abstract configurations, we will perform the same calculation as before, but entirely within the constructive Galois connection framework, and in abstraction-directed form.

**The Calculation** We show the calculation for the abstract interpretation of conditional expressions using constructive Galois connections in Figure 21. Steps 1–3 unfold semantic function and relation definitions; Step 4 applies commutativity of set union; Step 5 pushes abstraction through the set comprehension; Step 6 introduces adjacent concretization and abstraction functions, justified by Galois connection expansiveness (an explicit loss in precision); Step 7 applies the constructive Galois connection correspondence; Step 8 applies a sound abstract interpreter for boolean expressions; Step 9 pulls abstraction out of the set comprehension; Step 10 collapses adjacent abstraction and concretization functions; and Step 11 declares the final state of the calculation as the definition of the algorithm.

What this calculation shows is that constructive Galois connections support manipulating multivalued abstractions and algorithms, *via* an explicit finite set construction, which carries algorithmic content in a constructive logic setting. What classically was just a powerset with finite elements becomes an explicit finite set, and what classically was an undecidable specification of potentially infinite elements remains a powerset. Supporting relational abstraction can be done in this way as well, for example a relational abstraction for environments would have the shape of $\overset{rel}{\eta^r} : \mathfrak{p}(env) \nearrow env^\sharp$ and $\overset{rel}{\mu^r} : env^\sharp \nearrow \wp(\mathfrak{p}(env))$.

## 12 Related Work

This work connects two long strands of research: abstract interpretation *via* Galois connections and mechanized verification *via* dependently typed functional programming. The former is founded on the pioneering work of Cousot & Cousot (1977; 1979); the latter on that of Martin-Löf (1984), embodied in Norell's Agda (2007). Our key technical insight is to use a monadic structure for Galois connections, following the example of Moggi (1989) for the $\lambda$-calculus.

$$\lfloor \eta^{\Sigma} \rfloor^{*}(\mathcal{C}[\texttt{if}\ be\ \texttt{then}\ ce_1\ \texttt{else}\ ce_2]^{*}(\mu^{r}(\rho^{\sharp})))$$

$$(1)\quad = \wr\ \text{defn. of } \mathcal{C}[\texttt{if}\ be\ \texttt{then}\ ce_1\ \texttt{else}\ ce_2]\ \int$$

$$\lfloor \eta^{\Sigma} \rfloor^{*}\Big(\bigcup_{\rho \in \mu^{r}(\rho^{\sharp})} \{\langle \rho', ce\rangle \mid \langle \rho, \texttt{if}\ be\ \texttt{then}\ ce_1\ \texttt{else}\ ce_2\rangle \mapsto^{c} \langle \rho', ce\rangle\}\Big)$$

$$(2)\quad = \wr\ \text{defn. of } \langle \rho, \texttt{if}\ be\ \texttt{then}\ ce_1\ \texttt{else}\ ce_2\rangle \mapsto^{c} \langle \rho', ce'\rangle\ \int$$

$$\lfloor \eta^{\Sigma} \rfloor^{*}\Big(\bigcup_{\rho \in \mu^{r}(\rho^{\sharp})} \{\langle \rho, ce_1\rangle \mid \rho \vdash be \Downarrow^{b} true\} \cup \{\langle \rho, ce_2\rangle \mid \rho \vdash be \Downarrow^{b} false\}\Big)$$

$$(3)\quad = \wr\ \text{defn. of } \rho \vdash be \Downarrow^{b} b\ \int$$

$$\lfloor \eta^{\Sigma} \rfloor^{*}\Big(\bigcup_{\rho \in \mu^{r}(\rho^{\sharp})} \{\langle \rho, ce_1\rangle \mid true = \mathcal{B}[be](\rho)\} \cup \{\langle \rho, ce_2\rangle \mid false = \mathcal{B}[be](\rho)\}\Big)$$

$$(4)\quad = \wr\ \text{set equality (union commutativity)}\ \int$$

$$\lfloor \eta^{\Sigma} \rfloor^{*}\left(\bigcup \begin{cases} \bigcup_{\rho \in \mu^{r}(\rho^{\sharp})} \{\langle \rho, ce_1\rangle \mid true = \mathcal{B}[be](\rho)\} \\ \bigcup_{\rho \in \mu^{r}(\rho^{\sharp})} \{\langle \rho, ce_2\rangle \mid false = \mathcal{B}[be](\rho)\} \end{cases}\right)$$

$$(5)\quad = \wr\ \text{set equality}\ \int$$

$$\bigcup \begin{cases} \bigcup_{\rho \in \mu^{r}(\rho^{\sharp})} \{\langle \eta^{r}(\rho), \{\!\!\{ce_1\}\!\!\}\rangle \mid true = \mathcal{B}[be](\rho)\} \\ \bigcup_{\rho \in \mu^{r}(\rho^{\sharp})} \{\langle \eta^{r}(\rho), \{\!\!\{ce_2\}\!\!\}\rangle \mid false = \mathcal{B}[be](\rho)\} \end{cases}$$

$$(6)\quad \sqsubseteq \wr\ \mu^{b} \circledast \lfloor \eta^{b} \rfloor \text{ expansive } (ret \sqsubseteq \mu^{b} \circledast \lfloor \eta^{b} \rfloor)\ \int$$

$$\bigcup \begin{cases} \bigcup_{\rho \in \mu^{r}(\rho^{\sharp})} \{\langle \eta^{r}(\rho), \{\!\!\{ce_1\}\!\!\}\rangle \mid true \in \mu^{b}(\eta^{b}(\mathcal{B}[be](\rho)))\} \\ \bigcup_{\rho \in \mu^{r}(\rho^{\sharp})} \{\langle \eta^{r}(\rho), \{\!\!\{ce_2\}\!\!\}\rangle \mid false \in \mu^{b}(\eta^{b}(\mathcal{B}[be](\rho)))\} \end{cases}$$

$$(7)\quad = \wr\ \text{constructive GC correspondence } (b \in \mu^{b}(b^{\sharp}) \Leftrightarrow \eta^{b}(b) \sqsubseteq b^{\sharp})\ \int$$

$$\bigcup \begin{cases} \bigcup_{\rho \in \mu^{r}(\rho^{\sharp})} \{\langle \eta^{r}(\rho), \{\!\!\{ce_1\}\!\!\}\rangle \mid true \sqsubseteq \eta^{b}(\mathcal{B}[be](\rho))\} \\ \bigcup_{\rho \in \mu^{r}(\rho^{\sharp})} \{\langle \eta^{r}(\rho), \{\!\!\{ce_2\}\!\!\}\rangle \mid false \sqsubseteq \eta^{b}(\mathcal{B}[be](\rho))\} \end{cases}$$

$$(8)\quad \sqsubseteq \wr\ \mathcal{B}^{\sharp}[_] \text{ sound } (\eta^{b} \circ \mathcal{B}[be] \sqsubseteq \mathcal{B}^{\sharp}[be] \circ \eta^{r})\ \int$$

$$\bigcup \begin{cases} \bigcup_{\rho \in \mu^{r}(\rho^{\sharp})} \{\langle \eta^{r}(\rho), \{\!\!\{ce_1\}\!\!\}\rangle \mid true \sqsubseteq \mathcal{B}^{\sharp}[be](\eta^{r}(\rho))\} \\ \bigcup_{\rho \in \mu^{r}(\rho^{\sharp})} \{\langle \eta^{r}(\rho), \{\!\!\{ce_2\}\!\!\}\rangle \mid false \sqsubseteq \mathcal{B}^{\sharp}[be](\eta^{r}(\rho))\} \end{cases}$$

$$(9)\quad = \wr\ \text{set equality}\ \int$$

$$\bigcup \begin{cases} \bigcup_{\rho^{\sharp\prime} \in \lfloor \eta^{r} \rfloor^{*} \mu^{r}(\rho^{\sharp})} \{\langle \rho^{\sharp\prime}, \{\!\!\{ce_1\}\!\!\}\rangle \mid true \sqsubseteq \mathcal{B}^{\sharp}[be](\rho^{\sharp\prime})\} \\ \bigcup_{\rho^{\sharp\prime} \in \lfloor \eta^{r} \rfloor^{*} \mu^{r}(\rho^{\sharp})} \{\langle \rho^{\sharp\prime}, \{\!\!\{ce_2\}\!\!\}\rangle \mid false \sqsubseteq \mathcal{B}^{\sharp}[be](\rho^{\sharp\prime})\} \end{cases}$$

$$(10)\quad \sqsubseteq \wr\ \lfloor \eta^{r} \rfloor \circledast \mu^{r} \text{ reductive } (\lfloor \eta^{r} \rfloor \circledast \mu^{b} \sqsubseteq ret)\ \int$$

$$\left\{\left\langle \rho^{\sharp}, \bigcup \begin{cases} \{\!\!\{ce_1\}\!\!\} & \textit{if} \quad true \sqsubseteq \mathcal{B}^{\sharp}[be](\rho^{\sharp}) \\ \{\!\!\{ce_2\}\!\!\} & \textit{if} \quad false \sqsubseteq \mathcal{B}^{\sharp}[be](\rho^{\sharp}) \end{cases} \right\rangle\right\}$$

$$(11)\quad \triangleq \wr\ \text{by defining } \mathcal{C}^{\sharp}[\texttt{if}\ be\ \texttt{then}\ ce_1\ \texttt{else}\ ce_2](\rho^{\sharp}) := \left\langle \rho^{\sharp}, \bigcup \begin{cases} \{\!\!\{ce_1\}\!\!\} & \textit{if} \quad true \sqsubseteq \mathcal{B}^{\sharp}[be](\rho^{\sharp}) \\ \{\!\!\{ce_2\}\!\!\} & \textit{if} \quad false \sqsubseteq \mathcal{B}^{\sharp}[be](\rho^{\sharp}) \end{cases} \right\rangle\ \int$$

$$\lfloor \mathcal{C}^{\sharp}[\texttt{if}\ be\ \texttt{then}\ ce_1\ \texttt{else}\ ce_2] \rfloor(\rho^{\sharp})$$

Fig. 21. Conditional Expressions Constructive Calculation

**Calculational Abstract Interpretation** Cousot describes calculational abstract interpretation by example in his lecture notes (2005) and monograph (1999), and Cousot & Cousot recently introduced a unifying Galois connection calculus (2004). Our work mechanizes Cousot's calculations and provides a foundation for mechanizing other instances of calculational abstract interpretation (*e.g.*, (Midtgaard & Jensen, 2008; Sergey *et al.*, 2012)). We expect our work to have applications to the mechanization of calculational program design (Bird & de Moor, 1996; Bird, 1990) by employing only Galois *retractions*, *i.e.*, $\alpha \circ \gamma$ is an identity (Cousot & Cousot, 2014). There is prior work on mechanized program calculation (Tesson *et al.*, 2011), but it is not based on abstract interpretation.

**Verified Static Analyzers** Efforts in verified abstract interpretation have shown many promising results (Pichardie, 2005; Cachera & Pichardie, 2010; Blazy *et al.*, 2013; Barthe *et al.*, 2007), scaling up to large-scale real-world static analyzers (Jourdan *et al.*, 2015). However, mechanized abstract interpretation has yet to benefit from the Galois connection framework. Until now, approaches use classical axioms or "$\gamma$-only" encodings of soundness and (sometimes) completeness. Our techniques for mechanizing Galois connections should complement these approaches.

**Galculator** The Galculator (Silva & Oliveira, 2008) is a proof assistant founded on an algebra of Galois connections. This tool is similar to ours in that it mechanically verifies Galois connection calculations. Our approach is more general, supporting arbitrary set-theoretic reasoning and embedded within a general purpose proof assistant, however their approach is fully automated for the small set of derivations which reside within their supported theory.

**Deductive Synthesis** Fiat (Delaware *et al.*, 2015) is a library for the Coq proof assistant which supports semi-automated synthesis of programs as refinements of their specifications. Fiat uses the same powerset type and monad as we do, and their "deductive synthesis" process similarly derives correct-by-construction programs by calculus. Fiat derivations start with a user-defined specification and calculate towards an *under*-approximation ($\sqsupseteq$), whereas calculational abstract interpretation starts with an optimal specification and calculates towards an *over*-approximation ($\sqsubseteq$). It should be possible to generalize their framework to use partial orders to recover aspects of our work, or to invert the lattice used in our abstract interpretation framework to recover aspects of theirs. A notable difference in approach is that Fiat makes heavy use of Coq's tactic programming language to automate rewrites inside respectful contexts, whereas our system provides no interactive proof automation and each calculational step must be notated explicitly.

**Monadic Abstract Interpretation** Monads in abstract interpretation have recently been applied to good effect for modularity (Sergey *et al.*, 2013; Darais *et al.*, 2015). However, that work uses monads to structure the semantics, not the Galois connections and proofs.

**Future Directions** Now that we have established a foundation for constructive Galois connections, we see value in verifying larger calculations (*e.g.*, Midtgaard & Jensen (2008); Sergey *et al* (2012)). Furthermore we would like to explore whether or not our techniques have any benefit in the space of general-purpose program calculations *à la* Bird.

Currently our framework requires the user to justify every detail of the program calculation, including monotonicity proofs and proof scoping for rewrites inside monotonic contexts. We imagine much of this can be automated, requiring the user to only provide the interesting parts of the proof, *à la* Fiat (Delaware *et al.*, 2015). Our experience has been that even Coq's tactic system slows down considerably when automating all of these details, and we foresee using proof by reflection in either Coq (*e.g.*, Rtac (Malecha & Bengtson, 2016)) or Agda to automate these proofs in a way that maintains proof-checker performance.

There have been recent developments on compositional abstract interpretation frameworks (Darais *et al.*, 2015) where abstract interpreters and their proofs of soundness are systematically derived side-by-side. That framework relies on correctness properties transported by *Galois transformers*, which we posit would benefit from mechanization since they hold both computational and specification content.

## 13 Perspectives on Foundations

In this paper we present *a* foundation for constructive Galois connections, but certainly not *the* foundation for constructive Galois connections. Just as the classical Galois connection framework is an instantiation of the more general framework of adjunctions between functors, our constructive (and Kleisli) Galois connection setup can also be seen as an instantiation more general category-theoretic definitions.

To generalize our framework, monotonic functions ($f : A \nearrow B$) become functors ($f : \mathcal{A} \twoheadrightarrow \mathcal{B}$), powersets ($X : \wp(A)$) become presheaves ($X : \mathcal{A}^{op} \twoheadrightarrow Set$), and monotonic powerset-monadic functions ($f : A \nearrow \wp(B)$) generalize to profunctors ($f : \mathcal{B}^{op} \times \mathcal{A} \twoheadrightarrow Set$), or equivalently functors into presheaves ($f : \mathcal{A} \twoheadrightarrow \mathcal{B}^{op} \twoheadrightarrow Set$). The fact that any functor $F : \mathcal{A} \twoheadrightarrow \mathcal{B}$ induces adjoint profunctors $L \dashv R$ where $L(b,a) \mapsto hom(b, F(a))$ and $R(a,b) \mapsto hom(F(a), b)$ is well known, and corresponds to our lifting of $\eta : A \nearrow B$ to a Kleisli Galois connection $\kappa\alpha \dashv \kappa\gamma$ with Kleisli abstraction function $\kappa\alpha(x) := \{y \mid y \sqsubseteq \eta(x)\}$ and (inverse-image) induced Kleisli concretization function $\kappa\gamma(y) := \{x \mid \eta(x) \sqsubseteq y\}$. However, it is not clear to the authors what general conditions on categories is required to to recover our proof of constructive isomorphism between constructive and Kleisli Galois connections. It has been suggested by Max New[2] that the necessary restriction is for the base categories to be Cauchy complete; however this warrants further investigation (in particular its amenability to mechanized verification with program extraction) in future work.

[2] http://prl.ccs.neu.edu/blog/2016/11/16/understanding-constructive-galois-connections/

## 14 Conclusions

This paper realizes the vision of mechanized and constructive Galois connections foreshadowed by Cousot (1999, p. 85), giving the first mechanically verified proof by calculational abstract interpretation; once for his generic static analyzer and once for the semantics of gradual typing. Our proofs by calculus closely follow the originals. The primary discrepancy is the use of monads to isolate *specification effects*. By maintaining this discipline, we are able to verify calculations by Galois connections *and* extract computational content from pure results. The resulting artifacts are correct-by-verified-construction, thereby avoiding known bugs in the original.[3]

### *Acknowledgements*

We thank Ron Garcia and Éric Tanter for discussions of their work. Éric also helped with our French translation. We thank the Colony Club in D.C. and the Board & Brew in College Park for providing fruitful environments in which to work. We thank the anonymous reviewers of ICFP 2016 and JFP, and the first author's thesis committee for their helpful feedback. This material is partially based on research sponsored by DARPA under agreement number AFRL FA8750-15-2-0104.

[3] `http://www.di.ens.fr/~cousot/aisoftware/Marktoberdorf98/Bug_History`